\documentclass[prx,aps,twocolumn,reprint,amsmath,floatfix,amssymb,showpacs,superscriptaddress]{revtex4-1}

\usepackage{xfrac}
\usepackage[normalem]{ulem}
\usepackage{amsmath}
\usepackage{booktabs}
\usepackage{array}
\usepackage{ulem}

\usepackage{microtype}
\usepackage{wasysym}
\usepackage[varg]{txfonts}
\usepackage[percent]{overpic}
\usepackage{mathrsfs}
\usepackage{braket}
\usepackage{titlesec}
\usepackage{bm}
\usepackage{comment}
\usepackage{verbatim}
\usepackage{float}
\usepackage{tikz}
\usetikzlibrary{trees}

\usepackage{array}
\newcommand{\PreserveBackslash}[1]{\let\temp=\\#1\let\\=\temp}
\newcolumntype{C}[1]{>{\PreserveBackslash\centering}p{#1}}
\newcolumntype{R}[1]{>{\PreserveBackslash\raggedleft}p{#1}}
\newcolumntype{L}[1]{>{\PreserveBackslash\raggedright}p{#1}}

\usepackage{color}
\usepackage[colorlinks,bookmarks=false,citecolor=blue,linkcolor=blue,urlcolor=blue,pdfstartview=FitH]{hyperref}

\definecolor{darkred}{rgb}{0.7,0.0,0.0}

\definecolor{darkblue}{rgb}{0,0.02,0.45}

\definecolor{darkgreen}{rgb}{0.02,0.45,0.0}

\definecolor{violet}{rgb}{0.8,0.2,0.6}

\usepackage{amsmath}
\usepackage{amssymb}
\usepackage{graphicx}
\usepackage{subfigure}
\usepackage{pict2e}
\usepackage{multirow}
\usepackage{cancel}

\newcommand{\be}{\begin{equation}}
\newcommand{\ee}{\end{equation}}
\newcommand{\bea}{\begin{eqnarray}}
\newcommand{\eea}{\end{eqnarray}}

\newcommand{\sbe}{\small\begin{equation}}
\newcommand{\see}{\end{equation}\normalsize}
\newcommand{\sbea}{\small\begin{eqnarray}}
\newcommand{\seea}{\end{eqnarray}\normalsize}
\graphicspath{{Figures/}}

\def\bs{\boldsymbol}
\def\wt{\widetilde}
\def\vec{\mathbf}
\def\mc{\mathcal}

\begin{document}


\title{Non-Loudon-Fleury Raman scattering in spin-orbit coupled Mott insulators}

\author{Yang Yang}
\affiliation{School of Physics and Astronomy, University of Minnesota, Minneapolis, Minnesota 55455, USA}

\author{Mengqun Li}
\affiliation{School of Physics and Astronomy, University of Minnesota, Minneapolis, Minnesota 55455, USA}

\author{Ioannis Rousochatzakis}
\affiliation{Department of Physics, Loughborough University, Loughborough LE11 3TU, United Kingdom}

\author{Natalia B. Perkins}
\affiliation{School of Physics and Astronomy, University of Minnesota, Minneapolis, Minnesota 55455, USA}
\date{\today}

\begin{abstract}

We revisit the theory of magnetic Raman scattering in Mott insulators with strong spin-orbit coupling, with a major focus on Kitaev materials.   We show that Kitaev materials with bond-anisotropic interactions  are generally expected to show both one- and two-magnon responses. 
It is further shown that, in order to obtain the correct
leading contributions to the Raman vertex operator $\mc{R}$, one must take into account the precise, photon-assisted  microscopic hopping processes of the electrons and that, in systems with multiple hopping paths, $\mc{R}$ contains terms beyond those appearing in the traditional Loudon-Fleury theory.
Most saliently, a numerical implementation of the revised formalism to the case of the three-dimensional hyperhoneycomb Kitaev material $\beta$-Li$_2$IrO$_3$ reveals that the non-Loudon-Fleury scattering terms actually dominate the Raman intensity.  In addition, they induce a qualitative  modification of the polarization dependence, including, e.g., the emergence of a sharp one-magnon peak at low energies which is not expected in the traditional Loudon-Fleury theory. 
%
%
This peak is shown to arise from microscopic photon-assisted tunneling processes that are of similar type with the ones leading to the symmetric off-diagonal interaction $\Gamma$  (known to be present in many Kitaev materials), but take the form of a bond-directional magnetic dipole term in the Raman vertex.
These results are expected to apply across all Kitaev materials and mark a drastic change of paradigm for the understanding of Raman scattering in materials with strong spin-orbit coupling and multiple exchange paths.

\end{abstract}

\maketitle

\vspace*{-1cm}

\section{Introduction}
Raman scattering has proven to be a powerful experimental technique to understand and characterize the physics of strongly correlated systems~\cite{Devereaux2007}. Being a sensitive probe to single- and multi-particle  excitations over sufficiently wide ranges of temperatures and energies, Raman scattering has played an important role in elucidating ground state properties, symmetry and statistics of magnetic excitations, as well as the strength and nature of the exchange couplings in magnetic insulators with both magnetically ordered and spin liquid  ground states~\cite{Shastry1990,Shastry1991,Chubukov1995,Frenkel1995,Benfatto2006,Cepas2008a,Ko2010,Wulferding2010,Perkins2008,Perkins2013,Sen2019,Sandilands2015,Sandilands2016,Gretarsson2016,Glamazda2016,Sahasrabudhe2020,Dirk2020,Yiping2020,Knolle2014,Brent2015,Brent2016-short,Brent2016-long,Fu2017,Rousochatzaki2019}. 
In recent years, there has been a series of Raman studies (both experimental and theoretical) on a range of spin-orbit coupled (SOC) Mott insulators, with a view to elucidate the nature of their magnetic excitations (and lattice dynamics) and their proximity to the so-called Kitaev quantum spin liquid ground states~\cite{BookCao,Krempa2014,Rau2016,Sandilands2015,Sandilands2016,Nasu2016,Ulrich2015,Gim2016,Gretarsson2017,Souliou2017,Gretarsson2016,Glamazda2016,Sahasrabudhe2020,Dirk2020,Yiping2020,Knolle2014,Brent2015,Brent2016-short,Brent2016-long,Fu2017,Rousochatzaki2019,Metavitsiadis2021}. 
Most saliently, the reported Raman scattering data in the Kitaev candidate materials $\alpha$-RuCl$_3$  \cite{Sandilands2015,Sandilands2016,Yiping2020,Sahasrabudhe2020,Dirk2020} and the three-dimensional (3D) iridates $\beta$-Li$_2$IrO$_3$ and $\gamma$-Li$_2$IrO$_3$ \cite{Glamazda2016},  revealed signatures of both  multi-particle continua, characteristics of the proximate spin liquid phase, and sharp peaks, characteristic of magnon excitations of the low-temperature ordered phases. 
These results call for a close re-examination of the Raman scattering theory applied to strong spin-orbit coupled Mott insulators.

The history of understanding of the magnetic Raman scattering goes back to the seminal paper by Fleury and Loudon ~\cite{LF1968},  in which  they have identified  three main mechanisms for the coupling between light and magnetic excitations:
(i) direct coupling of photon  to  magnon through magnetic-dipole interaction, 
(ii) indirect electric-dipole coupling which mixes the spin and orbital motion of the electrons, and 
(iii) second-order electric-dipole coupling which is very similar to an exchange mechanism. 
The first mechanism (i) is very weak and is usually neglected.
The second (ii) is the Elliott-Loudon mechanism~\cite{Elliott1963}, in which the Raman scattering from the magnetic degrees of freedom on a single ion proceeds via a pair of allowed electric-dipole transitions through a spin-orbit active intermediate state.   In this process, the incident light excites an electron from the ground state to an excited state keeping the $z$-component of the spin unchanged. The spin states with different $z$-components are then mixed in the excited state via the spin-orbit coupling, and a transition back to the ground state but with opposite spin polarization can occur by emitting a Raman photon and a magnon with $\Delta S^z=\pm1$. Traditionally, the Elliott-Loudon process  is considered to be the main source of the one-magnon scattering response, and this is indeed the case in systems with weak SOC. 
The process (ii) also gives rise to two-magnon scattering, but its intensity is several orders of magnitude weaker compared to the one-magnon process. 

This brings us to the third mechanism (iii), which is the  exchange-scattering mechanism described by the well-known Loudon-Fleury theory of magnetic Raman scattering in Mott insulators~\cite{LF1968}. 
The basic idea of this theory is that the processes leading to the Raman response from Mott insulators are similar to those leading to the exchange interactions, except that the virtual electron hopping is (partly) assisted by photons. 
Consequently, the Loudon-Fleury Raman operator is proportional to the sum over the individual spin-exchange interactions, weighted by bond-specific, polarization-dependent factors that determine the ability of  photons to control the magnitude of the associated electron hopping~\cite{LF1968,Shastry1990,Shastry1991,Devereaux2007,Perkins2008,Perkins2013,Ko2010,Knolle2014,Fu2017}.
Traditionally, it is considered that the processes involved in this mechanism  contribute predominantly to the two-magnon scattering with $\Delta S^z=0$, in which a pair of magnons is created or destroyed. This perception follows in part from a concluding remark in the original paper of Loudon and Fleury~\cite{LF1968} that `the exchange mechanism discussed here (being proportional to $S_i^+S_j^-$) produces magnons in pairs and hence there is no exchange-scattering mechanism for one-magnon scattering'.
Now we understand that this statement is certainly far from being  general and, in particular, it doesn't  apply to the SOC Mott insulators with bond-dependent anisotropic interactions, which naturally give rise to one-magnon response \cite{Ulrich2015,Gim2016,Gretarsson2017,Souliou2017,Dirk2020,Sahasrabudhe2020,Yipingunpublished}.

Here we show that in the SOC Mott insulators, the  exchange-scattering mechanism (iii) leads to essential contributions beyond the Loudon-Fleury theory, and these non-Loudon-Fleury terms can give rise to a significant one-magnon Raman response, on top of the two-magnon response.
Quite remarkably, our numerical calculations for the representative case of $\beta$-Li$_2$IrO$_3$ shows that the Raman intensity (both in the one- and the two-magnon channels) is actually dominated by the contribution from the non-Loudon-Fleury terms by at least two orders of magnitude. In addition, these terms give rise to a qualitative modification of the scattering intensity including its polarization dependence. These include a distinctive, one-magnon low-energy peak in the ${\bf ac}$ polarization channel \cite{Yipingunpublished}, which is not expected in the traditional Loudon-Fleury theory. As we discuss below, similar results are expected across {\it all Kitaev materials}, given that they all share the same local geometry of virtual exchange paths and the same order of magnitude of microscopic hopping and interaction parameters. In this sense, the theoretical framework presented below calls for a general re-evaluation of Raman scattering in Kitaev materials of current interest and systems with strong spin-orbit coupling and multiple exchange paths more generally.

The remaining of the paper is organised as follows.   In Sec.~\ref{sec:generic}, we begin with a brief discussion of the relevant materials -- strong SOC Mott insulators in which the magnetic moment $j_{\text{eff}}\!=\!1/2$ comes from the five electrons (or one hole) on the $t_{2g}$ orbitals -- and their effective low-energy description.
In Sec.~\ref{sec:derivation}, we present the main steps of the $\mc{T}$-matrix formalism that lead to the microscopic  derivation of the Raman operator $\mc{R}$. 
In Sec.~\ref{sec:RamanKitaev},
we apply this framework to Kitaev materials, and establish the leading contributions to the Raman vertex $\mc{R}$ from the same microscopic processes that give rise to the minimal $J$-$K$-$\Gamma$ model. These include processes arising from direct hopping,  ligand-mediated hopping as well as processes involving both direct and ligand-mediated hopping. We then establish that the later two types of processes are the ones giving rise to the strong non-Loudon-Fleury contributions to the Raman operator.
In Sec.~\ref{sec:bosonic}, we
proceed with the application to magnetically ordered states. To that end, we express $\mc{R}$ in terms of magnon operators and obtain the expressions for the one- and two-magnon Raman intensities.
The numerical implementation of this theory to the three-dimensional Kitaev magnet $\beta$-Li${}_2$IrO${}_3$ is then presented in  Sec.~\ref{sec:Ramanbeta}, where it is demonstrated that the non-Loudon-Fleury terms dominate the Raman intensity by at least two orders of magnitude. 

Section~\ref{sec:discussion} provides a brief summary along with a general perspective of the results.  Some of the technical details and auxiliary information are relegated to App.~\ref{App:Ramandetails}.

\section{Relevant materials \& generic low-energy description}\label{sec:generic}

The theory developed below applies to SOC Mott insulators, such as the Kitaev materials with Ir${}^{4+}$ and Ru${}^{3+}$ ions, in which the magnetic moment $j_{\text{eff}}\!=\!1/2$ comes from the five electrons (or one hole) on the $t_{2g}$ orbitals due to the strong SOC~\cite{Rau2016,Trebst2017,BookCao,Takagi2019,Motome2020a}. 
These include, e.g., the layered compounds Na$_2$IrO$_3$~\cite{Singh2010,Chun2015}, $\alpha$-Li$_2$IrO$_3$~\cite{Williams2016}, and $\alpha$-RuCl$_3$~\cite{Plumb2014,Sears2015,Banerjee2017,Kasahara2018}, as well as the three-dimensional (3D) iridates $\beta$-Li$_2$IrO$_3$~\cite{Biffin2014a,Ruiz2017,Majumder2019} and $\gamma$-Li$_2$IrO$_3$~\cite{Biffin2014b,Modic2014}, for which most of the experimental Raman data has been reported so far.

The minimal electronic Hamiltonian of such SOC Mott insulators contains the following terms: 
\begin{align}
\mc{H}=\mc{H}_{\text{int}}+\mc{H}_{\text{pd}}+\mc{H}_{\text{SOC}}+\mc{H}_t,
\end{align}
where $\mc{H}_{\text{int}}$ is the interaction part of the three-orbital Hubbard Hamiltonian, $\mc{H}_{\text{pd}}=
 \Delta_{pd}\sum_{i,\sigma}n_{i\sigma}$ is the charge-transfer Hamiltonian (where $\Delta_{pd}$ stands for the charge-transfer energy of one electron from the magnetic ion to the ligand ion), and $\mc{H}_{\text{SOC}}$ described the onsite SOC,
and $\mc{H}_t$ stands for the hopping.
Specifically,
\begin{align}
\mc{H}_{\text{int}}=\sum_{i}&\left(U_1\sum_\alpha n_{i\alpha\uparrow}n_{i\alpha\downarrow}+\frac{1}{2}(U_2-J_H)\sum_{\alpha\neq\alpha',\sigma}n_{i\alpha\sigma}n_{i\alpha'\sigma}\right.\nonumber\\+&U_2\sum_{\alpha\neq\alpha'}n_{i\alpha\uparrow}n_{i\alpha'\downarrow}+J_H\sum_{\alpha\neq\alpha'}d_{i\alpha\uparrow}^\dagger d_{i\alpha\downarrow}^\dagger d_{i\alpha'\downarrow}d_{i\alpha'\uparrow}\nonumber\\-&\left. J_H\sum_{\alpha\neq\alpha'}d_{i\alpha\uparrow}^\dagger d_{i\alpha\downarrow}d_{i\alpha'\downarrow}^\dagger d_{i\alpha'\uparrow}\right)\,,
\end{align}
where $d_{i\alpha\sigma}^{\dagger}$  denotes the  creation operator of the $d$-electron 
of the magnetic ion on the $t_{2g}$ orbitals $\alpha=xy\, (Z),yz\, (X),zx\, (Y)$
(in the local axes bound to the oxygen octahedron) with spin $\sigma=\uparrow,\downarrow$.
The constants $U_1$ and $U_2$ denote the Coulomb repulsion among $d$-electrons on the same and on the different $t_{2g}$ orbitals, respectively, $J_H$ denotes the Hund's coupling constant, and $U_1=U_2+2J_H$, due to the cubic symmetry.
The spin-orbit coupling (SOC) is given by 
\be
\mc{H}_{\text{SOC}}=\lambda\sum_i {\bf s}_i\cdot\vec{l}_i\,,
\ee
where ${\bf s}_i$ is the spin of  the $i$-th electron, and  the SOC has been projected into the $t_{2g}$  manifold, leading to the effective orbital angular momentum  $l=1$.
Finally, the hopping term $\mc{H}_t$ has the general form
\begin{align}
{\mc H}_t =& \sum_{i,j}
\sum_{\alpha,\beta}\sum_{\sigma} t_{ij,\sigma}^{\alpha\beta}
d_{i\alpha\sigma}^{\dagger}d^{}_{j\beta\sigma}\nonumber\\
&+
\widetilde{\sum_{i,j}}
\sum_{\alpha,\beta}\sum_{\sigma}\Big[ \tilde{t}_{ij,\sigma}^{\alpha\beta}
d_{i\alpha\sigma}^{\dagger} p^{}_{j\beta\sigma} + \text{h.c.}\Big]\,,
\end{align}
where the first line gives the direct hopping between magnetic ions, and the second line gives the hopping between magnetic ions and ligand ions, with $p_{j\beta\sigma}$ denoting the annihilation of an electron on the $\beta$-th $p$-orbital of the ligand ion at site $j$. The hopping amplitudes $t_{ij,\sigma}^{\alpha\beta}$ and $\tilde{t}_{ij,\sigma}^{\alpha\beta}$ are determined by the overlaps between the orbitals and are material dependent. 

A technical comment is in order here. Usually, the hole picture (in which  the magnetic degrees of freedom come from the one hole states in the $j_{\text{eff}}\!=\!1/2$ doublets) is used for the description of the magnetic properties of the Kitaev materials, which are in the main focus of this paper. To change all the formulas to the hole picture, one can replace   $d^\dagger_{i\alpha\sigma}$ with $\tilde{d}_{i\alpha\sigma}$, $d_{i\alpha\sigma}$ with $\tilde{d}^\dagger_{i\alpha\sigma}$, and $n_{i\alpha\sigma}$ with $1-n_{i\alpha\sigma}$. With these substitutions, the eigenenergies remain the same (up to the constant energy shift), but we obtain an overall negative sign on each hopping amplitude. Therefore, to keep the above formalism unchanged for the hole picture we simply absorb this negative sign in the hopping parameters. In the following we will omit the tilde of the hole operators for a simpler notation.

A standard super-exchange expansion \footnote{In case of Kitaev materials, the super-exchange expansion  usually does not include processes when two holes meet at the same oxygen site in the intermediate state since these  intermediate states are higher energy states and their inclusion  only slightly modifies the effective couplings but does not change the picture qualitatively. We note, however, that in systems with small spin-orbit coupling, such as the cuprates,  such processes can also lead to small anisotropic interactions \cite{Yushankhai1999}.} of the above extended Hubbard model delivers a low-energy effective spin Hamiltonian that, for Kitaev materials such as $\beta$-Li$_2$IrO$_3$, can be well described by the nearest-neighbor (NN) $J$-$K$-$\Gamma$ model,
\be\label{eqn:Heff}
\mc{H}_{\text{eff}}\!=\!\sum_{\langle ij\rangle_\nu}
J\,\vec{S}_i\cdot\vec{S}_j
\!+\!
K\, S_i^{\alpha_\nu}S_j^{\alpha_\nu}
\!+\!\sigma_\nu~\Gamma\,(S_i^{\beta_\nu}S_j^{\gamma_\nu}\!+\!S_i^{\gamma_\nu}S_j^{\beta_\nu})\,,
\ee
where ${\bf S}_i$ denotes the pseudo-spin $j_{\text{eff}}\!=\!1/2$ operator at site $i$, 
$(\alpha_\nu,\beta_\nu,\gamma_\nu)\!=\!(x,y,z)$, $(y,z,x)$, and $(z,x,y)$, respectively, for 
$\nu\in\{x,y,z\}$ labeling the three different types of NN Ir-Ir or Ru-Ru bonds; the prefactor $\sigma_\nu$  equals $+1$ for two-dimensional materials and can be $+1$ or $-1$ (depending on the bond) for the three-dimensional systems $\beta$- and $\gamma$-Li$_2$IrO$_3$. Here $K$ is the Kitaev coupling, $J$ is the Heisenberg coupling and $\Gamma$ is the so-called symmetric exchange anisotropy, which is present in many Kitaev materials~\cite{Katukuri2014,Rau2014,Lee2015,Lee2016,IoannisGamma,Winter2017}. 
These interactions should be thought of as a minimal starting model, as other terms may also be relevant for materials with lower symmetry.

Crucially, the $J$, $K$ and $\Gamma$ couplings originate from very different microscopic processes. 
Specifically, as  it has been shown in the literature~\cite{Jackeli2009,Jackeli2010,Rau2014,Perkins2014, Sizyuk2014}, the Heisenberg interaction $J$ arises from direct virtual hopping processes between $d$ orbitals of magnetic ions, whereas the dominant contribution to the Kitaev interaction $K$ arises from the ligand-mediated hopping. As for $\Gamma$, this arises from a combination of direct and ligand-mediated hopping. 
As we discuss below, the Raman operator stems from the same  underlying microscopic processes as the  super-exchange Hamiltonian, and each type of these processes gives rise to a different contribution to the  Raman response.



\section{Microscopic derivation of the Raman operator}\label{sec:derivation}

We first review a number of key steps in the derivation of the Raman operator $\mc{R}$  in Mott insulators, with a view on Kitaev materials with strong spin-orbit coupling (SOC). 
The first step is to write down the total microscopic Hamiltonian, 
\be
\mc{H}_{\text{tot}} =\mc{H}+\mc{H}_\gamma+\mc{H}_c\,,
\ee
consisting of the extended Hubbard Hamiltonian $\mc{H}$, the free photon Hamiltonian 
\be
\mc{H}_\gamma=\sum_{\bf{k},\bs{\varepsilon}} \omega_{\bf k} \alpha^\dagger_{{\bf k},\bs{\varepsilon}}\alpha_{{\bf k},\bs{\varepsilon}}\,,
\ee
where $\alpha^\dagger_{{\bf k},\bs{\varepsilon}}$ and $\alpha_{{\bf k},\bs{\varepsilon}}$ are the creation and destruction operators of a photon with wavevector ${\bf k}$ and polarization $\bs{\varepsilon}$, and $\omega_{\bf k}$ is the corresponding frequency, and the perturbation $\mc{H}_c$ that describes the interaction of the electrons (holes) with the electromagnetic (EM) field.
The latter arises  from the coupling of the light to the electric dipoles induced by the virtual charge transfers between different lattice sites.
This coupling can be described by the Peierls substitution, in which a ``Wilson line" operator  is attached to the  electron (hole) hopping term between magnetic ions as~ \cite{Shastry1990,Devereaux2007,Ko2010}
\be\label{Wilson}
d_{i\alpha\sigma}^{\dagger}d^{}_{j\beta\sigma}\rightarrow
d_{i\alpha\sigma}^{\dagger}d^{}_{j\beta\sigma}
e^{\frac{ie}{\hbar c}\int_{\vec{r}_j}^{\vec{r}_i} d\vec{r} \cdot \vec{\mc{A}}(\vec{r})}
\ee 
(and similarly for the hopping between magnetic and ligand ions), where $\vec{\mc{A}}(\vec{r})$ denotes the vector potential of the radiation field.
This substitution amounts to replacing $\mc{H}_t+\mc{H}_c$ with
\begin{align}\label{Wilson2}
\mc{H}_{t,\mc{A}}&=\sum_{ij}\sum_{\alpha\beta}\sum_{\sigma} t_{ij,\sigma}^{\alpha\beta} d_{i\alpha\sigma}^\dagger
d_{j\beta\sigma} 
e^{
\frac{ie}{\hbar c}
\int_{\vec{r}_j}^{\vec{r}_i} d\vec{r} \cdot \vec{\mc{A}}(\vec{r})}\nonumber\\
&+\widetilde{\sum_{ij}}\sum_{\alpha\beta}\sum_{\sigma}\Big[ \tilde{t}_{ij,\sigma}^{\alpha\beta} d_{i\alpha\sigma}^\dagger
p_{j\beta\sigma} 
e^{
\frac{ie}{\hbar c}
\int_{\vec{r}_j}^{\vec{r}_i} d\vec{r} \cdot \vec{\mc{A}}(\vec{r})} + \text{h.c.}\Big]\,.
\end{align}
As usual, we consider the case where the wavelengths of the incoming and outgoing photons are much longer than the lattice constant, which allows us to safely replace 
\be
\frac{ie}{\hbar c}\int_{\vec{r}_{j}}^{\vec{r}_i} d\vec{r} \cdot \vec{\mc{A}}(\vec{r}) \simeq
\frac{ie}{\hbar c}\vec{\mc{A}}\cdot \delta\vec{r}_{ij}
\,,~~~~
\delta\vec{r}_{ij}\equiv\vec{r}_i-\vec{r}_j\,,
\ee
and then perform an expansion of $\mc{H}_{t,\mc{A}}$ in powers of the vector potential (which is appropriate for the weak EM fields of Raman experiments),namely
\be\label{photoinducedHt2} 
\mc{H}_{t,\mc{A}}
=\mc{H}_{t}+\mc{H}_{t,\mc{A}}^{(1)}+\cdots\,.
\ee
Here $\mc{H}_{t}$ is the hopping in the absence of light, and $\mc{H}_{t,\mc{A}}^{(1)}$ is the leading photon-induced hopping, 
\begin{align}
\mc{H}_{t,\mc{A}}^{(1)} = & \sum_{ij}\sum_{\alpha\beta}\sum_{\sigma} t_{ij,\sigma}^{\alpha\beta}
d_{i\alpha\sigma}^{\dagger}d^{}_{j\beta\sigma} \Big(\frac{i e}{\hbar c} \vec{\mc{A}}\cdot \delta\vec{r}_{ij}\Big)\nonumber\\
&+\widetilde{\sum_{ij}}\sum_{\alpha\beta}\sum_{\sigma}\Big[\tilde{t}^{\alpha\beta}_{ij,\sigma}d^\dagger_{i\alpha\sigma}p_{j\beta\sigma}\Big(\frac{i e}{\hbar c} \vec{\mc{A}}\cdot \delta\vec{r}_{ij}\Big)+ \text{h.c.}\Big]\,.
\end{align}
We can then express the vector potential in terms of creation and annihilation photon operators, and to that end, it suffices to keep only the terms referring to the incoming and outgoing photons, namely
\bea
\vec{\mc{A}} &=&  
g_{\text{in}}
\bs{\varepsilon}_{\text{in}} \alpha_{\vec{k}_{\text{in}},\bs{\varepsilon}_{\text{in}}} 
e^{i\vec{k}_{\text{in}}\cdot\delta\vec{r}_{ij}}
+
g_{\text{out}}
\bs{\varepsilon}_{\text{out}}
\alpha_{\vec{k}_{\text{out}},\bs{\varepsilon}_{\text{out}}}^\dagger 
e^{i\vec{k}_{\text{out}}\cdot \delta\vec{r}_{ij}}
\nonumber\\
&\simeq&
g_{\text{in}}
\bs{\varepsilon}_{\text{in}} \alpha_{\vec{k}_{\text{in}},\bs{\varepsilon}_{\text{in}}} 
+
g_{\text{out}}
\bs{\varepsilon}_{\text{out}}
\alpha_{\vec{k}_{\text{out}},\bs{\varepsilon}_{\text{out}}}^\dagger \,.
\eea
Here, ${\bf k}_{\text{in}}$ and  $\bs{\varepsilon}_{\text{in}}$ (respectively, ${\bf k}_{\text{out}}$ and $\bs{\varepsilon}_{\text{out}}$) denote the wavevectors and polarizations of the incoming (respectively, outgoing) photons, and $g_{\text{in}}$ and $g_{\text{out}}$ are constants depending on the photon frequencies~\cite{Shastry1990,Ko2010}.
Furthermore, in the second line we replaced $e^{i\vec{k}_{\text{in}}\cdot\delta\vec{r}_{ij}}\sim e^{i\vec{k}_{\text{out}}\cdot \delta\vec{r}_{ij}}\sim 1$, which is accurate in our long-wavelength limit.

Following Refs.~\cite{Shastry1990,Shastry1991,Ko2010}, in which the Raman scattering is treated in the framework of the $\mc{T}$-matrix formalism, with the photon-induced hopping terms  $\mc{H}_{t,\mc{A}}^{(1)}+\mc{H}_{t,\mc{A}}^{(2)}+\cdots$ treated as a perturbation, one arrives at the leading contribution to the Raman operator which is second order in $\mc{A}$ (describing a one photon
in, one photon out process)~\cite{Shastry1990,Ko2010}:
\be
\mc{R} =  
\mc{H}_{t,\mc{A}}^{(1)}~
\frac{1}{E-(\mc{H}+\mc{H}_{\gamma})+i\eta} ~
\mc{H}_{t,\mc{A}}^{(1)}\,,
\ee
where $\eta\to 0^+$ and $E=2E_{1\text{h}}+\omega_{\text{in}}$ is the eigenenergy of the initial state, in which all magnetic ions have  one hole in the $j_{\text{eff}}\!=\!1/2$ doublet. The next step is to treat $\mc{H}_t$ as a weak perturbation compared to
\be
\mc{H}_0 \equiv \mc{H}_{\text{int}}+\mc{H}_{\text{SOC}}+\mc{H}_{\text{pd}}+\mc{H}_\gamma\,.
\ee 
This allows to expand $\mc{R}$ as follows
\be\label{eqn:series}
\mc{R} = 
\mc{H}_{t,\mc{A}}^{(1)}~
\mc{G}~
\sum_{n=0}^\infty
\left(
\mc{H}_{t} ~\mc{G} \right)^n ~\mc{H}_{t,\mc{A}}^{(1)}\,,
\ee
where we have defined the resolvent 
\be\label{eq:resolvent}
\mc{G} = \left(E-\mc{H}_0+i\eta\right)^{-1}\,.
\ee
Note that both $\mc{H}_{t,\mc{A}}^{(1)}$ and $\mc{H}_t$ include hopping terms on all bonds of the lattice. However, since both the initial and final states belong to the ground-state manifold, only pathways consisting of closed loops contribute to the Raman operator. 

Incidentally, replacing $\mc{H}_{t,\mc{A}}^{(1)}$ by $\mc{H}_t$ in Eq.~(\ref{eqn:series}) gives the leading contributions to the effective spin Hamiltonian $\mc{H}_{\text{eff}}$. More specifically, \be\label{eqn:seriesForHeff}
\mc{H}_{\text{eff}} = \mc{H}_t~\mc{G}~\sum_{n=0}^\infty\left(\mc{H}_{t} ~\mc{G} \right)^n~\mc{H}_t + \text{other terms}\,,
\ee
where the `other terms' in the formal expansion~\cite{Klein1974} of $\mc{H}_{\text{eff}}$ can be safely disregarded for our purposes.

\section{Raman operator in Kitaev materials}\label{sec:RamanKitaev}
We are now ready to apply the above general formalism to the case of the Kitaev materials and highlight the main new insights of this work. 
In particular, we  will explicitly demonstrate that despite the fact that the microscopic processes underlying  the Raman operator and the super-exchange Hamiltonian are very similar in the Kitaev materials~ \cite{Jackeli2009,Jackeli2010,Sizyuk2014, Rau2014}, the presence of multiple non-equivalent super-exchange paths contributing to the coupling between the magnetic moments on a given bond leads to the contributions to the Raman operator that goes beyond the Loudon-Fleury  theory~\cite{LF1968}.
Recall that in the Loudon-Fleury  theory, the contribution $\mc{R}_{ij}$ to the total Raman operator from a given bond $(ij)$, is simply given by the super-exchange interactions $\mc{H}_{\text{eff},ij}$ on that bond, weighted by a {\it bond-specific} polarization-dependent factor. However, in systems with multiple non-equivalent super-exchange paths,
the polarization factors that come from the operators ${\mc H}_{t,\mc{A}}^{(1)}$ appearing at the first and last steps of the perturbative expansion of Eq.~(\ref{eqn:series}) give {\it unequal} weights to different paths. Hence, the summation over these paths leads to a Raman operator $\mc{R}_{ij}$ (on the given bond) which is, in general, {\it not proportional} to $\mc{H}_{\text{eff},ij}$ (obtained
by summing up the contributions from all possible  paths with {\it equal} weight).

\begin{figure}[!t]
\centering
\includegraphics[width=0.5\linewidth]{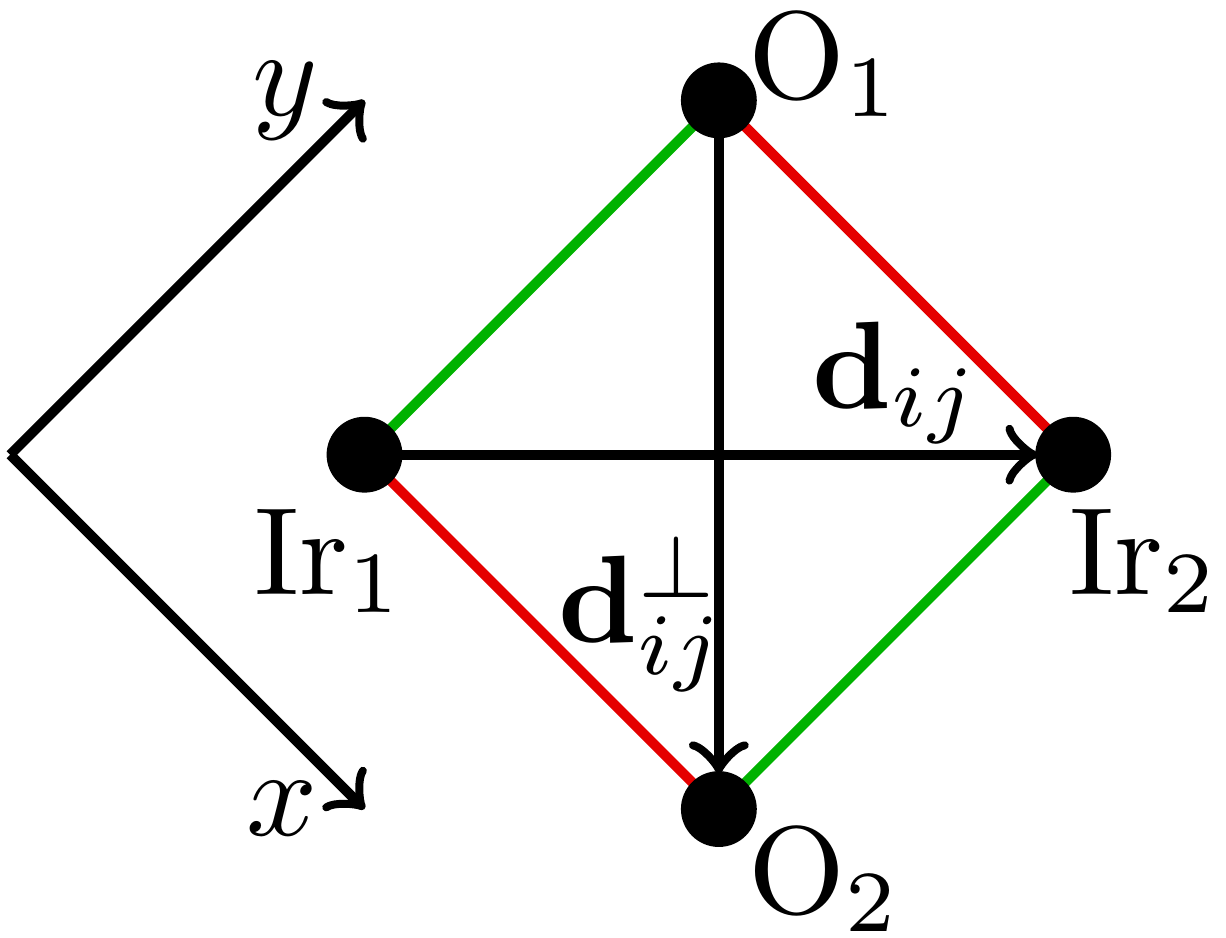}
\caption{
The square plaquette that is relevant for the super-exchange processes between  two magnetic ions sharing a `$z$-bond' in A${}_2$IrO${}_3$ compounds (e.g., $\beta$-Li${}_2$IrO${}_3$). The same plaquette provides the super-exchange processes in $\alpha$-RuCl${}_3$ with substitution Ir $\rightarrow$ Ru, O $\rightarrow$ Cl.
 \label{fig:Plaquette}}	
\end{figure}

Let us now begin with re-examining the various super-exchange paths contributing to the effective spin Hamiltonian of the Kitaev materials. 
In all of them, the local  environment of the magnetic ions is that of an octahedron ligand cage, see, e.g., Fig.~\ref{fig:lattice} for the case of $\beta$-Li${}_2$IrO${}_3$. 
The virtual hopping processes leading to the $J$-$K$-$\Gamma$ model (and the ones contributing to the Raman operator) are confined to a  plaquette consisting of two magnetic ions and two ligand ions.  
For iridium Kitaev materials, for example, the plaquette is formed by two iridium and two oxygen ions (see, e.g., Fig.~\ref{fig:Plaquette} for the case of two Ir$^{4+}$ ions sharing a `$z$-bond' in $\beta$-Li${}_2$IrO${}_3$), while for $\alpha$-RuCl$_3$ it is formed by two ruthenium and two chlorine ions. 
For concreteness, in the following discussion we will use notations for the  iridates,  but the final results will be exactly the same for  $\alpha$-RuCl$_3$ as well. 

\begin{table}[!b]
\renewcommand{\arraystretch}{1.5}
\centering
\begin{tabular}{|c|c|} 
\hline
transfer path & hopping amplitude\\
\hline
Ir($xz (Y)$ or $yz (X)$) $\rightarrow$ O($p_z$) & $t$\\
\hline
Ir${}_1$ ($xz(Y)$)$\rightarrow$ Ir${}_2$ ($xz(Y)$) & $t_1$\\
\hline
Ir${}_1$ ($yz(X)$)$\rightarrow$ Ir${}_2$ ($yz(X)$) & $t_1$\\
\hline
Ir${}_1$ ($xy(Z)$)$\rightarrow$ Ir${}_2$ ($xy(Z)$) & $t_3$\\
\hline
\end{tabular}
\caption{\label{Hop-tab} The matrix elements of $\mc{H}_t$ (in the hole picture) related to the $z$-bond for $\beta$-Li${}_2$IrO${}_3$. All matrix elements are real.}
\end{table}

 We will carry out our analysis for a `$z$-bond' formed by two iridium ions, Ir$_1$ and Ir$_2$. The results for other types of bonds can be obtained in a similar way (or simply by symmetry, if present). 
Using the frame of Fig.~\ref{fig:Plaquette}, the vector connecting these two ions is $\vec{d}_{ij}={\bf x} +{\bf y}$ (in appropriate length units), while the  vector connecting the two oxygen sites, O$_1$ and O$_2$, is ${\bf d}^\perp_{ij}={\bf x} -{\bf y}$. Finally, the vectors connecting Ir$_1$ with  O$_1$ and O$_2$ are , respectively, ${\bf y}$ and ${\bf x}$. 
The hopping matrix elements corresponding to this `$z$-bond' are listed in Table~\ref{Hop-tab}.

\begin{figure}[!t]
\includegraphics[width=0.4\linewidth]{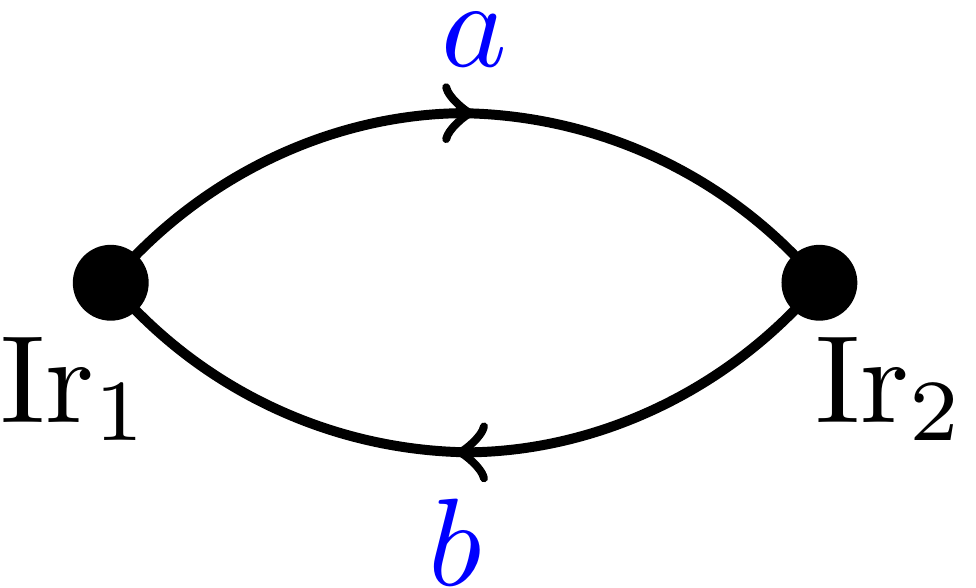}
\caption{ Direct hopping ($a$-$b$ and $b$-$a$)}\label{fig:directpaths}
\end{figure}

\begin{figure}[!b]
\centering
\subfigure[path 1 ($a$-$b$-$c$-$d$) and 2 ($c$-$d$-$a$-$b$) ]{\includegraphics[width=0.4\linewidth]{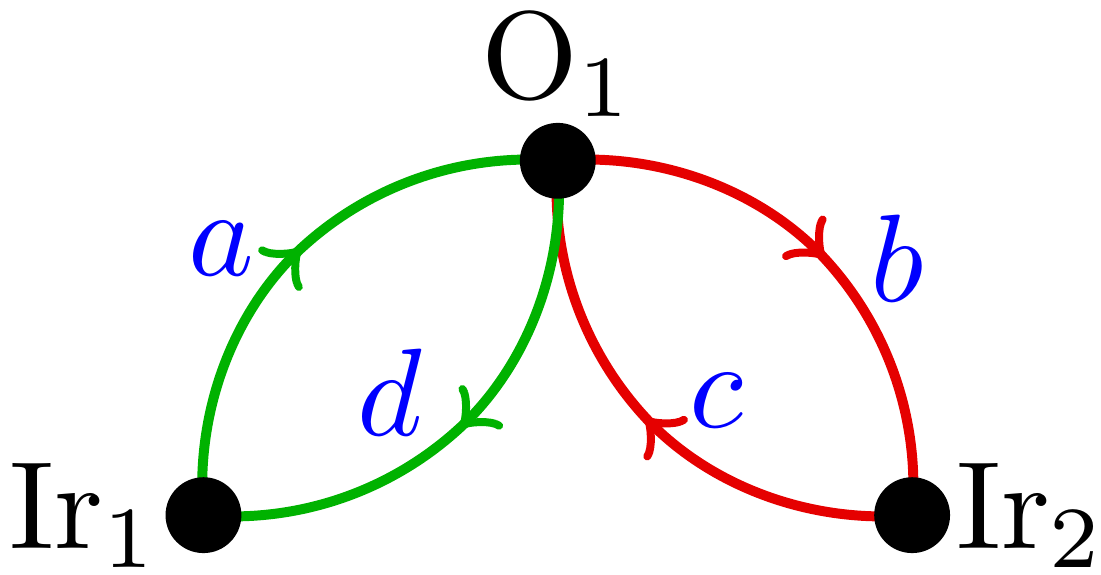}}\quad
\subfigure[path 3 ($a$-$b$-$c$-$d$) and 4 ($c$-$d$-$a$-$b$)]{\includegraphics[width=0.4\linewidth]{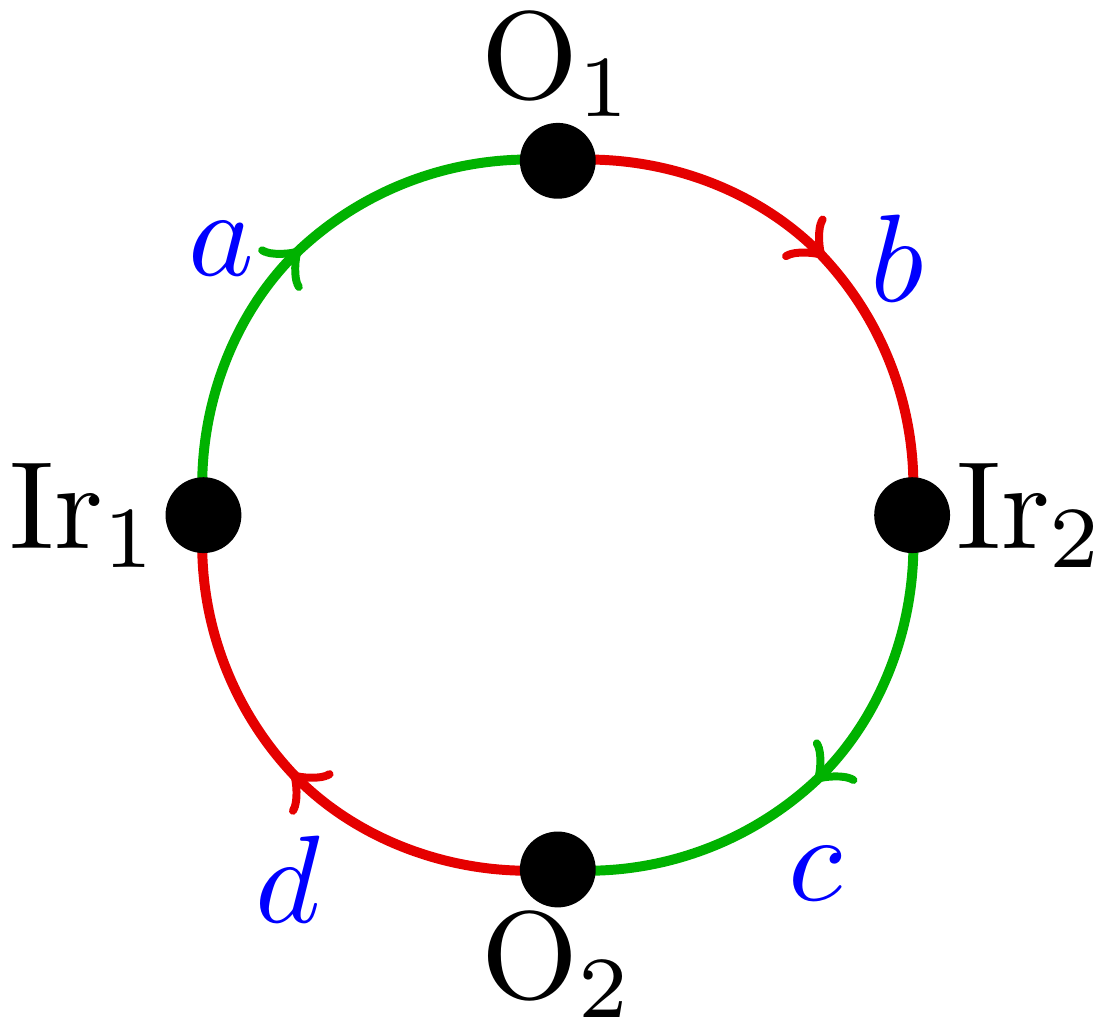}}\\
\subfigure[path 5 ($a$-$b$-$c$-$d$) and 6 ($c$-$d$-$a$-$b$)]{\includegraphics[width=0.4\linewidth]{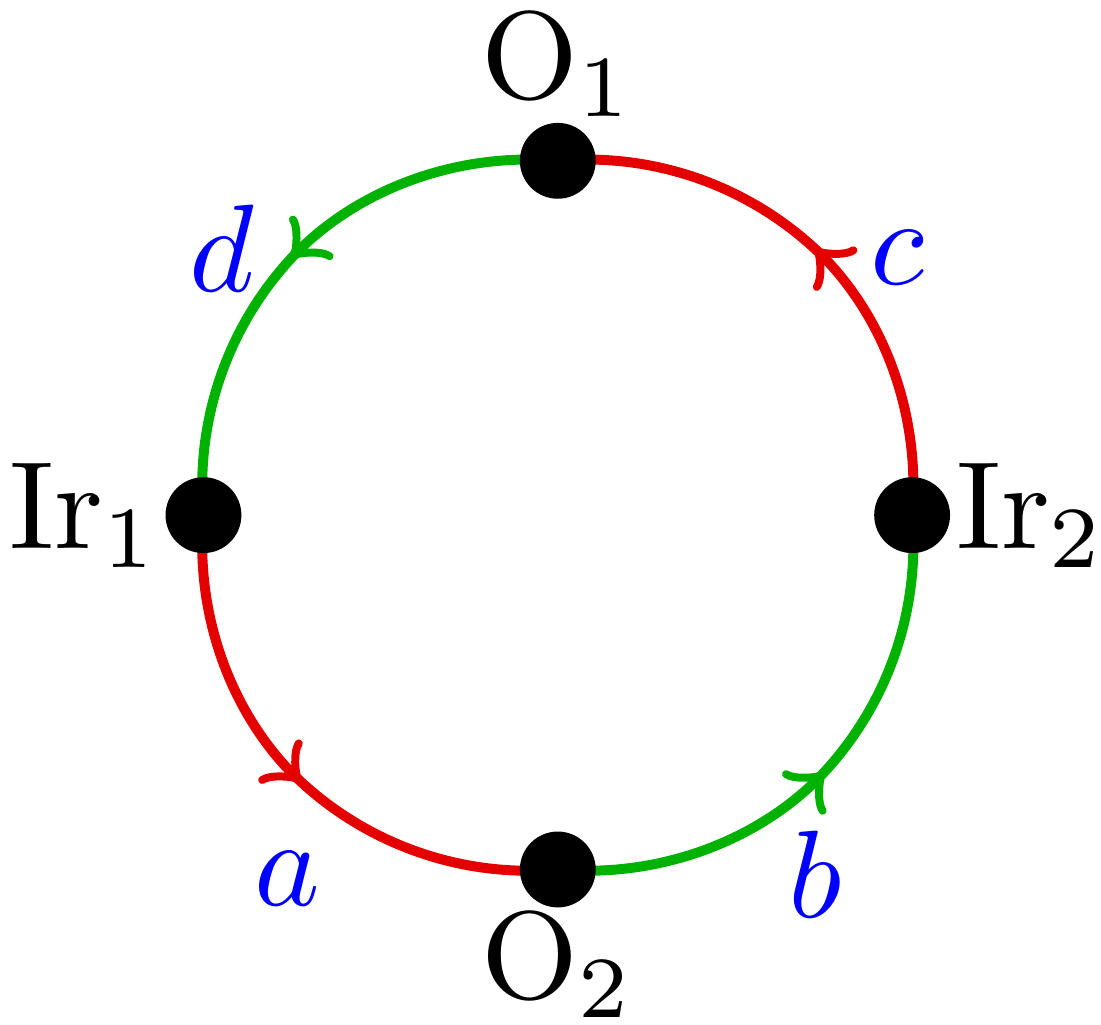}}\quad
\subfigure[path 7 ($a$-$b$-$c$-$d$) and 8 ($c$-$d$-$a$-$b$)]{\includegraphics[width=0.4\linewidth]{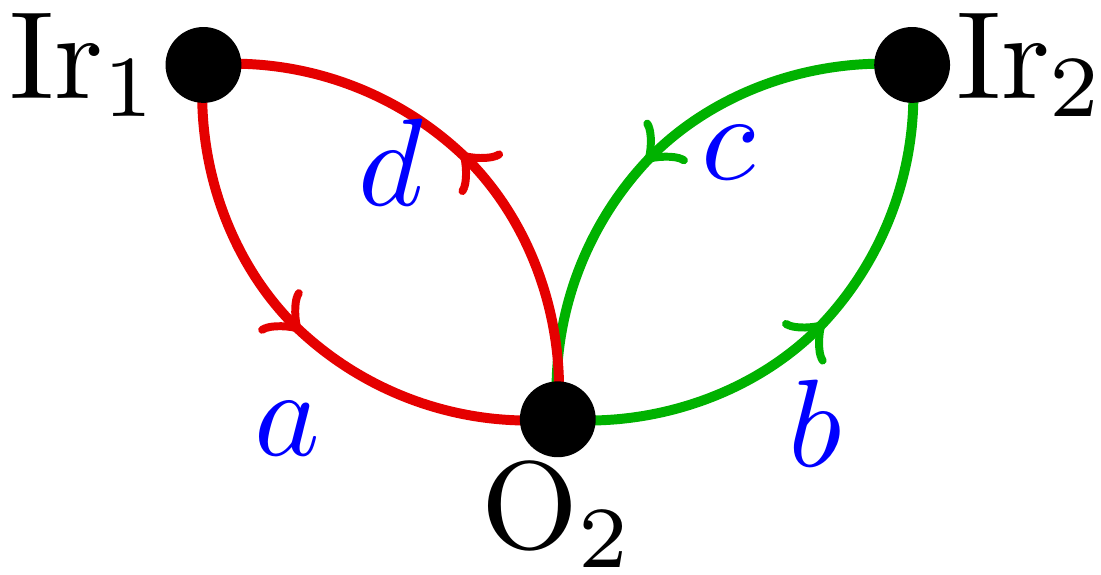}}
\caption{The eight different oxygen-mediated hopping paths connecting Ir${}_{1}$ and Ir${}_{2}$ in the plaquette of Fig.~\ref{fig:Plaquette}.}
\label{fig:oxygen-hopping-paths}
\end{figure}

There are three different types of paths on the plaquette of Fig.~\ref{fig:Plaquette}: (a) direct hopping (Fig.~\ref{fig:directpaths}), (b) oxygen-mediated hopping (Fig.~\ref{fig:oxygen-hopping-paths}), and (c) mixed hopping (Fig.~\ref{fig:mixed-hopping-paths}).
The direct hopping contributes at the lowest order of the perturbation [$n=0$ in Eq.~(\ref{eqn:series})], the oxygen-mediated hopping processes arise at fourth order ($n=2$), and the mixed direct/oxygen-mediated hopping processes arise at third order ($n=1$). 
Mathematically, the corresponding amplitudes for each of these types of processes can be obtained by performing a spectral decomposition of the resolvent $\mc{G}$ of Eq.~(\ref{eq:resolvent}) in terms of the relevant virtual excitations. These include the intermediate two-hole states on Iridium sites, $|D_\mu\rangle$, where $\mu=1,2...15$, obtained by the diagonalization of $\mc{H}_{\text{int}}$ ~\cite{Perkins2014,Sizyuk2014} (see also Table~\ref{tbl:twohole_orbital} in App.~\ref{App:Ramandirect}), and the intermediate one-hole states on oxygen sites, $|O_\nu\rangle$,where $\nu=1,2$ labels one of the two oxygen ions:
\bea
\!\!\!\mc{G} \!&=&\! 
\sum_{\mu=1}^{15}
\frac{|D_\mu\rangle\langle D_\mu|}{(2E_{1\text{h}}\!-\!E_{2\text{h}}\!-\!E_{0\text{h}})\!+\!\omega_{\text{in}}\!+\!i\eta}
\!+\!
\sum_{\nu=1}^2\frac{|O_\nu\rangle\langle O_\nu|}{\omega_{\text{in}}\!-\!\Delta_{\text{pd}}\!+\!i\eta}
\nonumber\\
&\equiv& \mc{G}_D + \mc{G}_O\,.
\eea
Here $\omega_{\text{in}}$ is the incoming photon frequency, $\Delta_{\text{pd}}$ is the charge transfer energy between Ir${}^{4+}$ and O${}^{2-}$, and we have also defined $E_{0\text{h}}$, $E_{1\text{h}}$ and $E_{2\text{h}}$ to be the zero-, one- and two-hole eigenenergies.

The effective Raman operator lives in the low-energy sector with magnetic ions being in their $j_{\text{eff}}=1/2$ ground state manifold. For two neighbouring Ir ions, this space is spanned by the four configurations written in the $|j^z_{\text{eff},1},j^z_{\text{eff},2}\rangle$ representation:
\be
\Big\{|\psi_1\rangle\,, |\psi_2\rangle, |\psi_3\rangle, |\psi_4\rangle\Big\} \equiv \Big\{
|\frac{1}{2},\frac{1}{2}\rangle,
|\frac{1}{2},-\frac{1}{2}\rangle,
|-\frac{1}{2},\frac{1}{2}\rangle,
|-\frac{1}{2},-\frac{1}{2}\rangle
\Big\}\,.\nonumber
\ee
One then  evaluates the matrix elements $\langle \psi_n| \mc{R} |\psi_{n'}\rangle$ of the Raman operator in this $4\times 4$ basis, and then expresses the resulting matrix in terms of the pseudospin operators ${\bf S}_i$ and ${\bf S}_j$ to obtain the effective spin representation of $\mc{R}$. 

For what follows, it is also expedient to define the following generic polarization factors 
that arise from the coupling of the incoming and outgoing photon to, respectively, the first and last bond of the virtual hopping paths involved in Eq.~(\ref{eqn:series}) as
\begin{align}\label{eqn:PPPP}
P_{dd}\,&\equiv~\zeta
(\bs{\varepsilon}_{\text{in}}\cdot\mathbf{d}_{ij})~(\bs{\varepsilon}_{\text{out}}\cdot\mathbf{d}_{ij})\,,\nonumber\\\nonumber
P_{d^\perp d^\perp}&\equiv~\zeta
(\bs{\varepsilon}_{\text{in}}\cdot\mathbf{d}_{ij}^\perp)~(\bs{\varepsilon}_{\text{out}}\cdot\mathbf{d}_{ij}^\perp)\,,
\\
P_{d d^\perp}\,&\equiv~\zeta
(\bs{\varepsilon}_{\text{in}}\cdot\mathbf{d}_{ij})~(\bs{\varepsilon}_{\text{out}}\cdot\mathbf{d}_{ij}^\perp)\,,\\
P_{d^\perp d}\,&\equiv~\zeta
(\bs{\varepsilon}_{\text{in}}\cdot\mathbf{d}_{ij}^\perp)~(\bs{\varepsilon}_{\text{out}}\cdot\mathbf{d}_{ij})\,,\nonumber
\end{align}
where $\zeta=-\frac{e^2}{\hbar^2c^2} g_{\text{in}}g_{\text{out}}$.

\subsection{Raman operator from direct hopping}
There are two direct hopping processes (see Fig.~\ref{fig:directpaths}) connecting the two Ir$^{4+}$ ions of the basic plaquette of Fig.~\ref{fig:Plaquette}. Since these processes are of second order in the hopping [i.e., $n=0$ in Eq.~(\ref{eqn:series})] and do not involve the oxygen sites, the corresponding Raman operator is given by $\mc{R}^{\text{dir}} = \mc{H}_{t,\mc{A}}^{(1)} ~\mc{G}_D~ \mc{H}_{t,\mc{A}}^{(1)}$, or, equivalently,
\be\label{eqn:Rdirdir}
\mc{R}^{\text{dir}}
= \sum_{\mu}
\frac{\mc{H}_{t,\mc{A}}^{(1)}|D_\mu\rangle \langle D_\mu|\mc{H}_{t,\mc{A}}^{(1)}}{(2E_{1\text{h}}-E_{2\text{h}}-E_{0\text{h}})+\omega_{\text{in}}+i\eta}\,.
\ee
The details of the computation of the matrix elements of $\mc{R}^{\text{dir}}$  are provided in App.~\ref{App:Ramandirect}. The resulting expression for $\mc{R}^{\text{dir}}_{ij}$ on the bond $\langle ij\rangle_z$ in terms of spin operators is
\be\label{eqn:Rdir}
\mc{R}^{\text{dir}}_{ \langle ij\rangle_z} =  -P_{dd} \left(
J^{(2)}~{\bf S}_i \cdot{\bf S}_j 
+ K^{(2)} ~S^z_i S^z_j \right)\,,
\ee
where $K^{(2)}$ and $J^{(2)}$ are  coupling constants (explicit analytic expressions of them are given in App.~\ref{App:Analytic}, and we shall comment on their numerical values for the case of $\beta$-Li${}_2$IrO${}_3$ in Sec.~\ref{sec:Ramanbeta}), and the superscript $(2)$ specifies that they are obtained in second order perturbation theory. 
We should note here that $K^{(2)}$ and $J^{(2)}$ depend on the frequency $\omega_{\text{in}}$ of the incoming light, but in the limit $\omega_{\text{in}}\rightarrow 0$, they reduce to the second order contributions to the effective couplings $K$ and $J$ of the effective  $J$-$K$-$\Gamma$ model, as they arise from the same microscopic processes. Indeed, starting from Eq.~(\ref{eqn:seriesForHeff}) one can show that, for $\omega_{\text{in}}\rightarrow 0$, the spin terms inside the bracket of Eq.~(\ref{eqn:Rdir}) are precisely the contributions to $\mc{H}_{\text{eff},\langle ij\rangle_z}$ from direct hopping, namely
\be
\omega_{\text{in}}\mapsto 0:~~~~~
\mc{R}^{\text{dir}}_{\langle ij\rangle_z} = - P_{dd}~ \mc{H}_{\text{eff},\langle ij\rangle_z}^{\text{dir}}\,.
\ee
Hence, the leading Raman operator coming from direct hopping processes has a Loudon-Fleury form. 

\begin{table}[!t]
\renewcommand{\arraystretch}{1.5}
\centering
\begin{tabular}{|c|c|} 
\hline
path \#, $\ell$ & polarization factor $p^O_\ell$\\
\hline
 1&$\zeta (\bs{\varepsilon}_{\text{in}}\cdot\mathbf{y})~(\bs{\varepsilon}_{\text{out}}\cdot(-\mathbf{y})) = -(P_{dd}+P_{d^\perp d^\perp}-P_{d d^\perp}-P_{d^\perp d})/4$\\
\hline
 2&$\zeta (\bs{\varepsilon}_{\text{in}}\cdot(-\mathbf{x}))~(\bs{\varepsilon}_{\text{out}}\cdot\mathbf{x}) = -(P_{dd}+P_{d^\perp d^\perp}+P_{d d^\perp}+P_{d^\perp d})/4$\\
\hline
 3&$\zeta (\bs{\varepsilon}_{\text{in}}\cdot\mathbf{y})~(\bs{\varepsilon}_{\text{out}}\cdot(-\mathbf{x})) = -(P_{dd}-P_{d^\perp d^\perp}+P_{d d^\perp}-P_{d^\perp d})/4$\\
\hline
 4&$\zeta (\bs{\varepsilon}_{\text{in}}\cdot(-\mathbf{y}))~(\bs{\varepsilon}_{\text{out}}\cdot\mathbf{x}) = -(P_{dd}-P_{d^\perp d^\perp}+P_{d d^\perp}-P_{d^\perp d})/4$\\
\hline
 5&$\zeta (\bs{\varepsilon}_{\text{in}}\cdot\mathbf{x})~(\bs{\varepsilon}_{\text{out}}\cdot(-\mathbf{y})) = -(P_{dd}-P_{d^\perp d^\perp}-P_{d d^\perp}+P_{d^\perp d})/4$\\
\hline
 6&$\zeta (\bs{\varepsilon}_{\text{in}}\cdot(-\mathbf{x}))~(\bs{\varepsilon}_{\text{out}}\cdot\mathbf{y}) = -(P_{dd}-P_{d^\perp d^\perp}-P_{d d^\perp}+P_{d^\perp d})/4$\\
\hline
 7&$\zeta (\bs{\varepsilon}_{\text{in}}\cdot\mathbf{x})~(\bs{\varepsilon}_{\text{out}}\cdot(-\mathbf{x})) = -(P_{dd}+P_{d^\perp d^\perp}+P_{d d^\perp}+P_{d^\perp d})/4$\\
\hline
 8&$\zeta (\bs{\varepsilon}_{\text{in}}\cdot(-\mathbf{y}))~(\bs{\varepsilon}_{\text{out}}\cdot\mathbf{y}) = -(P_{dd}+P_{d^\perp d^\perp}-P_{d d^\perp}-P_{d^\perp d})/4$\\
\hline
\end{tabular}
\caption{Polarization factors for the eight oxygen-mediated hopping paths of Fig.~\ref{fig:oxygen-hopping-paths}, and $P_{dd}$, $P_{d^\perp d^\perp}$, $P_{d d^\perp}$ and $P_{d^\perp d}$ are defined in Eq.~(\ref{eqn:PPPP}).}
\label{tbl:oxy_factor}
\end{table}

\subsection{Raman operator from oxygen-mediated hopping}
Turning to oxygen-mediated hopping processes, their leading contribution to the Raman operator appears at fourth order in the hopping [$n=2$ in Eq.~(\ref{eqn:series})] and has the form 
\be\label{eqn:oxygen}
\mc{R}^{\text{med}} =
\mc{H}_{t,\mc{A}}^{(1)}~
\mc{G}_O~
\mc{H}_t~
\mc{G}_D~
\mc{H}_t~
\mc{G}_O~
\mc{H}_{t,\mc{A}}^{(1)}\,.
\ee
In total, for the bond $\langle ij\rangle_z$, there are eight different paths contributing to this operator, labeled by $\ell=1,...8$, four of which begin from Ir${}_1$ and the other four from Ir${}_2$, see Fig.~\ref{fig:oxygen-hopping-paths}. 
Each path $\ell$ gives rise to a polarization factor $p^O_\ell$ (provided in Table~\ref{tbl:oxy_factor}) multiplying an effective spin operator $\mc{H}_{\ell}^{O}$, namely
\be
\mc{R}^{\text{med}}_{\langle ij\rangle_z} = \sum\nolimits_{\ell=1}^8 p^O_\ell \mc{H}_\ell^O
\ee
(see details in App.~\ref{App:Ramanmediated}). The final form  of the Raman operator from oxygen-mediated hopping is given by
\be\label{eqn:Rmed}
\!\!\mc{R}^{\text{med}}_{\langle ij\rangle_z} \!=\!
-\frac{P_{dd}}{4} K^{(4)} S_i^zS_j^z
-\frac{P_{d^\perp d^\perp}}{4} \left( J'^{(4)} {\bf S}_i\cdot{\bf S}_j\!+\!K'^{(4)} S_i^z S_j^z \right)\,,
\ee
where the frequency-dependent constants $K^{(4)}$, $J'^{(4)}$ and $K'^{(4)}$ can be obtained numerically for the convenience of calculation (see App.~\ref{App:Analytic} for their analytic expressions).
Comparing with the corresponding contributions to the effective spin Hamiltonian (computed with $\omega_{\text{in}}=0$),
\be\label{eqn:Heffmed}
\mc{H}_{\text{eff},\langle ij\rangle_z}^{\text{med}} = \sum\nolimits_{\ell=1}^8 \mc{H}_\ell^O = K S_i^z S_j^z\,,
\ee
shows that the Raman operator from oxygen-mediated hopping {\it does not} take a Loudon-Fleury form, i.e., $\mc{R}^{\text{med}}_{\langle ij\rangle_z}$  computed on a given bond is {\it not} proportional to $\mc{H}_{\text{eff},\langle ij\rangle_z}^{\text{med}}$.

\begin{figure}[!b]
\centering
\subfigure[path 1 ($a$-$b$-$c$) and 2 ($c$-$a$-$b$)]{\includegraphics[width=0.4\linewidth]{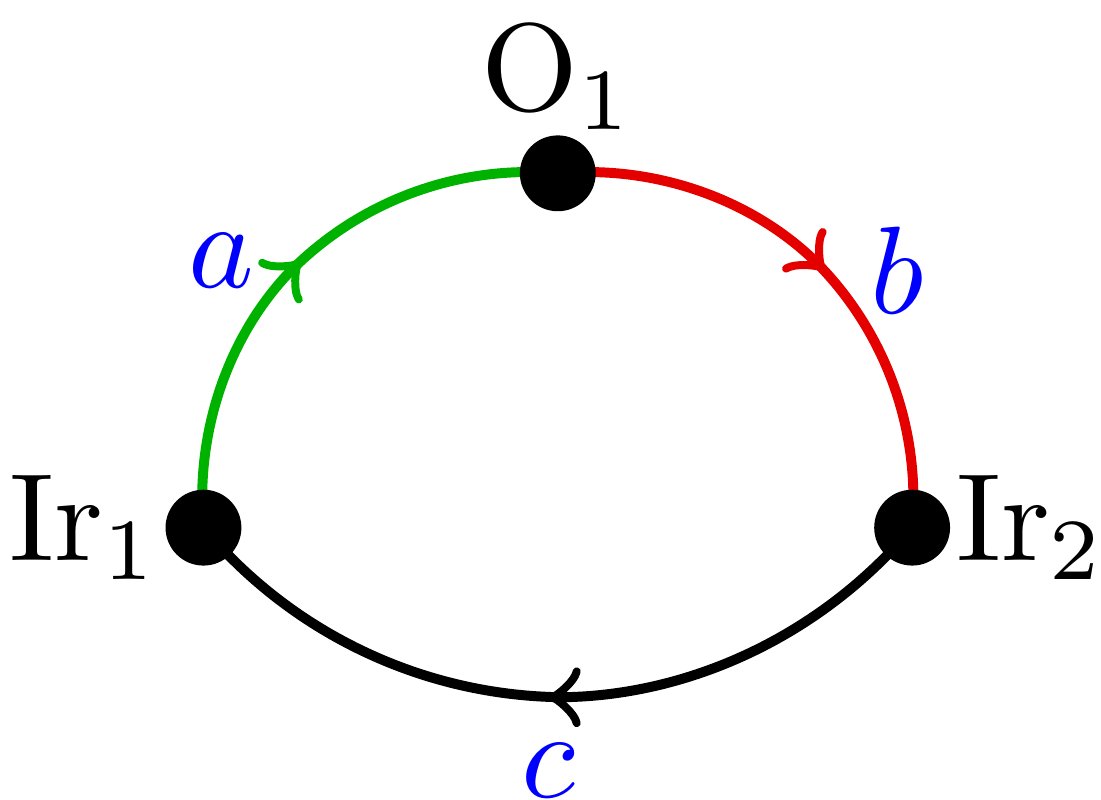}}\quad
\subfigure[path 3 ($a$-$b$-$c$) and 4 ($c$-$a$-$b$)]{\includegraphics[width=0.4\linewidth]{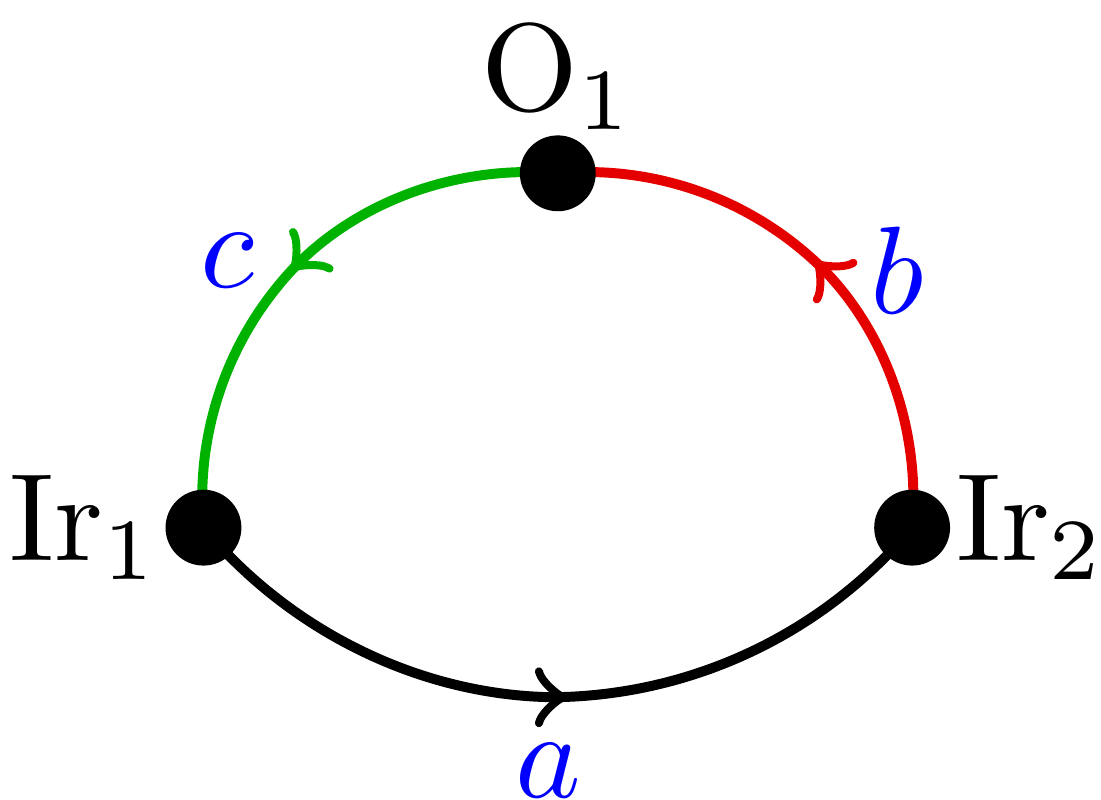}}\\
\subfigure[path 5 ($a$-$b$-$c$) and 6 ($c$-$a$-$b$)]{\includegraphics[width=0.4\linewidth]{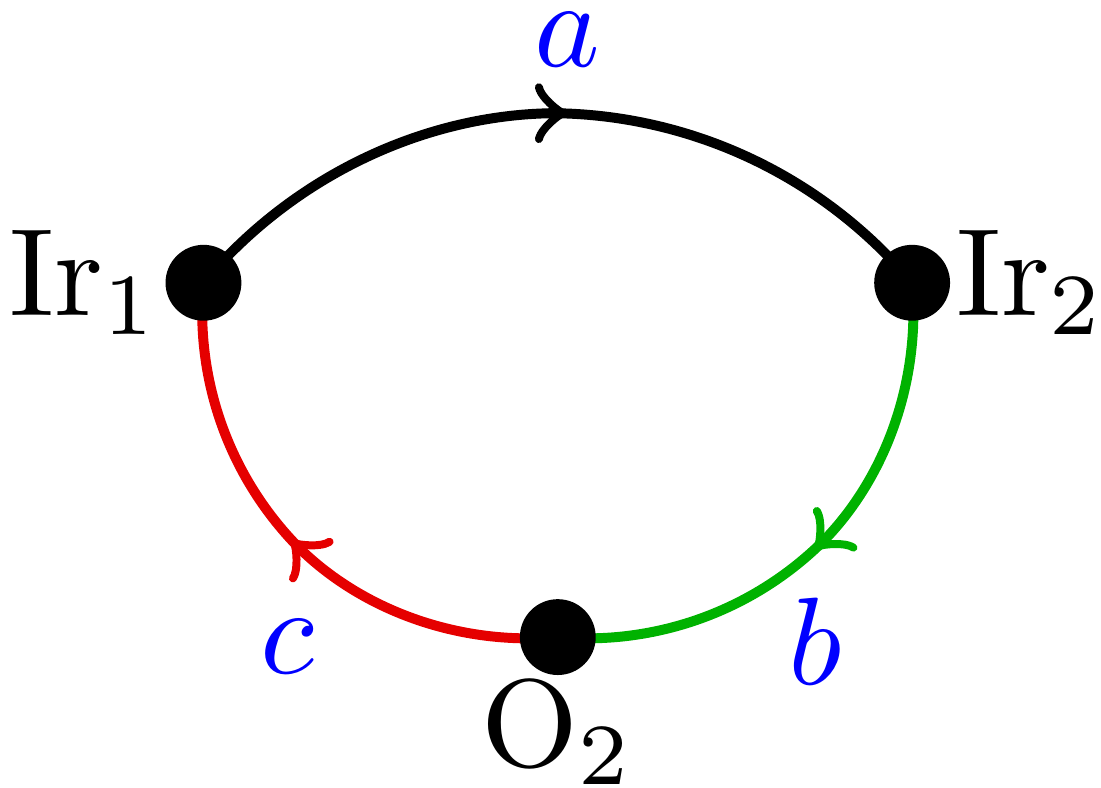}}\quad
\subfigure[path 7 ($a$-$b$-$c$) and 8 ($c$-$a$-$b$)]{\includegraphics[width=0.4\linewidth]{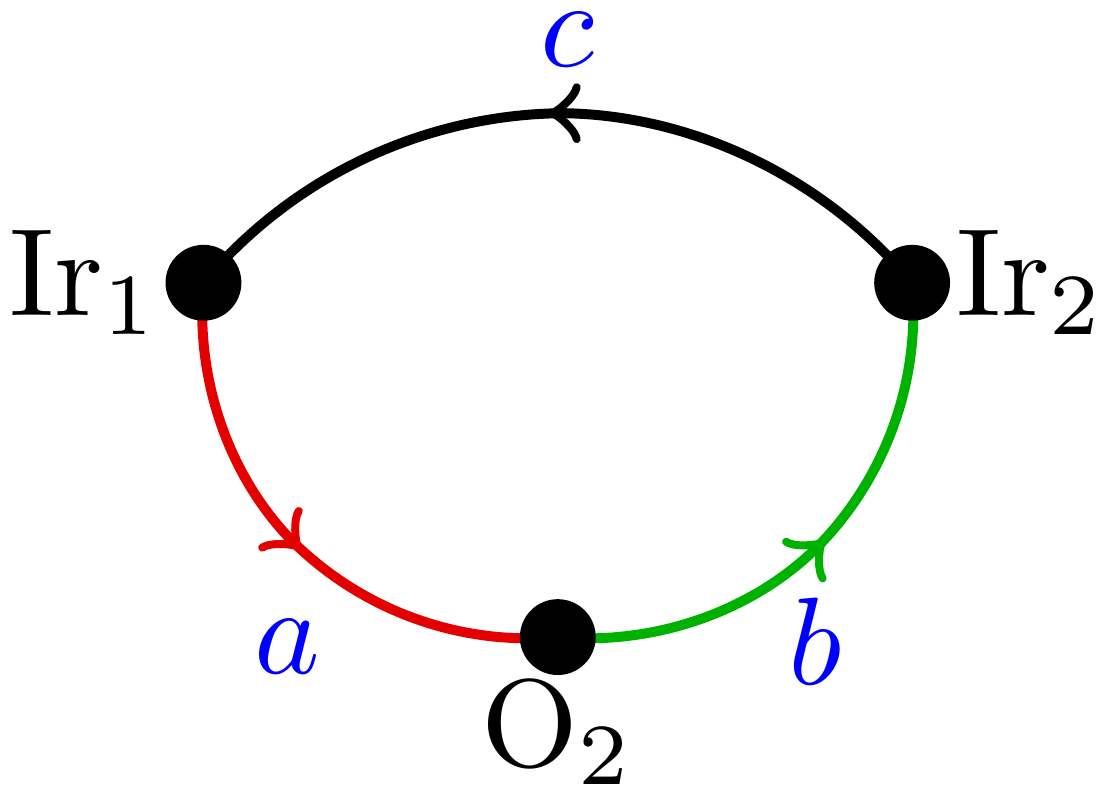}}
\caption{The eight different mixed hopping paths connecting Ir${}_{1}$ and Ir${}_{2}$ in the basic plaquette of Fig.~\ref{fig:Plaquette}.}
\label{fig:mixed-hopping-paths}
\end{figure}

\subsection{Raman operator from mixed hopping}
Let us now discuss virtual processes that involve both direct and oxygen-mediated hopping. The leading contributions to the corresponding Raman operator appear first at third order in the hopping [$n=1$ in Eq.~(\ref{eqn:series})], and take the form
\be\label{eqn:mixed}
\mc{R}^{\text{mix}}=
\mc{H}_{t,\mc{A}}^{(1)}~
\mc{G}~
\mc{H}_t~
\mc{G}~
\mc{H}_{t,\mc{A}}^{(1)}.
\ee
In total, for the bond $\langle ij\rangle_{z}$, there are again eight different paths contributing to this operator (see Fig.~\ref{fig:mixed-hopping-paths}), labeled by $\ell=1$-$8$.
As in the case of oxygen-mediated hopping, here too, each path $\ell$ gives rise to a polarization factor (provided in  Table~\ref{tbl:mixed_factor}) multiplying an effective spin operator $\mc{H}_\ell^{\text{m}}$, namely
\be
\mc{R}^{\text{mix}}_{\langle ij\rangle_z}=\sum\nolimits_{\ell=1}^8 p^m_\ell~\mc{H}_\ell^{\text{m}}
\ee
(see details in App.~\ref{App:Ramanmixed}), which takes the form
\be\label{eqn:Rmix}
\!\mc{R}^{\text{mix}}_{\langle ij\rangle_z} \!=\!  -\frac{P_{dd}}{2}\Gamma^{(3)} (S_i^xS_j^y \!+\! S_i^yS_j^x)
-\frac{P_{d^\perp d}\!-\!P_{d d^\perp}}{2}i\,h_\Gamma^{(3)}(S_i^z \!+\! S_j^z),
\ee
where the frequency-dependent constants $\Gamma^{(3)}$ and $h_\Gamma^{(3)}$ are  determined numerically again (see App.~\ref{App:Analytic} for their analytic expressions), and the additional factor $i$ in front of the real parameter $h_\Gamma^{(3)}$ ensures time-reversal symmetry.	
Comparing again with the corresponding contributions to the effective spin Hamiltonian (computed with $\omega_{\text{in}}=0$),
\be\label{eqn:Heffmix}
\mc{H}_{\text{eff},\langle ij\rangle_z}^{\text{mix}} = \sum\nolimits_{\ell=1}^8 \mc{H}_\ell^{\text{m}} 
=\Gamma~(S_i^xS_j^y\!+\!S_i^yS_j^x)\,,
\ee
we see that, as in the case of oxygen-mediated hopping, the Raman operator from mixed processes {\it does not} take a Loudon-Fleury form, i.e., $\mc{R}^\text{mix}_{\langle ij\rangle_z}$ is not proportional to $\mc{H}_{\text{eff},\langle ij\rangle_z}^{\text{mix}}$. 
In particular, the extra, non-Loudon-Fleury term [second term in Eq.~(\ref{eqn:Rmix})] takes the form of an effective, local magnetic field term $\propto h_\Gamma^{(3)}$, along the quantization axis $z$  associated with the Ir$_1$-Ir$_2$ bond of Fig.~\ref{fig:Plaquette}.

\begin{table}[!t]
\renewcommand{\arraystretch}{1.5}
\centering
\begin{tabular}{|c|c|} 
\hline
path \#, $\ell$ & polarization factor $p^{m}_\ell$\\
\hline
 1&$\zeta (\bs{\varepsilon}_{\text{in}}\cdot\mathbf{y})~(\bs{\varepsilon}_{\text{out}}\cdot(-\mathbf{d}_{ij})) = -(P_{dd}-P_{d^\perp d})/2$\\
\hline
 2&$\zeta (\bs{\varepsilon}_{\text{in}}\cdot(-\mathbf{d}_{ij}))(\bs{\varepsilon}_{\text{out}}\cdot\mathbf{x})	 = -(P_{dd}+P_{d d^\perp})/2$\\
\hline
 3&$\zeta (\bs{\varepsilon}_{\text{in}}\cdot\mathbf{d}_{ij})(\bs{\varepsilon}_{\text{out}}\cdot(-\mathbf{y})) = -(P_{dd}-P_{d d^\perp})/2$\\
\hline
 4&$\zeta (\bs{\varepsilon}_{\text{in}}\cdot(-\mathbf{x}))(\bs{\varepsilon}_{\text{out}}\cdot\mathbf{d}_{ij}) = -(P_{dd}+P_{d^\perp d})/2$\\
\hline
 5&$\zeta (\bs{\varepsilon}_{\text{in}}\cdot\mathbf{d}_{ij})(\bs{\varepsilon}_{\text{out}}\cdot(-\mathbf{x})) = -(P_{dd}+P_{d d^\perp})/2$\\
\hline
 6&$\zeta (\bs{\varepsilon}_{\text{in}}\cdot(-\mathbf{y}))(\bs{\varepsilon}_{\text{out}}\cdot\mathbf{d}_{ij}) = -(P_{dd}-P_{d^\perp d})/2$\\
\hline
 7&$\zeta (\bs{\varepsilon}_{\text{in}}\cdot\mathbf{x})(\bs{\varepsilon}_{\text{out}}\cdot(-\mathbf{d}_{ij}))
 ~=~ -(P_{dd}+P_{d^\perp d})/2$\\
\hline
 8&$\zeta (\bs{\varepsilon}_{\text{in}}\cdot(-\mathbf{d}_{ij}))(\bs{\varepsilon}_{\text{out}}\cdot\mathbf{y}) = -(P_{dd}-P_{d d^\perp})/2$\\
\hline
\end{tabular}
\caption{Polarization factors of the eight mixed hopping paths of Fig.~\ref{fig:mixed-hopping-paths}. For the definitions of $P_{dd}$, $P_{d^\perp d^\perp}$, $P_{d d^\perp}$ and $P_{d^\perp d}$ see Eq.~(\ref{eqn:PPPP}).}
\label{tbl:mixed_factor}
\end{table}

\subsection{Total Raman operator}
Collecting the various contributions to the Raman operator on `$z$-bonds', and the analogous expressions for `$x$-bonds' and `$y$-bonds' gives the total Raman operator
\bea\label{eqn:Raman1}
\mc{R} 
\!&=&\!\sum_{\langle ij\rangle_\nu}\!
\Bigg\{
\mc{P}_{ij,J}~{\bf S}_i \cdot{\bf S}_j \!+\!\mc{P}_{ij,K}~{S}_i^{\alpha_\nu}{S}_j^{\alpha_\nu}
+\mc{P}_{ij,\Gamma}~\big(S_i^{\beta_\nu}S_j^{\gamma_\nu}\!+\!S_i^{\gamma_\nu}S_j^{\beta_\nu}\big)
\nonumber\\
&&~~~~~
+\mc{P}_{ij, h_\Gamma} S ~\big(S_i^{\alpha_\nu}+S_j^{\alpha_\nu}\big)
\Bigg\}\,, 
\eea
where notations for $(\alpha_\nu,\beta_\nu,\gamma_\nu)$ are the same as in Eq.~(\ref{eqn:Heff}) and
\be\label{eqn:PijI}
\renewcommand{\arraystretch}{1.5}
\begin{array}{l}
\mc{P}_{ij,J}\equiv -P_{dd} J^{(2)}-\frac{1}{4}P_{d^\perp d^\perp} J'^{(4)}\,,
\\
\mc{P}_{ij,K}\equiv  -P_{dd} (K^{(2)}
+\frac{1}{4} K^{(4)}) -\frac{1}{4}P_{d^\perp d^\perp} K'^{(4)}\,,
\\
\mc{P}_{ij,\Gamma} \equiv  
-\frac{1}{2}P_{dd} \Gamma^{(3)}\,,
\\
\mc{P}_{ij,h_\Gamma} \equiv  -\frac{1}{2}(P_{d^\perp d}-P_{d d^\perp}) i \,h_\Gamma^{(3)}/S\,.
\end{array}
\ee
We repeat here that the various constants entering the Raman operator, i.e., $J^{(2)}$, $J'^{(4)}$, $K^{(2)}$, $K^{(4)}$, $K^{'(4)}$, $\Gamma^{(3)}$ and $h_\Gamma^{(3)}$, are frequency dependent and are not directly related to the effective couplings in the original super-exchange Hamiltonian (\ref{eqn:Heff}). However, the following relations hold, up to fourth order in $\mc{H}_t$, 
\bea\label{eqn:JKGterms}
\!\!\omega_{\text{in}}\to 0:~~
J^{(2)} =  J\,,~~
K^{(2)}+K^{(4)} = K\,,~~
\Gamma^{(3)} = \Gamma\,.
\eea
Note further that the constants $J'^{(4)}$, $K'^{(4)}$ and $h_\Gamma^{(3)}$ do not appear in the effective spin Hamiltonian  and are the ones that are responsible for the non-Loudon-Fleury Raman scattering.

\section{Bosonic representation of $\mc{R}$ in magnetically ordered states}\label{sec:bosonic}
Having established the leading contributions to the Raman operator, we can now turn to its magnon representation in the low-temperature, magnetically ordered states of iridates, such as $\beta$-Li$_2$IrO$_3$.
To describe the magnon excitations above an ordered state we first need to re-label the positions of the spins $i \to ({\bf R},\mu)$, where ${\bf R}$ is the position of the magnetic unit cell, and $\mu$ labels the different spin sublattices in the given magnetic state around which we wish to perform the $1/S$ semiclassical expansion.
The relabeling allows for the substitutions
\be
{\bf S}_i \!\to\! {\bf S}_{{\bf R},\mu},~
{\bf S}_{j} \!\to\! {\bf S}_{{\bf R}+{\bf t}_{\mu\mu'},\mu'},~
\mc{P}_{ij,I} \!\to\! \mc{P}_{\mu\mu',I},~
\sum_{\langle ij\rangle_\nu} \!\to\! \frac{1}{2}\!\sum_{\bf{R},(\mu\mu')_\nu},\nonumber
\ee
where ${\bf t}_{\mu\mu'}$ is a primitive translation of the magnetic superlattice that connects the sites $i$ and $j$.
Next, we rotate the  spin operators from the global laboratory frame to local reference frames
\be
\wt{\vec{S}}_{{\bf R},\mu}=\vec{U}_\mu \cdot \vec{S}_{{\bf R},\mu}\,,
\ee
where $\vec{U}_\mu$ is a rotation matrix which depends on the direction of the $\mu$-th spin sublattice in the classical configuration, and express the operators $\wt{\mathbf{S}}_{{\bf R},\mu}$ 
in terms of bosonic operators $a_{{\bf R},\mu}$ via the standard Holstein-Primakoff expansion (to leading order), 
\be
\renewcommand{\arraystretch}{1.5}
\begin{array}{l}
\wt{S}_{{\bf R},\mu}^x \simeq \sqrt{S/2}~ \Big(a_{{\bf R},\mu}^\dagger + a_{{\bf R},\mu}\Big)\,,\\
\wt{S}_{{\bf R},\mu}^y \simeq -i\sqrt{S/2}~ \Big(a_{{\bf R},\mu} - a_{{\bf R},\mu}^\dagger\Big)\,,\\
\wt{S}_{{\bf R},\mu}^z = S-a_{{\bf R},\mu}^\dagger a_{{\bf R},\mu}\,.
\end{array}
\ee
Replacing in Eq.~(\ref{eqn:Raman1}) and expanding in powers of $1/\sqrt{S}$ gives
\be\label{eq:ExpandR}
\mc{R}=\mc{R}_0+\mc{R}_1+\mc{R}_2+\mc{O}(S^{1/2})\,,
\ee
where $\mc{R}_0$ corresponds to a constant term and does not contribute to the scattering, whereas $\mc{R}_1$ and $\mc{R}_2$ describe, respectively, one-magnon and two-magnon scattering.

Knowing the Raman operator, we can then compute the Raman intensity as
\be\label{eqn:Intensity}
\mc{I}(\Omega)
=\frac{1}{2\pi} \int_{-\infty}^{\infty} \! d t~ e^{i \Omega t} ~\langle \mc{R}(t)\mc{R}(0)\rangle\,,
\ee
where $\Omega=\omega_{\rm in}-\omega_{\rm out}$ is the total energy transferred to the system (in units of $\hbar=1$) and $\langle \cdots\rangle$ denotes the ground state average. 
In the following, we shall assume that $\Omega \ll \omega_{\rm in,\, out}$.\\

\subsection{One-magnon scattering}\label{subsec:one}
The one-magnon Raman operator in Eq.~(\ref{eq:ExpandR}) reads
\be\label{eq:R1}
\mc{R}_1=
\sum_{\vec{R},(\mu\mu')_\nu}
\Big\lbrace V_{\mu\mu'}^{(1)}~ (a_{\mathbf{R},\mu}+a_{\mathbf{R}+\mathbf{t}_{\mu\mu'},\mu'})+h.c.\Big\rbrace,
\ee
where
\bea
V_{\mu\mu'}^{(1)} &=&
\frac{S^{3/2}}{2\sqrt{2}}
\Big\{
\mc{P}_{\mu\mu',J}
\big[\mathbf{U}_\mu\cdot\mathbf{U}_{\mu'}^{-1}\big]_{x-iy, z}
+
\mc{P}_{\mu\mu',K}
\big[\mathbf{U}_\mu\big]_{x-iy, \alpha_\nu}
\big[\mathbf{U}_{\mu'}^{-1}\big]_{\alpha_\nu z}
\nonumber\\
&+&
\mc{P}_{\mu\mu',\Gamma}
\left(
\big[\mathbf{U}_\mu\big]_{x-iy, \beta_\nu}~
\big[\mathbf{U}_{\mu'}^{-1}\big]_{\gamma_\nu z}
+\big[\mathbf{U}_\mu\big]_{x-iy, \gamma_\nu}~
\big[\mathbf{U}_{\mu'}^{-1}\big]_{\beta_\nu z}\right)
\nonumber\\
&+&
\mc{P}_{\mu\mu',h_\Gamma}
\big[\mathbf{U}_\mu^{-1}\big]_{ \alpha_\nu,x-iy}
\Big\}\,,
\eea
where we use the notation 
$[\cdots]_{x-iy,\alpha} \equiv [\cdots]_{x\alpha}-i [\cdots]_{y\alpha}$.

Next we switch to momentum space via Fourier transform, 
\be
a_{{\bf R},\mu}=\frac{1}{\sqrt{\mc{N}/\mc{N}_m}}\sum_{\vec{q}} e^{i\vec{q}\cdot(\vec{R}+\bs{\rho}_\mu)} a_{\mu,{\bf q}}\,,
\ee 
where $\mc{N}$ is the total number of sites, $\mc{N}_m$ is the number of sites inside the magnetic unit cell, and $\bs{\rho}_\mu$ denotes the position of the $\mu$-th sublattice inside the unit cell. Keeping only the ${\bf q}=0$ components, we can write 
\be\label{eq:R1q0}
\mc{R}_{1,{\bf q}=0} = {\bf V}^{(1)} \cdot {\bf x}_\vec{q=0}, 
\ee
where ${\bf x}_\vec{q}=(a_{1,\vec{q}}\, ,...\, ,a_{\mc{N}_m,\vec{q}}\, ,a_{1,-\vec{q}}^\dagger\, ,...\, ,a_{\mc{N}_m,-\vec{q}}^\dagger)^{\text{T}}$ and ${\bf V}^{(1)}$ is a $1\times(2\mc{N}_m)$ vector with elements $V^{(1)}_{\mu}=\sum_{\mu'}V^{(1)}_{\mu\mu'}$. 

Note that ${\bf x}_{\bf q}$ appears also explicitly in the quadratic part of the effective spin Hamiltonian, in the form
\begin{align}\label{H2}
\mc{H}_{\rm{LSW}} = \frac{S}{2} \sum_{\bf q} {\bf x}^\dagger_{\bf q} \cdot {\bf H}_{\bf q} \cdot {\bf x}_{\bf q}~,
\end{align}
where ${\bf H}_{\vec{q}}$ is a $(2\mc{N}_m)\times(2\mc{N}_m)$ coupling matrix. This Hamiltonian is diagonalized via a standard Bogoliubov transformation, ${\bf x}_{\bf q}={\bf T}_{\bf q}\cdot{\bf y}_{\bf q}$, where ${\bf T}_{\bf q}$ is the canonical transformation matrix, ${\bf y}_{\bf q} = (b_{1,{\bf q}}\, ,...\, ,b_{\mc{N}_m,{\bf q}}\, ,b_{1,{\bf q}}^\dagger\, ,...\, ,b_{\mc{N}_m,{\bf q}}^\dagger)^{\text{T}}$, which leads to
\be
\mc{H}_{\rm{LSW}} = \sum_{\bf q}\sum_{\mu=1}^{\mc{N}_m} \omega_{\mu,{\bf q}} \Big(
b_{\mu,{\bf q}}^\dagger b_{\mu,{\bf q}} + \frac{1}{2}
\Big)\,,
\ee
where the new bosons $b^\dagger_{\mu,{\bf q}}$ describe the magnon excitations with frequencies $\omega_{\mu,\bf q}$.

Coming back to the Raman operator and expressing ${\bf x}_{\vec{q}=0}=\vec{T_{{\bf q}=0}}\cdot {\bf y}_{\vec{q=0}}$ in Eq.~(\ref{eq:R1q0}) leads to the following expression for the relevant, ${\bf q}=0$ part of the one-magnon Raman operator:
\be\label{eq:R1-bosons}
\mc{R}_{1,{\bf q}=0} =  {\bf M}^{(1)}({\vec{q}}=0)  \cdot {\bf y}_\vec{q=0},
\ee
where ${\bf M}^{(1)}({\vec{q}}=0) = \bf{V}^{(1)}\cdot \vec{T}_\vec{q=0}$.
At zero temperature, it suffices to keep the terms involving $b^\dagger_{\mu,{\bf q}=0}$, since only the processes with a magnon creation on the $\mc{N}_m$ modes $(\mu,{\bf q}=0)$ are allowed. Therefore, the zero temperature one-magnon Raman intensity [Eq.~(\ref{eqn:Intensity}) with $\mc{R}\to \mc{R}_{1,{\bf q}=0}$] can be written as
\be\label{eq:R1_intensity}
\mc{I}_1(\Omega) \propto \sum_{\mu} |{\bf M}_{\mc{N}_m+\mu}^{(1)}({\vec{q}}=0)|^2 ~\delta(\Omega-\omega_{\mu,\vec{q}=0})\,.
\ee
The above equation is the basis for the numerical calculation of the one-magnon Raman scattering intensity. The delta functions are treated by allowing for a small, but otherwise arbitrary Lorentzian broadening  $\delta(x)\to\frac{1}{\pi}\frac{\eta}{x^2+\eta^2}$  with a small enough value for $\eta$.

\subsection{Two-magnon scattering}\label{subsec:two}
The two-magnon scattering involves the second-order term in Eq.~(\ref{eq:ExpandR}), which reduces to
\begin{align}\label{Ram2}
\mc{R}_2\!=&
\!\!\sum_{\mathbf{R},(\mu\mu')_\nu}\!\Bigg\lbrace V^{(2)}_{\mu+\mc{N}_m,\mu'}a_{\mathbf{R},\mu}a_{\mathbf{R}+\mathbf{t}_{\mu\mu'},\mu'}
\!+\!
V^{(2)}_{\mu+\mc{N}_m,\nu+\mc{N}_m}a_{\mathbf{R},\mu}a_{\mathbf{R}+\mathbf{t}_{\mu\mu'},\mu'}^\dagger
\nonumber\\
&~~+ V^{(2)}_{\mu,\mu}a_{\mathbf{R},\mu}^\dagger a_{\mathbf{R},\mu}
+
V^{(2)}_{\mu',\mu'}a_{\mathbf{R}+\mathbf{t}_{\mu\mu'},\mu'}^\dagger a_{\mathbf{R}+\mathbf{t}_{\mu\mu'},\mu'}+h.c.\Bigg\rbrace
\,,
\end{align}
where the various prefactors $V^{(2)}$ can be obtained from Eq.~(\ref{eqn:Raman1}), similarly to the one-magnon case (we do not, however, write them down here since the respective expressions are rather cumbersome).
Using again the Fourier transform and symmetrizing  with respect to $\vec{q}\to-\vec{q}$, we obtain
\be
\mc{R}_2 = \sum_{\vec{q}} {\bf x}_\vec{q}^\dagger \cdot {\bf V}^{(2)}(\vec{q})\cdot {\bf x}_\vec{q} = \sum_{\vec{q}} {\bf y}_\vec{q}^\dagger \cdot {\bf M}^{(2)}(\vec{q})\cdot {\bf y}_\vec{q}, 
\ee
where ${\bf V}^{(2)}(\vec{q})$ and ${\bf M}^{(2)}(\vec{q})={\bf T}_{\bf q}^\dagger\cdot{\bf V}^{(2)}\cdot{\bf T}_{\bf q}$ are $(2\mc{N}_m)\times(2\mc{N}_m)$ matrices. 

While the operator $\mc{R}_2$ contains  all  combinations of the bilinear terms  $b_{\mu,\vec{q}} b_{\mu',\vec{q}}^\dagger,~b_{\mu,\vec{q}} b_{\mu',-\vec{q}},~b_{\mu,-\vec{q}}^\dagger b_{\mu',\vec{q}}^\dagger,~ b_{\mu,-\vec{q}}^\dagger b_{\mu',-\vec{q}}$, at zero temperature only those corresponding to two creation operators contribute to two-magnon scattering. Also, momentum conservation requires that the  momenta of the two magnons must be opposite to each other, i.e., $\vec{q}'=-\vec{q}$. This leads to
\be\label{eq:R2q}
\mc{R}_2 = \sum_{\vec{q}} \sum_{\mu,\mu'=1}^{\mc{N}_m} M^{(2)}_{\mu,~\mu'+\mc{N}_m}(\vec{q})~
b^\dagger_{\mu,\vec{q}} b^\dagger_{\mu',-\vec{q}} + h.c.
\ee
Replacing in Eq.~(\ref{eqn:Intensity}) [with $\mc{R}\to \mc{R}_{2}$] and using a Lehmann spectral representation into the relevant two-magnon space leads to the (zero temperature) two-magnon Raman intensity 
\be\label{eq:R2_intensity}
\mc{I}_2(\Omega)\propto
\sum_{{\bf q},\mu\mu'} 
|M^{(2)}_{\mu,~\mu'+\mc{N}_m}(\vec{q})|^2~ \delta\big(\Omega-\omega_\mu(\vec{q})-\omega_{\mu'}(\vec{q})\big)\,,
\ee
which is the basis for our numerical calculations (with the appropriate Lorentzian Broadening as in the one-magnon case).

\section{Application to $\beta$-Li$_2$IrO$_3$}\label{sec:Ramanbeta}
\begin{figure}[!t]
{\includegraphics[width=0.95\columnwidth]{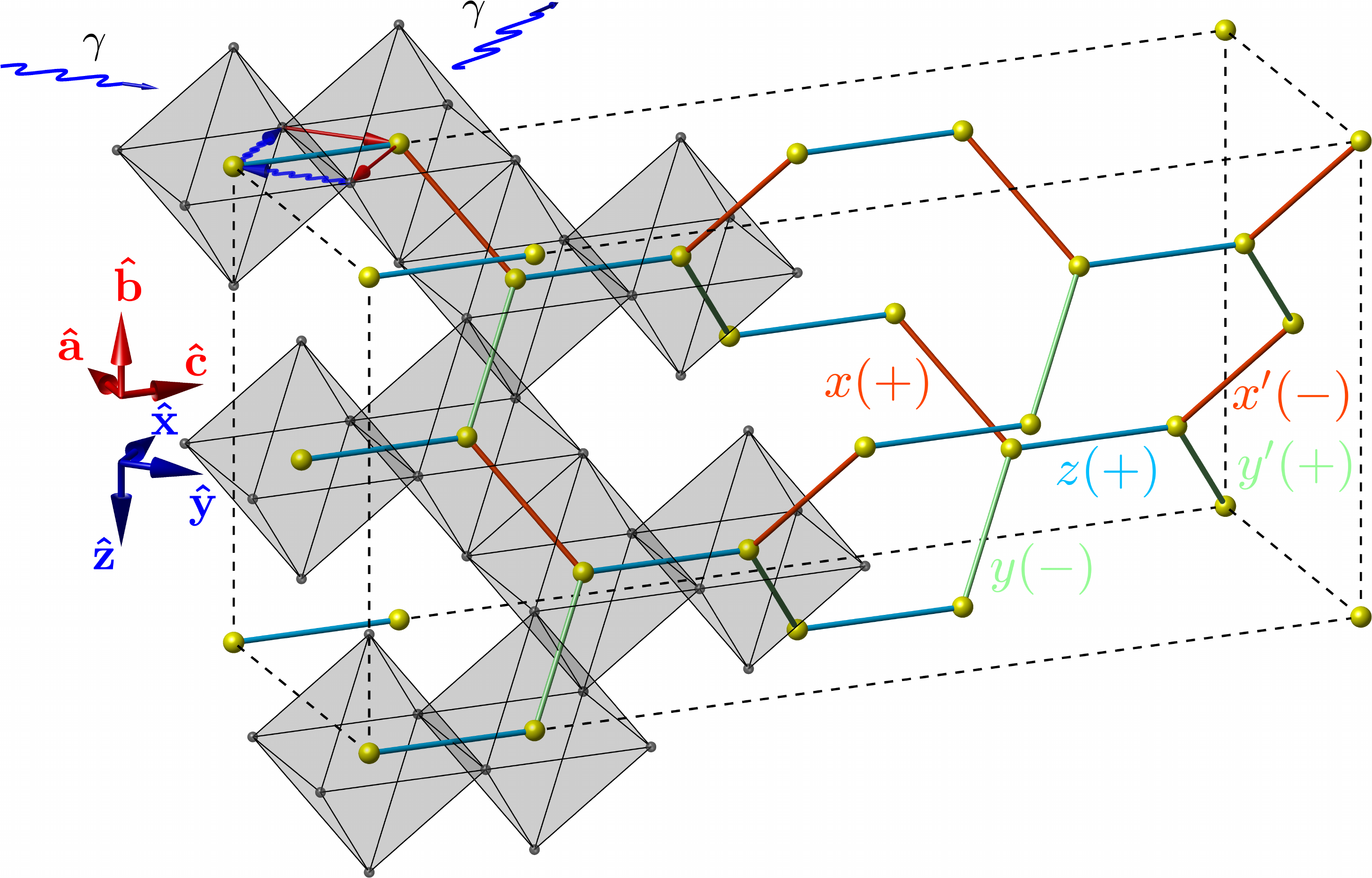}}
\caption{Sketch of a hyperhoneycomb lattice of $\beta$-$\text{Li}_2\text{IrO}_3$.  The orthorhombic unit cell is defined by the crystallographic axes $\{\hat{{\bf a}}, \hat{{\bf b}},\hat{{\bf c}}\}$ related to the Cartesian axes $\{\hat{{\bf x}}, \hat{{\bf y}}, \hat{{\bf z}}\}$ appearing in the spin Hamiltonian by the following relations:
$\hat{{\bf x}}=(\hat{{\bf a}}+\hat{{\bf c}})/\sqrt{2}\,,~~~
\hat{{\bf y}}=(\hat{{\bf c}}-\hat{{\bf a}})/\sqrt{2}\,,~~~
\hat{{\bf z}}=-\hat{{\bf b}}.
$
The five NN bonds of the $J$-$K$-$\Gamma$ model are marked in red for $\bs{d}\in\{x,x'\}$,	green for $\bs{d}\in\{y,y'\}$, and blue for $\bs{d}\in\{z\}$. Each octahedral denotes to the $\text{IrO}_6$ cage. }\label{fig:lattice}
\end{figure}

We are now ready to apply the theory developed in the previous sections to compute the Raman intensity for $\beta$-Li${}_2$IrO${}_3$. 
This compound crystallizes in  a hyperhoneycomb structure, with a conventional orthorhombic unit cell defined by the crystallographic axes $\{\hat{{\bf a}}, \hat{{\bf b}},\hat{{\bf c}}\}$, see Fig.~\ref{fig:lattice}.

\begin{figure}[!b]
{\includegraphics[width=1\columnwidth]{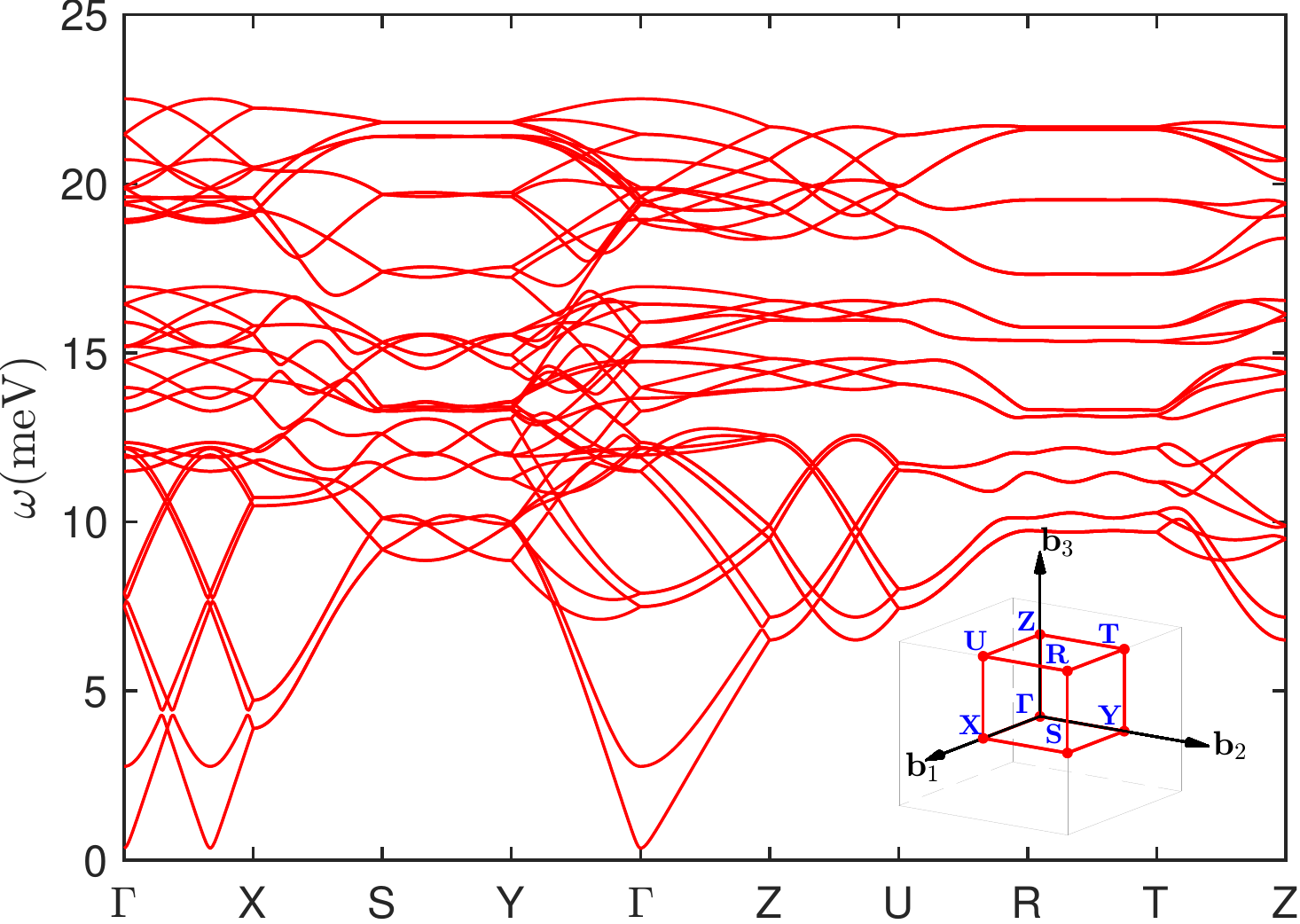}}
\caption{Linear spin wave  spectrum along a high symmetry path in the Brillouin zone of the orthorhombic unit cell (inset).}\label{fig:spinwave}
\end{figure}

At zero field, $\beta$-Li$_2$IrO$_3$ orders magnetically below $T_N\!=\!38$~K. The magnetic structure is characterized by a non-coplanar, incommensurate (IC) modulation, with propagation wavevector ${\bf Q}\!=\!(0.57,0,0)$ (in orthorhombic frame units), and two counter-rotating sets of moments~\cite{Biffin2014a}. According to previous theoretical works, the magnetism of $\beta$-Li$_2$IrO$_3$ can be accurately described by the $J$-$K$-$\Gamma$ model of Eq.~(\ref{eqn:Heff}) with $J\!=\!0.4~\text{meV}$, $K\!=\!-18~\text{meV} $ and $\Gamma\!=\!-10~\text{meV}$ \cite{Lee2015,Lee2016,Ducatman2018, Rousochatzakis2018,Li2019,Li2020,Ruiz2020}.
Furthermore, it has been shown~\cite{Ducatman2018,Rousochatzakis2018} that the IC order of $\beta$-Li$_2$IrO$_3$ can be treated as a long-distance twisting of a nearby commensurate period-3 state with ${\bf Q}\!=\!\frac{2}{3}\hat{\bf a}$ (in units $\frac{2\pi}{a}$). This state is amenable to a semi-analytical treatment, which delivers a very accurate representation of the ground state properties and the magnon excitation spectrum~\cite{Ducatman2018,Rousochatzakis2018,Li2019,Li2020}.
For the latter, we take a magnetic supercell composed of three orthorhombic unit cells along the ${\bf a}$-axis, and thus $\mc{N}_m=48$ spin sites~\cite{Ducatman2018}.
The ensuing 48 magnon branches delivered by the numerical diagonalization of $\mc{H}_{\rm{LSW}}$ is shown in Fig.~\ref{fig:spinwave} along a high symmetry path in the Brillouin zone of the orthorhombic unit cell~\cite{Ducatman2018,Li2019}. 
Note that the spectrum features a nonzero spin gap, which reflects the presence of  anisotropic exchange interactions and the absence of continuous translational symmetry.
 
Turning to the Raman operator and Eq.~(\ref{eqn:Raman1}), we will need the numerical values of the various quantities appearing in Eq.~(\ref{eqn:PijI}). 
These include the vectors ${\bf d}_{ij}$ and ${\bf d}^\perp_{ij}$ of Eq.~(\ref{eqn:PPPP}), as well as the parameters $J^{(2)}$, $J'^{(4)}$,  $K^{(2)}$, $K^{(4)}$, $K'^{(4)}$, $\Gamma^{(3)}$ and $h_\Gamma^{(3)}$.
For the former, there are five types of bonds in $\beta$-Li${}_2$IrO${}_3$, labeled by $x$, $x'$, $y$, $y'$, and $z$ (see Fig.~\ref{fig:lattice}), with   
\be\label{eq:bonds}
\renewcommand{\arraystretch}{1.5}
\begin{array}{c}
\vec{d}_{x}={\vec{d}}^\perp_{x'} =  \frac{1}{2} [1,\sqrt{2},-1],~~~ 
{\vec{d}}^\perp_{x}={\vec{d}}_{x'}=\frac{1}{2} [1,-\sqrt{2},-1], 
\\
\vec{d}_{y}={\vec{d}}^\perp_{y'} =-\frac{1}{2} [1,\sqrt{2},1],~~~
{\vec{d}}^\perp_{y}={\vec{d}}_{y'}=\frac{1}{2} [-1,\sqrt{2},-1], 
\\
\vec{d}_{z} = [0,0,1],~~ \vec{d}^\perp_{z}=[1,0,0]\,,
\end{array}
\ee
in the orthorhombic frame. 
The remaining parameters appearing in Eq.~(\ref{eqn:Raman1}) depend on the hopping matrix elements $t_1$, $t_2=t^2/\Delta_{\text{pd}}$, and $t_3$, where $\Delta_{\text{pd}}$ is the charge transfer energy, and the interaction terms $U_2$, $J_H$ and $\lambda$.
Fixing the latter to the typical values of $U_2=1.8$ eV, $J_H=0.4$ eV and $\lambda=0.4$ eV, allows to adjust $t_1$, $t_2$ and $t_3$ so that we reproduce the values of $J$, $K$, $\Gamma$ mentioned above.  This gives $t_1=-0.042$ eV, $t_2=0.332$ eV, and $t_3=0.190$ eV, which are within the typical range of density functional theory (DFT) calculations, see, e.g., [\onlinecite{Kim2015}].
Furthermore, we assume that the incoming light is off-resonance and neglect the frequency of the incoming light, i.e., we set $\omega_{\text {in}}=0$ in the expressions for the Raman operator. 
With these assumptions and numerical estimates we arrive at: $J^{(2)}=0.4$ meV, $J'^{(4)}=-101.797$ meV, $K^{(2)}=3.49$ meV, $K^{(4)}=-21.49$ meV, $K^{'(4)}=210.757$ meV, $\Gamma^{(3)}=-10$ meV and $h_\Gamma^{(3)}=14.615$ meV.
Quite remarkably, the parameters $J'^{(4)}$ and $K'^{(4)}$ that are partly responsible for the non-Loudon-Fleury Raman scattering, have much larger magnitude compared to the corresponding values of $J^{(2)}$ and $K^{(2)}+K^{(4)}$ of the Loudon-Fleury terms. Similarly, the magnitude of $h_\Gamma^{(3)}$, which does not have any analogue in the spin Hamiltonian, is appreciably high as well. This tells us that the Raman intensity (which scales quadratically with the parameters) is dominated by the non-Loudon-Fleury scattering terms. This significant result will be demonstrated explicitly below.

Having the numerical values of the various quantities appearing in Eq.~(\ref{eqn:Raman1}) we can now calculate the one- and two-magnon Raman intensity of $\beta$-Li$_2$IrO$_3$ using Eqs.~(\ref{eq:R1_intensity}) and (\ref{eq:R2_intensity}), respectively.
In particular, we shall focus on scattering geometries corresponding to incoming and outgoing light polarizations along the orthorhombic crystal axes. 
Among these are the diagonal polarization channels where $\bs{\varepsilon}_{\text{in}}=\bs{\varepsilon}_{\text{out}}= {\bf a}$ or ${\bf b}$ or ${\bf c}$ (which we shall label by $\mc{R}^{aa}$, $\mc{R}^{bb}$ and $\mc{R}^{cc}$, respectively), as well as off-diagonal polarization channels where, e.g., $\bs{\varepsilon}_{\text{in}}={\bf a}$ and $\bs{\varepsilon}_{\text{out}}={\bf b}$ (which we label as $\mc{R}^{ab}$), etc.
A symmetry analysis based on the $D_{2h}$ point group~\footnote{The zero-field ground state of  $\beta$-$\text{Li}_2\text{IrO}_3$~\cite{Biffin2014a,Ruiz2017,Majumder2019} breaks some of the symmetries of the lattice~\cite{Ducatman2018,Li2019}, but the point group of this state is isomorphic to $D_{2h}$ so we can still use this group for the analysis of the Raman scattering channels.} shows that $\mc{R}^{aa}$, $\mc{R}^{bb}$ and $\mc{R}^{cc}$ transform according to the $A_g$ irreducible representation, while $\mc{R}^{ab}$, $\mc{R}^{ac}$ and $\mc{R}^{bc}$ transform as $B_{1g}$, $B_{2g}$ and $B_{3g}$,respectively~\cite{Brent2015,Brentthesis}.

\begin{figure*}
{\includegraphics[width=1\linewidth]{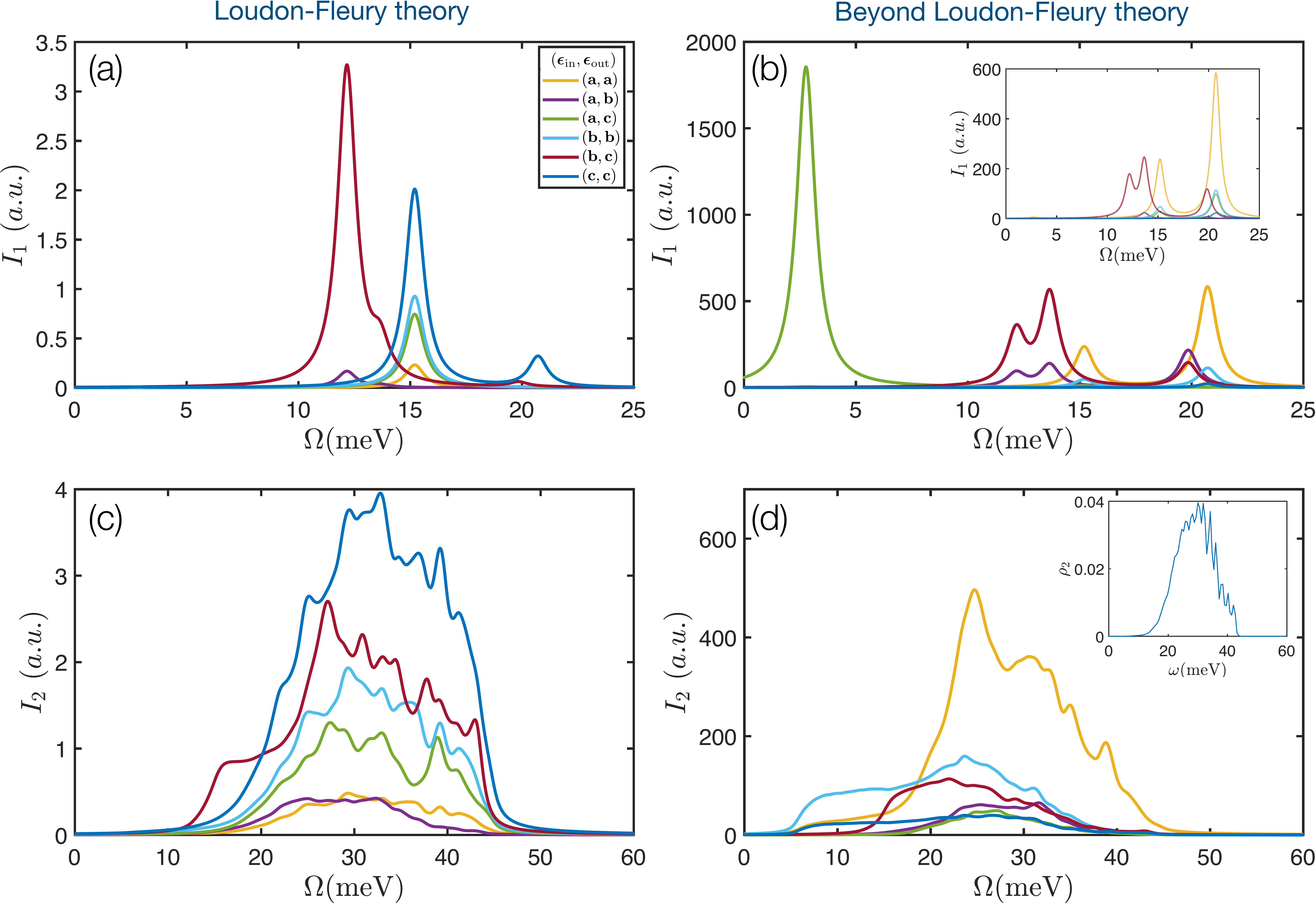}}
\caption{
One-magnon (a-b) and two-magnon (c-d) Raman intensities, computed with [(b) and (d)] and without [(a) and (c)] taking into account the non-Loudon-Fleury scattering terms, see main text.
Lines with different colors correspond to different polarization channels (with the color scheme being consistent across all the panels). 
The inset of (b) shows the Raman response without the non-Loudon-Fleury,  magnetic-dipole term $\propto h_\Gamma^{(3)}$. %
The inset of (d) shows the two-magnon density of states $\rho_2$ of Eq.~(\ref{eqn:2dos}).}\label{fig:OneMagnonTwoMagnon}
\end{figure*}

{\it One-magnon response.}  Figs.~\ref{fig:OneMagnonTwoMagnon}\,(a-b) show the one-magnon Raman scattering intensities in the polarization channels mentioned above, as obtained from numerical calculations based on Eq.~(\ref{eq:R1_intensity}) with a Lorentzian broadening parameter $\eta=0.5$~meV. 
To disentangle the contributions coming from the non-Loudon-Fleury terms we perform calculations with (panel b) and without (panel a) these terms. 
A quick inspection of the intensity scales in the two panels demonstrates the dramatic impact of the non-Loudon-Fleury terms announced above, namely that these terms dominate the scattering.
Another significant ramification of these terms is that the sharp peak at $\Omega\sim 3$ meV, appearing in the $ac$-polarization channel in Fig.~\ref{fig:OneMagnonTwoMagnon}\,(b), is absent from Fig.~\ref{fig:OneMagnonTwoMagnon}\,(a). This one-magnon $\mathbf{q}=0$ peak, therefore,  originates from the non-Loudon-Fleury terms. We have checked, in particular, that this peak stems from the magnetic dipole-active terms $\propto h_\Gamma^{(3)}$, see inset of Figs.~\ref{fig:OneMagnonTwoMagnon}\,(c). It is furthermore noteworthy that this peak is absent in the remaining polarization channels shown in Fig.~\ref{fig:OneMagnonTwoMagnon}\,(b), which can be used as a smoking-gun diagnostic feature in experiments~\footnote{This peak has, in fact, been observed experimentally \cite{Yipingunpublished} and will be discussed elsewhere.}.

The strong polarization dependence is not special to the low-energy peak, but manifests in the higher-energy part of the response as well, as shown in Fig.~\ref{fig:OneMagnonTwoMagnon}\,(b). Comparing with panel (a), the non-Loudon-Fleury terms play a decisive role, as they modify significantly the relative intensity and overall shape of the high-energy peaks.

{\it Two-magnon response.}
We now turn to the  two-magnon intensities shown in Figs.~\ref{fig:OneMagnonTwoMagnon}\,(c-d). 
As above, we disentangle the contributions from the non-Loudon-Fleury terms by performing calculations with (panel d) and without (panel c) these terms.
The intensities are computed using  Eq.~(\ref{eq:R2_intensity}), where the sum over ${\bf q}$ in Eq.~(\ref{eq:R2_intensity}) is carried out on a finite-size grid of 25200 ${\bf q}$ points within the magnetic Brillouin zone, and the Lorentzian broadening parameter is chosen to be $\eta=0.6$~ meV.

Quite generally, the 2-magnon response features a broad continuum, mainly due to the fact that Eq.~(\ref{eq:R2_intensity}) involves a sum over all ${\bf q}$-modes. This sum has the form of a convolution between a polarization-dependent weight $|M^{(2)}({\bf q})|^2$ and the two-magnon density of states, defined as \be\label{eqn:2dos}
\rho_2(\omega)=\sum_{\mu,\mu',{\bf q}}\delta\left(\omega-\omega_{\mu}(\bf q)-\omega_{\mu'}(-{\bf q})\right)\,.
\ee
The latter is calculated using  a histogram method  and is shown for comparison in the inset of Fig.~\ref{fig:OneMagnonTwoMagnon}\,(d) and reproduces well the  bandwidth and overall shape of the response. 

A quick inspection of the intensity scales in panels (c) and (d) shows that the two-magnon intensity too is dominated by the non-Loudon-Fleury terms.
Furthermore, these terms change significantly the relative intensities of the various polarization channels. For example, the $\mc{R}^{aa}$ channel features the largest response, unlike the computed intensities based on the Loudon-Fleury terms alone (panel c). These significant changes come with distinctive features which can again be tested experimentally.

  Here we also note that the  two-magnon intensities shown in Figs.~\ref{fig:OneMagnonTwoMagnon}\,(c-d) are obtained without taking into account  the effects of the final-state magnon-magnon interactions \cite{Elliott1963}, which might be not small  given the complex nature of the   magnetic ordering in $\beta$-Li$_2$IrO$_3$.  In principle, their effect can be taken into account by computing the vertex corrections to the bare Raman vertex in Eq.(49), cred although this is a technically  rather tedious task due to the large  number of magnon bands in $\beta$-Li$_2$IrO$_3$. Qualitatively, we expect that
 these corrections  can lead to a shifting of the two-magnon peaks to lower energies and to the formation of an even broader continuum at the higher energies, similarly to the cases considered in Refs.~\cite{Chubukov1995} and \cite{Perkins2008}.
 
 Finally, we point out that the numerical results presented here correspond to the case where the incoming light is off-resonance. 
 The case of resonance, i.e., when the  frequency of the incoming photon is comparable to the charge gap, requires further analysis~ \cite{Chubukov1995,Frenkel1995}.

\section{Discussion}\label{sec:discussion}
We have revisited the theory of magnetic Raman scattering in Mott insulators with strong spin-orbit coupling with a special focus on Kitaev materials.  
A detailed consideration of the precise photon-assisted, virtual hopping processes that contribute to the magnetic Raman sattering reveals that the Raman vertex $\mc{R}$ contains terms beyond those appearing in the traditional Loudon-Fleury theory.
Quite remarkably, these non-Loudon-Fleury terms are shown to dominate the scattering intensity in the three-dimensional Kitaev material $\beta$-Li$_2$IrO$_3$ by at least two orders of magnitude.
In addition, the non-Loudon-Fleury terms give rise to a qualitative  modification of the polarization dependence, with distinctive signatures that can be tested experimentally. 
Most saliently, in $\beta$-Li$_2$IrO$_3$ the non-Loudon-Fleury terms give rise to a sharp magnetic dipole-active magnon peak at low energies, which is absent in the traditional Loudon-Fleury theory. 
This peak has been observed recently in the predicted $ac$-polarization channel~ \cite{Yipingunpublished}, lending strong support to the importance of the non-Loudon-Fleury terms.
The peak is shown to arise from virtual tunneling processes involving both direct and ligand mediated paths. These processes are of similar type with the ones leading to the symmetric off-diagonal interaction $\Gamma$, but, in the Raman vertex, they take the form of a bond-directional magnetic dipole term. In particular, these processes involve an intermediate hopping to the ligand (oxygen in $\beta$-Li$_2$IrO$_3$), which does not conserve the projection of the total pseudo-spin along the corresponding axis (e.g., $S_i^z\!+\!S_j^z$ for the $z$-type of bonds). 


 On a broader perspective, we would also  like to emphasize that 
our theory is fully applicable to any strong spin-orbit-coupled Mott insulator, in which the magnetic moments $j_{\text{eff}}\!=\!1/2$ come from the five electrons (or one hole) on the $t_{2g}$ orbitals.
For Kitaev materials, in particular, we even expect similar quantitative results with the ones presented here for $\beta$-Li$_2$IrO$_3$, as the underlying local geometry (and the effective spin Hamiltonian description) of $\beta$-Li$_2$IrO$_3$ is common in all Kitaev materials. 
Specifically, we expect the same type of non-Loudon-Fleury terms (including the magnetic dipole term $\propto h_\Gamma^{(3)}$) to be present generically across all Kitaev materials, and we also anticipate that these will dominate the scattering intensity, given the similar order of magnitude of the microscopic parameters $U_2$, $J_H$, $\lambda$ and $\Delta_{\text{pd}}$.
The presented analysis therefore underpins a drastic change of paradigm for the understanding of Raman scattering in materials with strong spin-orbit coupling and multiple exchange paths. In addition, it calls for a general re-evaluation of Raman scattering in Kitaev materials of current interest, as this would help to elucidate their correct microscopic description 
and their relative proximity to the sought-after quantum spin liquid.

{\it Note added:} We recently became aware that a modification of exchange interactions similar to the one presented here for $\mc{R}$ has been discussed for Kitaev materials under magnetic field~ \cite{Natori2019} and circularly polarized light~ \cite{Arakawa2021}. In particular, Ref.~\cite{Natori2019} reports an effective magnetic field term similar to $h_{\Gamma}^{(3)}$, which also arises from mixed hopping terms.

\vspace*{0.3cm} 
\noindent{\it  Acknowledgments:} 
We thank Kenneth Burch and Yiping  Wang for helpful discussions and for sharing with us unpublished  Raman data on $\beta$-$\text{Li}_2\text{IrO}_3$. 
The work by Y.Y, M.L. and N.B.P.  was supported by the U.S. Department of Energy, Office of Science, Basic Energy Sciences under Award No. DE-SC0018056. We also acknowledge the support of the Minnesota Supercomputing Institute (MSI) at the University of Minnesota. N.B.P. and Y.Y.  are thankful for the hospitality of Kavli Institute for Theoretical Physics and the support of
 the National Science Foundation under Grant No. NSF PHY-1748958.

\appendix
\titleformat{\section}{\normalfont\bfseries\filcenter}{Appendix~\thesection:}{0.25em}{}


\section{Technical details of the derivation of the Raman operator}\label{App:Ramandetails} 

\subsection {The Raman operator from the processes involving the direct hoppings} \label{App:Ramandirect} 
\begin{figure*}
\includegraphics[width=0.7\linewidth]{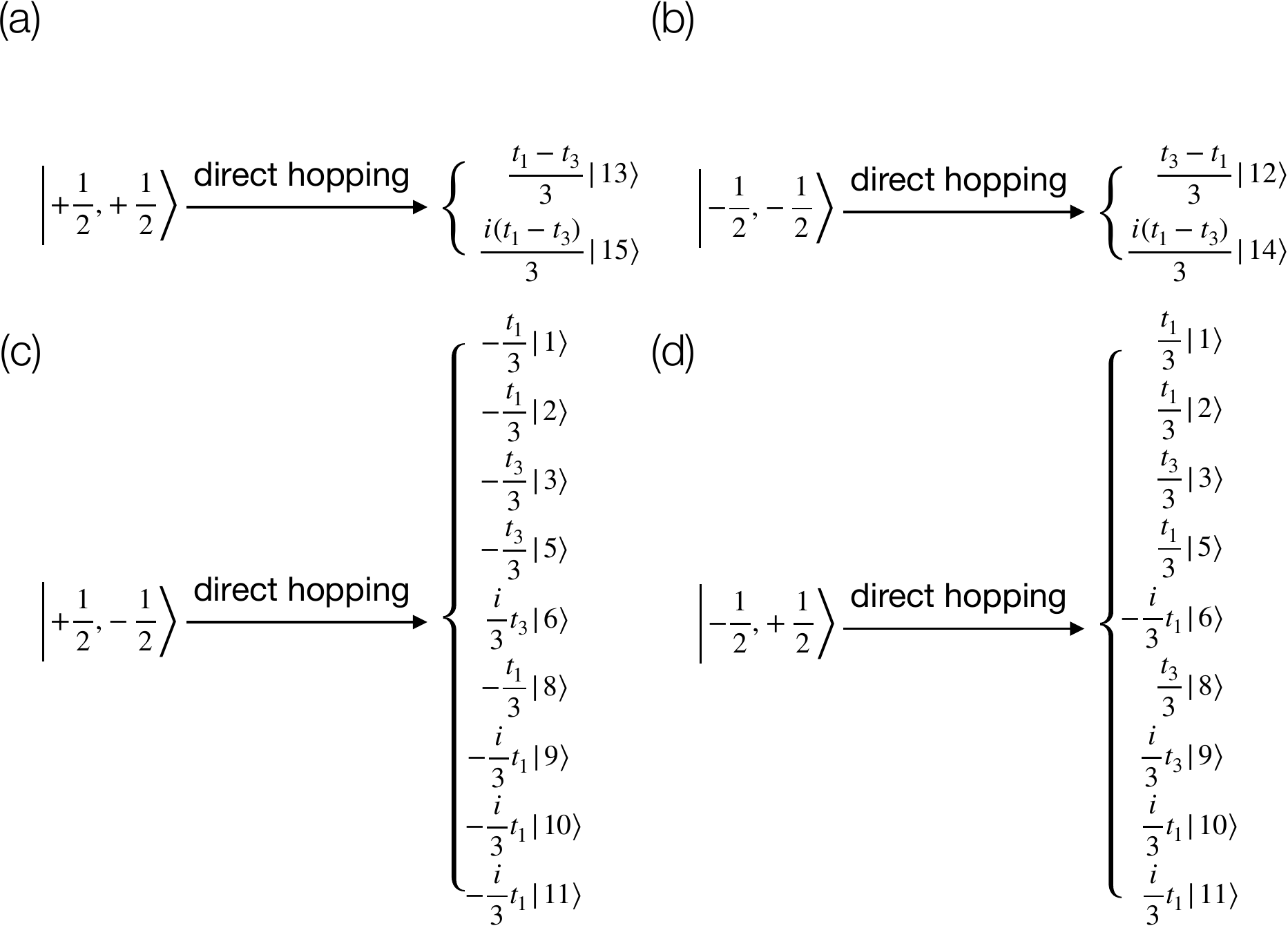}
\caption{Direct hopping from one-hole state on Ir${}_1$ and Ir${}_2$ to the intermediate two-hole states (denoted in Table~\ref{tbl:twohole_orbital}) on Ir${}_2$ from one of four possible  $|\psi_n\rangle$ ground states: (a) $|\psi_1\rangle\equiv|+1/2;+1/2\rangle$, (b) $|\psi_4\rangle\equiv|-1/2;-1/2\rangle$, (c) $|\psi_3\rangle\equiv|+1/2;-1/2\rangle$,  and (d) $|\psi_2\rangle\equiv|-1/2;+1/2\rangle$.}\label{fig:direct_construction}
\end{figure*}

\begin{figure*}
\includegraphics[width=0.7\linewidth]{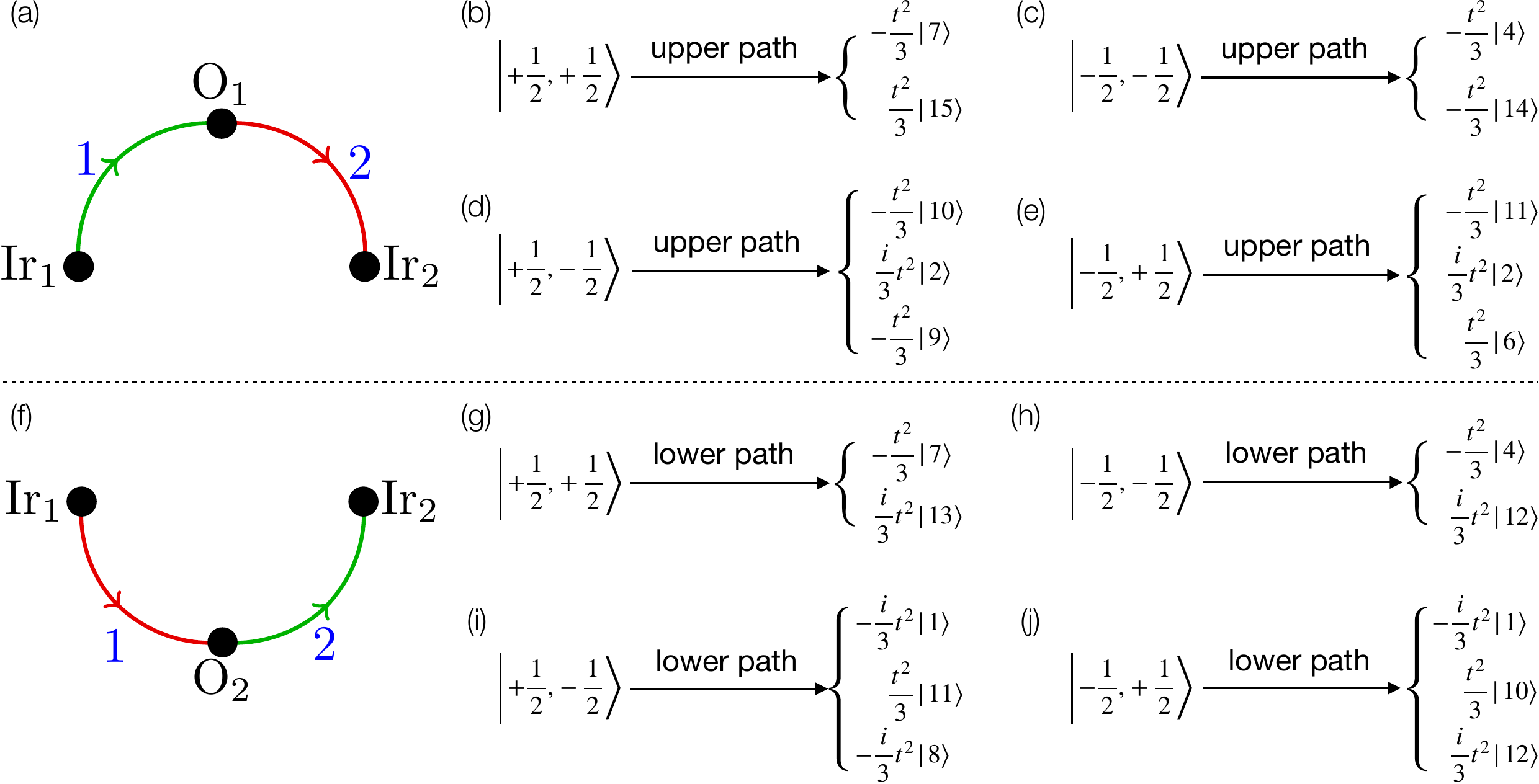}
\caption{Oxygen mediated hopping from one-hole state on Ir${}_1$ and Ir${}_2$ to the intermediate two-hole states (denoted in Table~\ref{tbl:twohole_orbital}) on Ir${}_2$ from one of four  possible  $|\psi_n\rangle$ ground states.   Panels  (b-e) represent  the hopping via the upper path (a)  and  panels (g-j) represent the  hopping via the lower path (f).}
\label{fig:oxy_hopping}
\end{figure*}
    
Since  it is more convenient to  represent the hopping matrix $\mc{H}_t$  in the orbital basis, we project the $j_{\text{eff}}=1/2$ pseudospin degree of freedom to the orbital basis using
\bea
\renewcommand{\arraystretch}{1.5}
\begin{array}{l}
|j^z_{\text{eff}} =+\frac{1}{2}\rangle=-\frac{1}{\sqrt{3}}\Big(
|X,\downarrow\rangle + i |Y,\downarrow\rangle + |Z,\uparrow\rangle
\Big)\\
|j^z_{\text{eff}} =-\frac{1}{2}\rangle=-\frac{1}{\sqrt{3}}\Big(
|X,\uparrow\rangle - i |Y,\uparrow\rangle - |Z,\downarrow\rangle
\Big)
\end{array}
\eea
Next we notice that there are fifteen intermediate  states $|D_\mu\rangle$  with two holes on the iridium ion,  which can be obtained by the diagonalization of $\mc{H}_{\text{int}}$ and can be easily written in the two-hole orbital basis at the zero SOC limit $\lambda\rightarrow 0$.
Explicitly, we denote the two-hole orbital basis in the way given in the Table~{\ref{tbl:twohole_orbital}}.

\begin{table}[!b]
\renewcommand{\arraystretch}{2}
\centering
\begin{tabular}{|c|c|c|} 
\hline
$|X\uparrow, X\downarrow\rangle\equiv|1\rangle$ & $|Y\uparrow, Z\uparrow\rangle\equiv|6\rangle$&$|X\downarrow, Y\uparrow\rangle\equiv|11\rangle$\\
\hline
$|Y\uparrow, Y\downarrow\rangle\equiv|2\rangle$ & $|X\downarrow, Y\downarrow\rangle\equiv|7\rangle$ & $|X\uparrow, Z\downarrow\rangle\equiv |12\rangle$\\
\hline
$|Z\uparrow, Z\downarrow\rangle\equiv|3\rangle$ & $|X\downarrow, Z\downarrow\rangle\equiv|8\rangle$ & $|X\downarrow, Z\uparrow\rangle\equiv|13\rangle$\\
\hline
$|X\uparrow, Y\uparrow\rangle\equiv|4\rangle$ & $|Y\downarrow, Z\downarrow\rangle\equiv|9\rangle$ & $|Y\uparrow, Z\downarrow\rangle\equiv |14\rangle$\\
\hline
$|X\uparrow, Z\uparrow\rangle\equiv|5\rangle$ & $|X\uparrow, Y\downarrow\rangle\equiv|10\rangle$ & $|Y\downarrow, Z\uparrow\rangle\equiv|15\rangle$\\
\hline
\end{tabular}
\caption{Intermediate, two-hole (Slater determinant) states.\label{tbl:twohole_orbital}}
\end{table}

There are several processes that contribute to the spin interaction and we consider those involving the direct  hopping only. These processes give rise  to the  effective spin coupling at the second order of perturbation theory. The  half of the direct hopping path   projected  into all possible orbital channels is shown in Fig.~\ref{fig:direct_construction}. In this process, the hopping starts from the Ir$_1$ ion  from one of the $j^z_{\text{eff}}=\pm 1/2$ states and ends on Ir$_2$ ion in one  of the fifteen  states belonging to the two-hole orbital basis. 
  
For example,  in Fig.~\ref{fig:direct_construction} (a), we show the action of the hopping term on the $|\psi_1\rangle$ state, which we denote as  $\mc{H}_t|\psi_1\rangle\equiv\mc{H}_t|+1/2;+1/2\rangle$. The hole on Ir${}_1$ is first projected to the orbital basis $|X\downarrow\rangle$, $|Y\downarrow\rangle$ and $|Z\uparrow\rangle$. Each of these states  overlaps with the two-hope states on  Ir${}_2$. Recalling that we are interested only on those two-hole states, in which  one hole is the original hope on the   $|j^z_{\text{eff}}=1/2\rangle_2$,  we  should take into account the projection of $|j^z_{\text{eff}}=1/2\rangle_2$ to the orbital basis. The product of weights from the projection gives the weights for each channel of hopping in the orbital basis. The sum of all the contributions allows us to explicitly compute the matrix element $\langle D_\mu|\mc{H}_t|\psi_1\rangle$, where  $|D_\mu\rangle$  denotes the two-hole intermediate state in the $\mu=1,2...15$-th state entering into Eq.~(\ref{eqn:Rdirdir}).
The same calculation can be performed for other states $\mc{H}_t|n\rangle$ as illustrated in Fig.~\ref{fig:direct_construction} (b)-(d). Next we compute $\langle D_\mu|{\mc H}_{t}|n\rangle$ and $\langle D_\mu|{\mc H}_{t}|m\rangle^*$ to form the complete hopping path. The process starting from Ir${}_2$ gives exactly the same result. 

When we consider the Raman operator, we should recall that the incoming light must couple to the hopping on the first bond, and the outgoing light must couple to the  hopping on the last bond of the path. With only two hopping bonds in the direct hopping path, the polarization factor can only be $(\bs{\varepsilon}_{\text{in}}\cdot\mathbf{d}_{ij})(\bs{\varepsilon}_{\text{out}}\cdot(-\mathbf{d}_{ij}))$, 
%
and so we can compute the Raman operator originated from the direct hopping as
\begin{align}\label{app:Rdir}
\mc{R}^{\text{dir}}_{\langle ij\rangle_z}= &-2\zeta(\bs{\varepsilon}_{\text{in}}\cdot\mathbf{d}_{ij})(\bs{\varepsilon}_{\text{out}}\cdot\mathbf{d}_{ij})\nonumber\\&\sum_{n,n'}\sum_{\mu}\frac{\langle D_\mu|\mc{H}_{t,ij}|\psi_n\rangle^*\cdot\langle D_\mu|\mc{H}_{t,ij}|\psi_{n'}\rangle}{2E_{1\text{h}}+\omega_{\text{in}}-(E_{2\text{h}}+E_{0\text{h}})}|\psi_n\rangle\langle \psi_{n'}|,
\end{align}
where $\bs{\varepsilon}_{\text{out}}$ as the outgoing light polarization, and  $|\psi_n\rangle$, $|\psi_{n'}\rangle$ again represent the four states of the low-energy sector of two magnetic ions, namely $|\frac{1}{2},\frac{1}{2}\rangle$, $|\frac{1}{2},-\frac{1}{2}\rangle$, $|-\frac{1}{2},\frac{1}{2}\rangle$, $|-\frac{1}{2},-\frac{1}{2}\rangle$. 
 
\subsection {The Raman operator from the processes involving the oxygen-mediated hopping} \label{App:Ramanmediated} 
There are  eight  paths that include the oxygen-mediated hopping. These processes give rise  to the  effective  spin coupling at the  fourth  order of perturbation theory, and the sum over all eight  contributions gives us the super-exchange Hamiltonian with the dominant Kitaev interaction \cite{Jackeli2010,Rau2014, Sizyuk2014}. Each of these paths also gives the contribution to the Raman operator (\ref{eqn:oxygen}), which apart from the corresponding polarization prefactor is proportional to \begin{widetext}
\begin{align}
\mc{H}_\ell^O=&\,\mc{H}_t\,\mc{G}_O\,\mc{H}_t\,\mc{G}_D\,\mc{H}_t\,\mc{G}_O\,\mc{H}_t=\mc{H}_t\frac{|O_{\nu'}\rangle\langle O_{\nu'}|}{\omega_{\text{in}}-\Delta_{\text{pd}}}\mc{H}_t\left(\sum_{\mu}\frac{|D_\mu\rangle\langle D_\mu|}{2E_{1\text{h}}+\omega_{\text{in}}-(E_{2\text{h}}+E_{0\text{h}})}\right)\mc{H}_t\frac{|O_{\nu} \rangle\langle O_{\nu}|}{\omega_{\text{in}}-\Delta_{\text{pd}}}\mc{H}_t
\nonumber\\=&
\sum_{n,n'}\sum_{\mu}\frac{\left(\langle \psi_n|\mc{H}_t|O_{\nu'}\rangle\langle O_{\nu'}|\mc{H}_t|D_\mu\rangle\right)\left(\langle D_\mu|\mc{H}_t|O_\nu\rangle\langle O_\nu|\mc{H}_t|\psi_{n'}\rangle\right)}{\left(2E_{1\text{h}}+\omega_{\text{in}}-(E_{2\text{h}}+E_{0\text{h}})\right)(\omega_{\text{in}}-\Delta_{\text{pd}})^2}|\psi_{n}\rangle\langle \psi_{n'}|.
\end{align}
\end{widetext}
The choice of $|O_\nu\rangle$ determines whether the upper path or the lower path is considered (see Fig.~\ref{fig:oxy_hopping} (a) and (f), respectively). The explicit construction of $\mc{H}_t|O_\nu\rangle\langle O_\nu|\mc{H}_t|\psi_{n'}\rangle$  with all projection factors is  obtained on the similar way as for the  paths with the direct hopping only and is shown
in Fig.~\ref{fig:oxy_hopping}. 

Once we have $\langle D_\mu|\mc{H}_t|O_\nu\rangle\langle O_\nu|\mc{H}_t|\psi_{n'}\rangle$ computed for all $\nu$ and $\beta$, we can assemble the hopping path for the oxygen-mediated hopping as
\begin{align*}
\left(\langle D_\mu|\mc{H}_t|O_{\nu'}\rangle\langle O_{\nu'}|\mc{H}_t|\psi_n\rangle\right)^*\cdot\left(\langle D_\mu|\mc{H}_t|O_\nu\rangle\langle O_\nu|\mc{H}_t|\psi_{n'}\rangle\right).
\end{align*}
This gives us the following expressions for  $\mc{H}^{O}_{\ell}$:
\begin{widetext}
\begin{align}
\mc{H}^{O}_{\ell}= \sum_{n,n'}\sum_{\mu}\frac{\left(\langle D_\mu|\mc{H}_t|O_{\nu'}\rangle\langle O_{\nu'}|\mc{H}_t|\psi_n\rangle\right)^*\cdot\left(\langle D_\mu|\mc{H}_t|O_\nu\rangle\langle O_\nu|\mc{H}_t|\psi_{n'}\rangle\right)}{(2E_{1\text{h}}+\omega_{\text{in}}-(E_{2\text{h}}+E_{0\text{h}}))(\omega_{\text{in}}-\Delta_{\text{pd}})^2}|\psi_n\rangle\langle \psi_{n'}|,
\end{align}	    
\end{widetext}
where the combination of  the indices $\nu$ and $\nu'$ determines the path involved in the process; e.g., when $\nu=1$ and $\nu'=1$, the hole only hops through $O_1$ hence giving $\mc{H}_1^O$. Symmetry leads to the following equivalence relations between $\mc{H}_\ell^O$ corresponding to the processes starting from Ir${}_2$ and processes starting from Ir${}_1$ ions:
\begin{align}\label{eqn:Equivalence1}
\mc{H}_2^O\sim\mc{H}_7^O\,,
\mc{H}_4^O\sim\mc{H}_3^O\,,
\mc{H}_6^O\sim\mc{H}_5^O\,,
\mc{H}_8^O\sim\mc{H}_1^O\,,
\end{align}
in which the equivalence relation means the equality of matrix elements involving $\psi_2$ and $\psi_3$ under interchanging $\psi_2\!\mapsto\!\psi_3$ and $\psi_3\!\mapsto\!\psi_2$. For example, $\langle \psi_2|\mc{H}_2^O|\psi_2\rangle=\langle \psi_3|\mc{H}_7^O|\psi_3\rangle$, $\langle \psi_2|\mc{H}_2^O|\psi_3\rangle=\langle \psi_3|\mc{H}_7^O|\psi_2\rangle$,	and so on. 
Summing up over all paths (both starting at Ir$_1$ and Ir$_2$) with the corresponding polarization prefactors in the way given in Eq.~(\ref{fig:oxygen-hopping-paths}) leads to the final expression Eq.~(\ref{eqn:Rmed}) for the Raman operator $\mc{R}_{\langle ij\rangle_z}^{\text{med}}$ 
coming from oxygen-mediated processes.

\subsection {The Raman operator from the processes involving mixed direct and the oxygen-mediated hopping} \label{App:Ramanmixed} 
 
The mixed hopping can be viewed as the combination of the direct hopping and the oxygen-mediated hopping, so we can compute $\mc{H}_i^m$ associated with each path as
\begin{widetext}
\begin{align}
\mc{H}_\ell^m=&\sum_{n,n'}|\psi_n\rangle\langle \psi_n|\mc{H}_t\,\mc{G}\,\mc{H}_t\,\mc{G}\,\mc{H}_t|\psi_{n'}\rangle\langle \psi_{n'}|\nonumber=
\begin{cases}
\displaystyle\sum_{n,n'}\sum_{\mu}\frac{\left(\langle D_\mu|\mc{H}_t|\psi_n\rangle\right)^*\cdot\left(\langle D_\mu|\mc{H}_t|O_\nu\rangle\langle O_\nu|\mc{H}_t|\psi_{n'}\rangle\right)}{(2E_{1\text{h}}+\omega_{\text{in}}-(E_{2\text{h}}+E_{0\text{h}}))(\omega_{\text{in}}-\Delta_{\text{pd}})}|\psi_n\rangle\langle \psi_{n'}|
\\[15pt]
\displaystyle\sum_{n,n'}\sum_{\mu}\frac{\left(\langle D_\mu|\mc{H}_t|O_\nu\rangle\langle O_\nu|\mc{H}_t|\psi_n\rangle\right)^*\cdot\left(\langle D_\mu|\mc{H}_t|\psi_{n'}\rangle\right)}{(\omega_{\text{in}}-\Delta_{\text{pd}})(2E_{1\text{h}}+\omega_{\text{in}}-(E_{2\text{h}}+E_{0\text{h}}))}|\psi_n\rangle\langle \psi_{n'}|  
\end{cases},\label{eqn:mixed_hopping_path}
\end{align}
\end{widetext} 
where the first case represent the path starting from the oxygen-mediated hopping and the second one from the direct hopping. The overlap matrix elements $\langle \dots\rangle$ in (\ref{eqn:mixed_hopping_path}) are computed either for the direct hopping process or for the oxygen-mediated hopping process, so no additional consideration needs to be taken here. The equivalence relations between the processes starting from Ir${}_2$ and from Ir${}_1$ are given by
\begin{align}
\mc{H}^m_2\sim\mc{H}^m_5\,,
\mc{H}^m_4\sim\mc{H}^m_7\,,
\mc{H}^m_6\sim\mc{H}^m_1\,,
\mc{H}^m_8\sim\mc{H}^m_3\,,
\end{align}
where the equivalence has the same meaning as in Eq.~(\ref{eqn:Equivalence1}) above. 
Finally, summing up over all paths with the corresponding polarization factors leads to the final expression of Eq.~(\ref{eqn:Rmix}) for the Raman operator $\mc{R}_{\langle ij\rangle_z}^{\text{mix}}$ coming from the mixed processes.

\subsection {Analytic expressions of the prefactor constants in the Raman operator}\label{App:Analytic} 
Here we provide the analytic expressions for the prefactor constants entering the Raman operator. For simpler algebra manipulation, the calculations are done using the total-$J$ basis (e.g. $|J,J_z\rangle$) instead of the orbital basis. The details of using this basis can be found at \cite{Perkins2014}. Denoting
\begin{widetext}
    \begin{align}
    f_1=&-\frac{1}{3}\left(\frac{J_H}{6J_H^2+J_H(U_1+4\lambda-\omega_{\text{in}})-(U_1-\omega_{\text{in}})(U_1+3\lambda-\omega_{\text{in}})}\right),\\
    f_2=&\frac{4}{3}\left(\frac{3J_H-U_1-3\lambda+\omega_{\text{in}}}{6J_H-2U_1-3\lambda+2\omega_{\text{in}}}\right)\frac{J_H}{6J_H^2-J_H(8U_1+17\lambda-8\omega_{\text{in}})+(2U_1+3\lambda-2\omega_{\text{in}})(U_1+3\lambda-\omega_{\text{in}})},\\
    f_3=&\frac{7J_H-3U_1-9\lambda+3\omega_{\text{in}}}{6J_H^2-J_H(8U_1+17\lambda-8\omega_{\text{in}})+(2U_1+3\lambda-2\omega_{\text{in}})(U_1+3\lambda-\omega_{\text{in}})},\\
    f_4=&\frac{1}{6J_H-2U_1-3\lambda+2\omega_{\text{in}}},
\end{align}
\end{widetext}
the various coupling constants take the following form
\begin{align}
    J^{(2)}&=\frac{4}{9}f_1\,(2t_1+t_3)^2-\frac{8}{9}f_2\,\left(9t_4^2+2(t_1-t_3)^2\right),\\
    K^{(2)}&=\frac{8}{3}f_2\,\left(3t_4^2+(t_1-t_3)^2\right),\\
    \Gamma^{(2)}&=8f_2\,t_4^2,~~~
    \Gamma'^{(2)}=-\frac{8}{3}f_2\, t_4(t_1-t_3),\\
    \Gamma^{(3)}&=\frac{16}{3}f_2\,t_2(t_1-t_3),~~~
    \Gamma'^{(3)}=8f_2\, t_2 t_4,\\
    K^{(4)}&=-8f_2\,t_2^2\,,
\end{align}
where we have introduced $t_4$ for the general case with lower bond symmetry (in the main text  $t_4=0$). In the limit $\omega_\text{in}\rightarrow 0$, these coupling constants reduce to the super-exchange coupling constants for the nearest neighbor $JK\Gamma\Gamma'$-model, and are in agreement with expressions given in Ref.~\cite{Winter2016}. The remaining coupling constants that are associated with the non-Loudon-Fleury processes take the form
\begin{align}
    J'^{(4)}&=\frac{16}{9}\left(f_2-f_1\right)t_2^2,\\
    K'^{(4)}&=\frac{8}{9}\left(4f_1-f_2\right)t_2^2,\\
    h_\Gamma^{(3)}&=\frac{8}{3}\left(\frac{1}{2}f_2+\frac{1}{3}f_4\right)t_2(t_1-t_3),\\
    h_\Gamma'^{(3)}&=-\frac{1}{3}(f_3+f_4)t_2t_4,\\
    \tilde{\Gamma}^{(3)}&=-\frac{2}{9}(f_3-3f_4)t_2 t_4\,.
\end{align}
The last two coupling constants, which are associated with $t_4$, give rise to the following additional terms in the Raman operator
\begin{widetext}
\begin{align}
   & -\frac{1}{2}\sum_{\langle ij \rangle_\nu}(P_{d^\perp d}-P_{dd^\perp}) i h_\Gamma'^{(3)}(S_i^{\beta_\nu}+S_i^{\gamma_\nu}+S_j^{\beta_\nu} +S_j^{\gamma_\nu})\\& -\frac{1}{2}\sum_{\langle ij \rangle_\nu}(P_{d^\perp d}+P_{dd^\perp})\tilde{\Gamma}^{(3)}\left(S_i^{\alpha_\nu}(S_j^{\beta_\nu}-S_j^{\gamma_\nu})
   +(S_i^{\beta_\nu}-S_i^{\gamma_\nu})S_j^{\alpha_\nu}\right),
\end{align}
which are ignored in the main text where $t_4=0$.
\end{widetext}

\bibliographystyle{apsrev4-1}
\bibliography{reference.bib}

\begin{thebibliography}{77}%
\makeatletter
\providecommand \@ifxundefined [1]{%
 \@ifx{#1\undefined}
}%
\providecommand \@ifnum [1]{%
 \ifnum #1\expandafter \@firstoftwo
 \else \expandafter \@secondoftwo
 \fi
}%
\providecommand \@ifx [1]{%
 \ifx #1\expandafter \@firstoftwo
 \else \expandafter \@secondoftwo
 \fi
}%
\providecommand \natexlab [1]{#1}%
\providecommand \enquote  [1]{``#1''}%
\providecommand \bibnamefont  [1]{#1}%
\providecommand \bibfnamefont [1]{#1}%
\providecommand \citenamefont [1]{#1}%
\providecommand \href@noop [0]{\@secondoftwo}%
\providecommand \href [0]{\begingroup \@sanitize@url \@href}%
\providecommand \@href[1]{\@@startlink{#1}\@@href}%
\providecommand \@@href[1]{\endgroup#1\@@endlink}%
\providecommand \@sanitize@url [0]{\catcode `\\12\catcode `\$12\catcode
  `\&12\catcode `\#12\catcode `\^12\catcode `\_12\catcode `\%12\relax}%
\providecommand \@@startlink[1]{}%
\providecommand \@@endlink[0]{}%
\providecommand \url  [0]{\begingroup\@sanitize@url \@url }%
\providecommand \@url [1]{\endgroup\@href {#1}{\urlprefix }}%
\providecommand \urlprefix  [0]{URL }%
\providecommand \Eprint [0]{\href }%
\providecommand \doibase [0]{http://dx.doi.org/}%
\providecommand \selectlanguage [0]{\@gobble}%
\providecommand \bibinfo  [0]{\@secondoftwo}%
\providecommand \bibfield  [0]{\@secondoftwo}%
\providecommand \translation [1]{[#1]}%
\providecommand \BibitemOpen [0]{}%
\providecommand \bibitemStop [0]{}%
\providecommand \bibitemNoStop [0]{.\EOS\space}%
\providecommand \EOS [0]{\spacefactor3000\relax}%
\providecommand \BibitemShut  [1]{\csname bibitem#1\endcsname}%
\let\auto@bib@innerbib\@empty
\bibitem [{\citenamefont {Devereaux}\ and\ \citenamefont
  {Hackl}(2007)}]{Devereaux2007}%
  \BibitemOpen
  \bibfield  {author} {\bibinfo {author} {\bibfnamefont {T.~P.}\ \bibnamefont
  {Devereaux}}\ and\ \bibinfo {author} {\bibfnamefont {R.}~\bibnamefont
  {Hackl}},\ }\href {\doibase 10.1103/RevModPhys.79.175} {\bibfield  {journal}
  {\bibinfo  {journal} {Rev. Mod. Phys.}\ }\textbf {\bibinfo {volume} {79}},\
  \bibinfo {pages} {175} (\bibinfo {year} {2007})}\BibitemShut {NoStop}%
\bibitem [{\citenamefont {Shastry}\ and\ \citenamefont
  {Shraiman}(1990)}]{Shastry1990}%
  \BibitemOpen
  \bibfield  {author} {\bibinfo {author} {\bibfnamefont {B.~S.}\ \bibnamefont
  {Shastry}}\ and\ \bibinfo {author} {\bibfnamefont {B.~I.}\ \bibnamefont
  {Shraiman}},\ }\href {\doibase 10.1103/PhysRevLett.65.1068} {\bibfield
  {journal} {\bibinfo  {journal} {Phys. Rev. Lett.}\ }\textbf {\bibinfo
  {volume} {65}},\ \bibinfo {pages} {1068} (\bibinfo {year}
  {1990})}\BibitemShut {NoStop}%
\bibitem [{\citenamefont {Shastry}\ and\ \citenamefont
  {Shraiman}(1991)}]{Shastry1991}%
  \BibitemOpen
  \bibfield  {author} {\bibinfo {author} {\bibfnamefont {B.~S.}\ \bibnamefont
  {Shastry}}\ and\ \bibinfo {author} {\bibfnamefont {B.~I.}\ \bibnamefont
  {Shraiman}},\ }\href
  {http://www.worldscientific.com/doi/abs/10.1142/S0217979291000237} {\bibfield
   {journal} {\bibinfo  {journal} {Int. J. Mod. Phys. B}\ }\textbf {\bibinfo
  {volume} {5}},\ \bibinfo {pages} {365} (\bibinfo {year} {1991})}\BibitemShut
  {NoStop}%
\bibitem [{\citenamefont {Chubukov}\ and\ \citenamefont
  {Frenkel}(1995{\natexlab{a}})}]{Chubukov1995}%
  \BibitemOpen
  \bibfield  {author} {\bibinfo {author} {\bibfnamefont {A.~V.}\ \bibnamefont
  {Chubukov}}\ and\ \bibinfo {author} {\bibfnamefont {D.~M.}\ \bibnamefont
  {Frenkel}},\ }\href {\doibase 10.1103/PhysRevLett.74.3057} {\bibfield
  {journal} {\bibinfo  {journal} {Phys. Rev. Lett.}\ }\textbf {\bibinfo
  {volume} {74}},\ \bibinfo {pages} {3057} (\bibinfo {year}
  {1995}{\natexlab{a}})}\BibitemShut {NoStop}%
\bibitem [{\citenamefont {Chubukov}\ and\ \citenamefont
  {Frenkel}(1995{\natexlab{b}})}]{Frenkel1995}%
  \BibitemOpen
  \bibfield  {author} {\bibinfo {author} {\bibfnamefont {A.~V.}\ \bibnamefont
  {Chubukov}}\ and\ \bibinfo {author} {\bibfnamefont {D.~M.}\ \bibnamefont
  {Frenkel}},\ }\href {\doibase 10.1103/PhysRevB.52.9760} {\bibfield  {journal}
  {\bibinfo  {journal} {Phys. Rev. B}\ }\textbf {\bibinfo {volume} {52}},\
  \bibinfo {pages} {9760} (\bibinfo {year} {1995}{\natexlab{b}})}\BibitemShut
  {NoStop}%
\bibitem [{\citenamefont {Benfatto}\ \emph {et~al.}(2006)\citenamefont
  {Benfatto}, \citenamefont {Silva~Neto}, \citenamefont {Gozar}, \citenamefont
  {Dennis}, \citenamefont {Blumberg}, \citenamefont {Miller}, \citenamefont
  {Komiya},\ and\ \citenamefont {Ando}}]{Benfatto2006}%
  \BibitemOpen
  \bibfield  {author} {\bibinfo {author} {\bibfnamefont {L.}~\bibnamefont
  {Benfatto}}, \bibinfo {author} {\bibfnamefont {M.~B.}\ \bibnamefont
  {Silva~Neto}}, \bibinfo {author} {\bibfnamefont {A.}~\bibnamefont {Gozar}},
  \bibinfo {author} {\bibfnamefont {B.~S.}\ \bibnamefont {Dennis}}, \bibinfo
  {author} {\bibfnamefont {G.}~\bibnamefont {Blumberg}}, \bibinfo {author}
  {\bibfnamefont {L.~L.}\ \bibnamefont {Miller}}, \bibinfo {author}
  {\bibfnamefont {S.}~\bibnamefont {Komiya}}, \ and\ \bibinfo {author}
  {\bibfnamefont {Y.}~\bibnamefont {Ando}},\ }\href {\doibase
  10.1103/PhysRevB.74.024416} {\bibfield  {journal} {\bibinfo  {journal} {Phys.
  Rev. B}\ }\textbf {\bibinfo {volume} {74}},\ \bibinfo {pages} {024416}
  (\bibinfo {year} {2006})}\BibitemShut {NoStop}%
\bibitem [{\citenamefont {C\'epas}\ \emph {et~al.}(2008)\citenamefont
  {C\'epas}, \citenamefont {Haerter},\ and\ \citenamefont
  {Lhuillier}}]{Cepas2008a}%
  \BibitemOpen
  \bibfield  {author} {\bibinfo {author} {\bibfnamefont {O.}~\bibnamefont
  {C\'epas}}, \bibinfo {author} {\bibfnamefont {J.~O.}\ \bibnamefont
  {Haerter}}, \ and\ \bibinfo {author} {\bibfnamefont {C.}~\bibnamefont
  {Lhuillier}},\ }\href {\doibase 10.1103/PhysRevB.77.172406} {\bibfield
  {journal} {\bibinfo  {journal} {Phys. Rev. B}\ }\textbf {\bibinfo {volume}
  {77}},\ \bibinfo {pages} {172406} (\bibinfo {year} {2008})}\BibitemShut
  {NoStop}%
\bibitem [{\citenamefont {Ko}\ \emph {et~al.}(2010)\citenamefont {Ko},
  \citenamefont {Liu}, \citenamefont {Ng},\ and\ \citenamefont {Lee}}]{Ko2010}%
  \BibitemOpen
  \bibfield  {author} {\bibinfo {author} {\bibfnamefont {W.-H.}\ \bibnamefont
  {Ko}}, \bibinfo {author} {\bibfnamefont {Z.-X.}\ \bibnamefont {Liu}},
  \bibinfo {author} {\bibfnamefont {T.-K.}\ \bibnamefont {Ng}}, \ and\ \bibinfo
  {author} {\bibfnamefont {P.~A.}\ \bibnamefont {Lee}},\ }\href {\doibase
  10.1103/PhysRevB.81.024414} {\bibfield  {journal} {\bibinfo  {journal} {Phys.
  Rev. B}\ }\textbf {\bibinfo {volume} {81}},\ \bibinfo {pages} {024414}
  (\bibinfo {year} {2010})}\BibitemShut {NoStop}%
\bibitem [{\citenamefont {Wulferding}\ \emph {et~al.}(2010)\citenamefont
  {Wulferding}, \citenamefont {Lemmens}, \citenamefont {Scheib}, \citenamefont
  {R\"oder}, \citenamefont {Mendels}, \citenamefont {Chu}, \citenamefont
  {Han},\ and\ \citenamefont {Lee}}]{Wulferding2010}%
  \BibitemOpen
  \bibfield  {author} {\bibinfo {author} {\bibfnamefont {D.}~\bibnamefont
  {Wulferding}}, \bibinfo {author} {\bibfnamefont {P.}~\bibnamefont {Lemmens}},
  \bibinfo {author} {\bibfnamefont {P.}~\bibnamefont {Scheib}}, \bibinfo
  {author} {\bibfnamefont {J.}~\bibnamefont {R\"oder}}, \bibinfo {author}
  {\bibfnamefont {P.}~\bibnamefont {Mendels}}, \bibinfo {author} {\bibfnamefont
  {S.}~\bibnamefont {Chu}}, \bibinfo {author} {\bibfnamefont {T.}~\bibnamefont
  {Han}}, \ and\ \bibinfo {author} {\bibfnamefont {Y.~S.}\ \bibnamefont
  {Lee}},\ }\href {\doibase 10.1103/PhysRevB.82.144412} {\bibfield  {journal}
  {\bibinfo  {journal} {Phys. Rev. B}\ }\textbf {\bibinfo {volume} {82}},\
  \bibinfo {pages} {144412} (\bibinfo {year} {2010})}\BibitemShut {NoStop}%
\bibitem [{\citenamefont {Perkins}\ and\ \citenamefont
  {Brenig}(2008)}]{Perkins2008}%
  \BibitemOpen
  \bibfield  {author} {\bibinfo {author} {\bibfnamefont {N.}~\bibnamefont
  {Perkins}}\ and\ \bibinfo {author} {\bibfnamefont {W.}~\bibnamefont
  {Brenig}},\ }\href {\doibase 10.1103/PhysRevB.77.174412} {\bibfield
  {journal} {\bibinfo  {journal} {Phys. Rev. B}\ }\textbf {\bibinfo {volume}
  {77}},\ \bibinfo {pages} {174412} (\bibinfo {year} {2008})}\BibitemShut
  {NoStop}%
\bibitem [{\citenamefont {Perkins}\ \emph {et~al.}(2013)\citenamefont
  {Perkins}, \citenamefont {Chern},\ and\ \citenamefont
  {Brenig}}]{Perkins2013}%
  \BibitemOpen
  \bibfield  {author} {\bibinfo {author} {\bibfnamefont {N.~B.}\ \bibnamefont
  {Perkins}}, \bibinfo {author} {\bibfnamefont {G.-W.}\ \bibnamefont {Chern}},
  \ and\ \bibinfo {author} {\bibfnamefont {W.}~\bibnamefont {Brenig}},\ }\href
  {\doibase 10.1103/PhysRevB.87.174423} {\bibfield  {journal} {\bibinfo
  {journal} {Phys. Rev. B}\ }\textbf {\bibinfo {volume} {87}},\ \bibinfo
  {pages} {174423} (\bibinfo {year} {2013})}\BibitemShut {NoStop}%
\bibitem [{\citenamefont {Sen}\ \emph {et~al.}(2019)\citenamefont {Sen},
  \citenamefont {Yao}, \citenamefont {Heid}, \citenamefont {Omoumi},
  \citenamefont {Hardy}, \citenamefont {Willa}, \citenamefont {Merz},
  \citenamefont {Haghighirad},\ and\ \citenamefont {Le~Tacon}}]{Sen2019}%
  \BibitemOpen
  \bibfield  {author} {\bibinfo {author} {\bibfnamefont {K.}~\bibnamefont
  {Sen}}, \bibinfo {author} {\bibfnamefont {Y.}~\bibnamefont {Yao}}, \bibinfo
  {author} {\bibfnamefont {R.}~\bibnamefont {Heid}}, \bibinfo {author}
  {\bibfnamefont {A.}~\bibnamefont {Omoumi}}, \bibinfo {author} {\bibfnamefont
  {F.}~\bibnamefont {Hardy}}, \bibinfo {author} {\bibfnamefont
  {K.}~\bibnamefont {Willa}}, \bibinfo {author} {\bibfnamefont
  {M.}~\bibnamefont {Merz}}, \bibinfo {author} {\bibfnamefont {A.~A.}\
  \bibnamefont {Haghighirad}}, \ and\ \bibinfo {author} {\bibfnamefont
  {M.}~\bibnamefont {Le~Tacon}},\ }\href {\doibase 10.1103/PhysRevB.100.104301}
  {\bibfield  {journal} {\bibinfo  {journal} {Phys. Rev. B}\ }\textbf {\bibinfo
  {volume} {100}},\ \bibinfo {pages} {104301} (\bibinfo {year}
  {2019})}\BibitemShut {NoStop}%
\bibitem [{\citenamefont {Sandilands}\ \emph {et~al.}(2015)\citenamefont
  {Sandilands}, \citenamefont {Tian}, \citenamefont {Plumb}, \citenamefont
  {Kim},\ and\ \citenamefont {Burch}}]{Sandilands2015}%
  \BibitemOpen
  \bibfield  {author} {\bibinfo {author} {\bibfnamefont {L.~J.}\ \bibnamefont
  {Sandilands}}, \bibinfo {author} {\bibfnamefont {Y.}~\bibnamefont {Tian}},
  \bibinfo {author} {\bibfnamefont {K.~W.}\ \bibnamefont {Plumb}}, \bibinfo
  {author} {\bibfnamefont {Y.-J.}\ \bibnamefont {Kim}}, \ and\ \bibinfo
  {author} {\bibfnamefont {K.~S.}\ \bibnamefont {Burch}},\ }\href {\doibase
  10.1103/PhysRevLett.114.147201} {\bibfield  {journal} {\bibinfo  {journal}
  {Phys. Rev. Lett.}\ }\textbf {\bibinfo {volume} {114}},\ \bibinfo {pages}
  {147201} (\bibinfo {year} {2015})}\BibitemShut {NoStop}%
\bibitem [{\citenamefont {Sandilands}\ \emph {et~al.}(2016)\citenamefont
  {Sandilands}, \citenamefont {Tian}, \citenamefont {Reijnders}, \citenamefont
  {Kim}, \citenamefont {Plumb}, \citenamefont {Kim}, \citenamefont {Kee},\ and\
  \citenamefont {Burch}}]{Sandilands2016}%
  \BibitemOpen
  \bibfield  {author} {\bibinfo {author} {\bibfnamefont {L.~J.}\ \bibnamefont
  {Sandilands}}, \bibinfo {author} {\bibfnamefont {Y.}~\bibnamefont {Tian}},
  \bibinfo {author} {\bibfnamefont {A.~A.}\ \bibnamefont {Reijnders}}, \bibinfo
  {author} {\bibfnamefont {H.-S.}\ \bibnamefont {Kim}}, \bibinfo {author}
  {\bibfnamefont {K.~W.}\ \bibnamefont {Plumb}}, \bibinfo {author}
  {\bibfnamefont {Y.-J.}\ \bibnamefont {Kim}}, \bibinfo {author} {\bibfnamefont
  {H.-Y.}\ \bibnamefont {Kee}}, \ and\ \bibinfo {author} {\bibfnamefont
  {K.~S.}\ \bibnamefont {Burch}},\ }\href {\doibase 10.1103/PhysRevB.93.075144}
  {\bibfield  {journal} {\bibinfo  {journal} {Phys. Rev. B}\ }\textbf {\bibinfo
  {volume} {93}},\ \bibinfo {pages} {075144} (\bibinfo {year}
  {2016})}\BibitemShut {NoStop}%
\bibitem [{\citenamefont {Gretarsson}\ \emph {et~al.}(2016)\citenamefont
  {Gretarsson}, \citenamefont {Sung}, \citenamefont {H\"oppner}, \citenamefont
  {Kim}, \citenamefont {Keimer},\ and\ \citenamefont
  {Le~Tacon}}]{Gretarsson2016}%
  \BibitemOpen
  \bibfield  {author} {\bibinfo {author} {\bibfnamefont {H.}~\bibnamefont
  {Gretarsson}}, \bibinfo {author} {\bibfnamefont {N.~H.}\ \bibnamefont
  {Sung}}, \bibinfo {author} {\bibfnamefont {M.}~\bibnamefont {H\"oppner}},
  \bibinfo {author} {\bibfnamefont {B.~J.}\ \bibnamefont {Kim}}, \bibinfo
  {author} {\bibfnamefont {B.}~\bibnamefont {Keimer}}, \ and\ \bibinfo {author}
  {\bibfnamefont {M.}~\bibnamefont {Le~Tacon}},\ }\href {\doibase
  10.1103/PhysRevLett.116.136401} {\bibfield  {journal} {\bibinfo  {journal}
  {Phys. Rev. Lett.}\ }\textbf {\bibinfo {volume} {116}},\ \bibinfo {pages}
  {136401} (\bibinfo {year} {2016})}\BibitemShut {NoStop}%
\bibitem [{\citenamefont {Glamazda}\ \emph {et~al.}(2016)\citenamefont
  {Glamazda}, \citenamefont {Lemmens}, \citenamefont {Do}, \citenamefont
  {Choi},\ and\ \citenamefont {Choi}}]{Glamazda2016}%
  \BibitemOpen
  \bibfield  {author} {\bibinfo {author} {\bibfnamefont {A.}~\bibnamefont
  {Glamazda}}, \bibinfo {author} {\bibfnamefont {P.}~\bibnamefont {Lemmens}},
  \bibinfo {author} {\bibfnamefont {S.-H.}\ \bibnamefont {Do}}, \bibinfo
  {author} {\bibfnamefont {Y.~S.}\ \bibnamefont {Choi}}, \ and\ \bibinfo
  {author} {\bibfnamefont {K.-Y.}\ \bibnamefont {Choi}},\ }\href {\doibase
  http://dx.doi.org/10.1038/ncomms12286} {\bibfield  {journal} {\bibinfo
  {journal} {Nat. Commun.}\ }\textbf {\bibinfo {volume} {7}},\ \bibinfo {pages}
  {12286} (\bibinfo {year} {2016})}\BibitemShut {NoStop}%
\bibitem [{\citenamefont {Sahasrabudhe}\ \emph {et~al.}(2020)\citenamefont
  {Sahasrabudhe}, \citenamefont {Kaib}, \citenamefont {Reschke}, \citenamefont
  {German}, \citenamefont {Koethe}, \citenamefont {Buhot}, \citenamefont
  {Kamenskyi}, \citenamefont {Hickey}, \citenamefont {Becker}, \citenamefont
  {Tsurkan}, \citenamefont {Loidl}, \citenamefont {Do}, \citenamefont {Choi},
  \citenamefont {Gr\"uninger}, \citenamefont {Winter}, \citenamefont {Wang},
  \citenamefont {Valent\'{\i}},\ and\ \citenamefont {van
  Loosdrecht}}]{Sahasrabudhe2020}%
  \BibitemOpen
  \bibfield  {author} {\bibinfo {author} {\bibfnamefont {A.}~\bibnamefont
  {Sahasrabudhe}}, \bibinfo {author} {\bibfnamefont {D.~A.~S.}\ \bibnamefont
  {Kaib}}, \bibinfo {author} {\bibfnamefont {S.}~\bibnamefont {Reschke}},
  \bibinfo {author} {\bibfnamefont {R.}~\bibnamefont {German}}, \bibinfo
  {author} {\bibfnamefont {T.~C.}\ \bibnamefont {Koethe}}, \bibinfo {author}
  {\bibfnamefont {J.}~\bibnamefont {Buhot}}, \bibinfo {author} {\bibfnamefont
  {D.}~\bibnamefont {Kamenskyi}}, \bibinfo {author} {\bibfnamefont
  {C.}~\bibnamefont {Hickey}}, \bibinfo {author} {\bibfnamefont
  {P.}~\bibnamefont {Becker}}, \bibinfo {author} {\bibfnamefont
  {V.}~\bibnamefont {Tsurkan}}, \bibinfo {author} {\bibfnamefont
  {A.}~\bibnamefont {Loidl}}, \bibinfo {author} {\bibfnamefont {S.~H.}\
  \bibnamefont {Do}}, \bibinfo {author} {\bibfnamefont {K.~Y.}\ \bibnamefont
  {Choi}}, \bibinfo {author} {\bibfnamefont {M.}~\bibnamefont {Gr\"uninger}},
  \bibinfo {author} {\bibfnamefont {S.~M.}\ \bibnamefont {Winter}}, \bibinfo
  {author} {\bibfnamefont {Z.}~\bibnamefont {Wang}}, \bibinfo {author}
  {\bibfnamefont {R.}~\bibnamefont {Valent\'{\i}}}, \ and\ \bibinfo {author}
  {\bibfnamefont {P.~H.~M.}\ \bibnamefont {van Loosdrecht}},\ }\href {\doibase
  10.1103/PhysRevB.101.140410} {\bibfield  {journal} {\bibinfo  {journal}
  {Phys. Rev. B}\ }\textbf {\bibinfo {volume} {101}},\ \bibinfo {pages}
  {140410} (\bibinfo {year} {2020})}\BibitemShut {NoStop}%
\bibitem [{\citenamefont {Wulferding}\ \emph {et~al.}(2020)\citenamefont
  {Wulferding}, \citenamefont {Choi}, \citenamefont {Do}, \citenamefont {Lee},
  \citenamefont {Lemmens}, \citenamefont {Faugeras},\ and\ \citenamefont
  {Gallais}}]{Dirk2020}%
  \BibitemOpen
  \bibfield  {author} {\bibinfo {author} {\bibfnamefont {D.}~\bibnamefont
  {Wulferding}}, \bibinfo {author} {\bibfnamefont {Y.}~\bibnamefont {Choi}},
  \bibinfo {author} {\bibfnamefont {S.-H.}\ \bibnamefont {Do}}, \bibinfo
  {author} {\bibfnamefont {C.~H.}\ \bibnamefont {Lee}}, \bibinfo {author}
  {\bibfnamefont {P.}~\bibnamefont {Lemmens}}, \bibinfo {author} {\bibfnamefont
  {C.}~\bibnamefont {Faugeras}}, \ and\ \bibinfo {author} {\bibfnamefont
  {K.-Y.}\ \bibnamefont {Gallais}, \bibfnamefont {Yann~andChoi}},\ }\href
  {\doibase 10.1038/s41467-020-15370-1} {\bibfield  {journal} {\bibinfo
  {journal} {Nat. Commun.}\ }\textbf {\bibinfo {volume} {11}},\ \bibinfo
  {pages} {1603} (\bibinfo {year} {2020})}\BibitemShut {NoStop}%
\bibitem [{\citenamefont {Wang}\ \emph {et~al.}(2020)\citenamefont {Wang},
  \citenamefont {Osterhoudt}, \citenamefont {Tian}, \citenamefont
  {Lampen-Kelley}, \citenamefont {Banerjee}, \citenamefont {Goldstein},
  \citenamefont {Yan}, \citenamefont {Knolle}, \citenamefont {Ji},
  \citenamefont {Cava}, \citenamefont {Nasu}, \citenamefont {Motome},
  \citenamefont {Nagler}, \citenamefont {Mandrus},\ and\ \citenamefont
  {Burch}}]{Yiping2020}%
  \BibitemOpen
  \bibfield  {author} {\bibinfo {author} {\bibfnamefont {Y.}~\bibnamefont
  {Wang}}, \bibinfo {author} {\bibfnamefont {G.~B.}\ \bibnamefont
  {Osterhoudt}}, \bibinfo {author} {\bibfnamefont {Y.}~\bibnamefont {Tian}},
  \bibinfo {author} {\bibfnamefont {P.}~\bibnamefont {Lampen-Kelley}}, \bibinfo
  {author} {\bibfnamefont {A.}~\bibnamefont {Banerjee}}, \bibinfo {author}
  {\bibfnamefont {T.}~\bibnamefont {Goldstein}}, \bibinfo {author}
  {\bibfnamefont {J.}~\bibnamefont {Yan}}, \bibinfo {author} {\bibfnamefont
  {J.}~\bibnamefont {Knolle}}, \bibinfo {author} {\bibfnamefont
  {H.}~\bibnamefont {Ji}}, \bibinfo {author} {\bibfnamefont {R.~J.}\
  \bibnamefont {Cava}}, \bibinfo {author} {\bibfnamefont {J.}~\bibnamefont
  {Nasu}}, \bibinfo {author} {\bibfnamefont {Y.}~\bibnamefont {Motome}},
  \bibinfo {author} {\bibfnamefont {S.~E.}\ \bibnamefont {Nagler}}, \bibinfo
  {author} {\bibfnamefont {D.}~\bibnamefont {Mandrus}}, \ and\ \bibinfo
  {author} {\bibfnamefont {K.~S.}\ \bibnamefont {Burch}},\ }\href {\doibase
  10.1038/s41535-020-0216-6} {\bibfield  {journal} {\bibinfo  {journal} {npj
  Quantum Materials}\ }\textbf {\bibinfo {volume} {5}},\ \bibinfo {pages} {14}
  (\bibinfo {year} {2020})}\BibitemShut {NoStop}%
\bibitem [{\citenamefont {Knolle}\ \emph {et~al.}(2014)\citenamefont {Knolle},
  \citenamefont {Chern}, \citenamefont {Kovrizhin}, \citenamefont {Moessner},\
  and\ \citenamefont {Perkins}}]{Knolle2014}%
  \BibitemOpen
  \bibfield  {author} {\bibinfo {author} {\bibfnamefont {J.}~\bibnamefont
  {Knolle}}, \bibinfo {author} {\bibfnamefont {G.-W.}\ \bibnamefont {Chern}},
  \bibinfo {author} {\bibfnamefont {D.~L.}\ \bibnamefont {Kovrizhin}}, \bibinfo
  {author} {\bibfnamefont {R.}~\bibnamefont {Moessner}}, \ and\ \bibinfo
  {author} {\bibfnamefont {N.~B.}\ \bibnamefont {Perkins}},\ }\href {\doibase
  10.1103/PhysRevLett.113.187201} {\bibfield  {journal} {\bibinfo  {journal}
  {Phys. Rev. Lett.}\ }\textbf {\bibinfo {volume} {113}},\ \bibinfo {pages}
  {187201} (\bibinfo {year} {2014})}\BibitemShut {NoStop}%
\bibitem [{\citenamefont {Perreault}\ \emph {et~al.}(2015)\citenamefont
  {Perreault}, \citenamefont {Knolle}, \citenamefont {Perkins},\ and\
  \citenamefont {Burnell}}]{Brent2015}%
  \BibitemOpen
  \bibfield  {author} {\bibinfo {author} {\bibfnamefont {B.}~\bibnamefont
  {Perreault}}, \bibinfo {author} {\bibfnamefont {J.}~\bibnamefont {Knolle}},
  \bibinfo {author} {\bibfnamefont {N.~B.}\ \bibnamefont {Perkins}}, \ and\
  \bibinfo {author} {\bibfnamefont {F.~J.}\ \bibnamefont {Burnell}},\ }\href
  {\doibase 10.1103/PhysRevB.92.094439} {\bibfield  {journal} {\bibinfo
  {journal} {Phys. Rev. B}\ }\textbf {\bibinfo {volume} {92}},\ \bibinfo
  {pages} {094439} (\bibinfo {year} {2015})}\BibitemShut {NoStop}%
\bibitem [{\citenamefont {Perreault}\ \emph
  {et~al.}(2016{\natexlab{a}})\citenamefont {Perreault}, \citenamefont
  {Knolle}, \citenamefont {Perkins},\ and\ \citenamefont
  {Burnell}}]{Brent2016-short}%
  \BibitemOpen
  \bibfield  {author} {\bibinfo {author} {\bibfnamefont {B.}~\bibnamefont
  {Perreault}}, \bibinfo {author} {\bibfnamefont {J.}~\bibnamefont {Knolle}},
  \bibinfo {author} {\bibfnamefont {N.~B.}\ \bibnamefont {Perkins}}, \ and\
  \bibinfo {author} {\bibfnamefont {F.~J.}\ \bibnamefont {Burnell}},\ }\href
  {\doibase 10.1103/PhysRevB.94.060408} {\bibfield  {journal} {\bibinfo
  {journal} {Phys. Rev. B}\ }\textbf {\bibinfo {volume} {94}},\ \bibinfo
  {pages} {060408} (\bibinfo {year} {2016}{\natexlab{a}})}\BibitemShut
  {NoStop}%
\bibitem [{\citenamefont {Perreault}\ \emph
  {et~al.}(2016{\natexlab{b}})\citenamefont {Perreault}, \citenamefont
  {Knolle}, \citenamefont {Perkins},\ and\ \citenamefont
  {Burnell}}]{Brent2016-long}%
  \BibitemOpen
  \bibfield  {author} {\bibinfo {author} {\bibfnamefont {B.}~\bibnamefont
  {Perreault}}, \bibinfo {author} {\bibfnamefont {J.}~\bibnamefont {Knolle}},
  \bibinfo {author} {\bibfnamefont {N.~B.}\ \bibnamefont {Perkins}}, \ and\
  \bibinfo {author} {\bibfnamefont {F.~J.}\ \bibnamefont {Burnell}},\ }\href
  {\doibase 10.1103/PhysRevB.94.104427} {\bibfield  {journal} {\bibinfo
  {journal} {Phys. Rev. B}\ }\textbf {\bibinfo {volume} {94}},\ \bibinfo
  {pages} {104427} (\bibinfo {year} {2016}{\natexlab{b}})}\BibitemShut
  {NoStop}%
\bibitem [{\citenamefont {Fu}\ \emph {et~al.}(2017)\citenamefont {Fu},
  \citenamefont {Rau}, \citenamefont {Gingras},\ and\ \citenamefont
  {Perkins}}]{Fu2017}%
  \BibitemOpen
  \bibfield  {author} {\bibinfo {author} {\bibfnamefont {J.}~\bibnamefont
  {Fu}}, \bibinfo {author} {\bibfnamefont {J.~G.}\ \bibnamefont {Rau}},
  \bibinfo {author} {\bibfnamefont {M.~J.~P.}\ \bibnamefont {Gingras}}, \ and\
  \bibinfo {author} {\bibfnamefont {N.~B.}\ \bibnamefont {Perkins}},\ }\href
  {\doibase 10.1103/PhysRevB.96.035136} {\bibfield  {journal} {\bibinfo
  {journal} {Phys. Rev. B}\ }\textbf {\bibinfo {volume} {96}},\ \bibinfo
  {pages} {035136} (\bibinfo {year} {2017})}\BibitemShut {NoStop}%
\bibitem [{\citenamefont {Rousochatzakis}\ \emph {et~al.}(2019)\citenamefont
  {Rousochatzakis}, \citenamefont {Kourtis}, \citenamefont {Knolle},
  \citenamefont {Moessner},\ and\ \citenamefont {Perkins}}]{Rousochatzaki2019}%
  \BibitemOpen
  \bibfield  {author} {\bibinfo {author} {\bibfnamefont {I.}~\bibnamefont
  {Rousochatzakis}}, \bibinfo {author} {\bibfnamefont {S.}~\bibnamefont
  {Kourtis}}, \bibinfo {author} {\bibfnamefont {J.}~\bibnamefont {Knolle}},
  \bibinfo {author} {\bibfnamefont {R.}~\bibnamefont {Moessner}}, \ and\
  \bibinfo {author} {\bibfnamefont {N.~B.}\ \bibnamefont {Perkins}},\ }\href
  {\doibase 10.1103/PhysRevB.100.045117} {\bibfield  {journal} {\bibinfo
  {journal} {Phys. Rev. B}\ }\textbf {\bibinfo {volume} {100}},\ \bibinfo
  {pages} {045117} (\bibinfo {year} {2019})}\BibitemShut {NoStop}%
\bibitem [{\citenamefont {Cao}\ and\ \citenamefont {DeLong}(2013)}]{BookCao}%
  \BibitemOpen
  \bibinfo {editor} {\bibfnamefont {G.}~\bibnamefont {Cao}}\ and\ \bibinfo
  {editor} {\bibfnamefont {L.}~\bibnamefont {DeLong}},\ eds.,\ \href@noop {}
  {\emph {\bibinfo {title} {{Frontiers of 4d- and 5d-Transition Metal
  Oxides}}}}\ (\bibinfo  {publisher} {World Scientific Publishing Co. Pte.
  Ltd.},\ \bibinfo {year} {2013})\BibitemShut {NoStop}%
\bibitem [{\citenamefont {Witczak-Krempa}\ \emph {et~al.}(2014)\citenamefont
  {Witczak-Krempa}, \citenamefont {Chen}, \citenamefont {Kim},\ and\
  \citenamefont {Balents}}]{Krempa2014}%
  \BibitemOpen
  \bibfield  {author} {\bibinfo {author} {\bibfnamefont {W.}~\bibnamefont
  {Witczak-Krempa}}, \bibinfo {author} {\bibfnamefont {G.}~\bibnamefont
  {Chen}}, \bibinfo {author} {\bibfnamefont {Y.~B.}\ \bibnamefont {Kim}}, \
  and\ \bibinfo {author} {\bibfnamefont {L.}~\bibnamefont {Balents}},\ }\href
  {\doibase 10.1146/annurev-conmatphys-020911-125138} {\bibfield  {journal}
  {\bibinfo  {journal} {Annual Review of Condensed Matter Physics}\ }\textbf
  {\bibinfo {volume} {5}},\ \bibinfo {pages} {57} (\bibinfo {year} {2014})},\
  \Eprint
  {http://arxiv.org/abs/https://doi.org/10.1146/annurev-conmatphys-020911-125138}
  {https://doi.org/10.1146/annurev-conmatphys-020911-125138} \BibitemShut
  {NoStop}%
\bibitem [{\citenamefont {Rau}\ \emph {et~al.}(2016)\citenamefont {Rau},
  \citenamefont {Lee},\ and\ \citenamefont {Kee}}]{Rau2016}%
  \BibitemOpen
  \bibfield  {author} {\bibinfo {author} {\bibfnamefont {J.~G.}\ \bibnamefont
  {Rau}}, \bibinfo {author} {\bibfnamefont {E.~K.-H.}\ \bibnamefont {Lee}}, \
  and\ \bibinfo {author} {\bibfnamefont {H.-Y.}\ \bibnamefont {Kee}},\ }\href
  {\doibase 10.1146/annurev-conmatphys-031115-011319} {\bibfield  {journal}
  {\bibinfo  {journal} {Ann. Rev. Cond. Matt. Phys.}\ }\textbf {\bibinfo
  {volume} {7}},\ \bibinfo {pages} {195} (\bibinfo {year} {2016})}\BibitemShut
  {NoStop}%
\bibitem [{\citenamefont {Nasu}\ \emph {et~al.}(2016)\citenamefont {Nasu},
  \citenamefont {Knolle}, \citenamefont {Kovrizhin}, \citenamefont {Motome},\
  and\ \citenamefont {Moessner}}]{Nasu2016}%
  \BibitemOpen
  \bibfield  {author} {\bibinfo {author} {\bibfnamefont {J.}~\bibnamefont
  {Nasu}}, \bibinfo {author} {\bibfnamefont {J.}~\bibnamefont {Knolle}},
  \bibinfo {author} {\bibfnamefont {D.}~\bibnamefont {Kovrizhin}}, \bibinfo
  {author} {\bibfnamefont {Y.}~\bibnamefont {Motome}}, \ and\ \bibinfo {author}
  {\bibfnamefont {R.}~\bibnamefont {Moessner}},\ }\href {\doibase
  10.1038/nphys3809} {\bibfield  {journal} {\bibinfo  {journal} {Nature
  Physics}\ }\textbf {\bibinfo {volume} {12}},\ \bibinfo {pages} {912}
  (\bibinfo {year} {2016})}\BibitemShut {NoStop}%
\bibitem [{\citenamefont {Ulrich}\ \emph {et~al.}(2015)\citenamefont {Ulrich},
  \citenamefont {Khaliullin}, \citenamefont {Guennou}, \citenamefont {Roth},
  \citenamefont {Lorenz},\ and\ \citenamefont {Keimer}}]{Ulrich2015}%
  \BibitemOpen
  \bibfield  {author} {\bibinfo {author} {\bibfnamefont {C.}~\bibnamefont
  {Ulrich}}, \bibinfo {author} {\bibfnamefont {G.}~\bibnamefont {Khaliullin}},
  \bibinfo {author} {\bibfnamefont {M.}~\bibnamefont {Guennou}}, \bibinfo
  {author} {\bibfnamefont {H.}~\bibnamefont {Roth}}, \bibinfo {author}
  {\bibfnamefont {T.}~\bibnamefont {Lorenz}}, \ and\ \bibinfo {author}
  {\bibfnamefont {B.}~\bibnamefont {Keimer}},\ }\href {\doibase
  10.1103/PhysRevLett.115.156403} {\bibfield  {journal} {\bibinfo  {journal}
  {Phys. Rev. Lett.}\ }\textbf {\bibinfo {volume} {115}},\ \bibinfo {pages}
  {156403} (\bibinfo {year} {2015})}\BibitemShut {NoStop}%
\bibitem [{\citenamefont {Gim}\ \emph {et~al.}(2016)\citenamefont {Gim},
  \citenamefont {Sethi}, \citenamefont {Zhao}, \citenamefont {Mitchell},
  \citenamefont {Cao},\ and\ \citenamefont {Cooper}}]{Gim2016}%
  \BibitemOpen
  \bibfield  {author} {\bibinfo {author} {\bibfnamefont {Y.}~\bibnamefont
  {Gim}}, \bibinfo {author} {\bibfnamefont {A.}~\bibnamefont {Sethi}}, \bibinfo
  {author} {\bibfnamefont {Q.}~\bibnamefont {Zhao}}, \bibinfo {author}
  {\bibfnamefont {J.~F.}\ \bibnamefont {Mitchell}}, \bibinfo {author}
  {\bibfnamefont {G.}~\bibnamefont {Cao}}, \ and\ \bibinfo {author}
  {\bibfnamefont {S.~L.}\ \bibnamefont {Cooper}},\ }\href {\doibase
  10.1103/PhysRevB.93.024405} {\bibfield  {journal} {\bibinfo  {journal} {Phys.
  Rev. B}\ }\textbf {\bibinfo {volume} {93}},\ \bibinfo {pages} {024405}
  (\bibinfo {year} {2016})}\BibitemShut {NoStop}%
\bibitem [{\citenamefont {Gretarsson}\ \emph {et~al.}(2017)\citenamefont
  {Gretarsson}, \citenamefont {Sauceda}, \citenamefont {Sung}, \citenamefont
  {H\"oppner}, \citenamefont {Minola}, \citenamefont {Kim}, \citenamefont
  {Keimer},\ and\ \citenamefont {Le~Tacon}}]{Gretarsson2017}%
  \BibitemOpen
  \bibfield  {author} {\bibinfo {author} {\bibfnamefont {H.}~\bibnamefont
  {Gretarsson}}, \bibinfo {author} {\bibfnamefont {J.}~\bibnamefont {Sauceda}},
  \bibinfo {author} {\bibfnamefont {N.~H.}\ \bibnamefont {Sung}}, \bibinfo
  {author} {\bibfnamefont {M.}~\bibnamefont {H\"oppner}}, \bibinfo {author}
  {\bibfnamefont {M.}~\bibnamefont {Minola}}, \bibinfo {author} {\bibfnamefont
  {B.~J.}\ \bibnamefont {Kim}}, \bibinfo {author} {\bibfnamefont
  {B.}~\bibnamefont {Keimer}}, \ and\ \bibinfo {author} {\bibfnamefont
  {M.}~\bibnamefont {Le~Tacon}},\ }\href {\doibase 10.1103/PhysRevB.96.115138}
  {\bibfield  {journal} {\bibinfo  {journal} {Phys. Rev. B}\ }\textbf {\bibinfo
  {volume} {96}},\ \bibinfo {pages} {115138} (\bibinfo {year}
  {2017})}\BibitemShut {NoStop}%
\bibitem [{\citenamefont {Souliou}\ \emph {et~al.}(2017)\citenamefont
  {Souliou}, \citenamefont {Chaloupka}, \citenamefont {Khaliullin},
  \citenamefont {Ryu}, \citenamefont {Jain}, \citenamefont {Kim}, \citenamefont
  {Le~Tacon},\ and\ \citenamefont {Keimer}}]{Souliou2017}%
  \BibitemOpen
  \bibfield  {author} {\bibinfo {author} {\bibfnamefont {S.-M.}\ \bibnamefont
  {Souliou}}, \bibinfo {author} {\bibfnamefont {J.~c.~v.}\ \bibnamefont
  {Chaloupka}}, \bibinfo {author} {\bibfnamefont {G.}~\bibnamefont
  {Khaliullin}}, \bibinfo {author} {\bibfnamefont {G.}~\bibnamefont {Ryu}},
  \bibinfo {author} {\bibfnamefont {A.}~\bibnamefont {Jain}}, \bibinfo {author}
  {\bibfnamefont {B.~J.}\ \bibnamefont {Kim}}, \bibinfo {author} {\bibfnamefont
  {M.}~\bibnamefont {Le~Tacon}}, \ and\ \bibinfo {author} {\bibfnamefont
  {B.}~\bibnamefont {Keimer}},\ }\href {\doibase
  10.1103/PhysRevLett.119.067201} {\bibfield  {journal} {\bibinfo  {journal}
  {Phys. Rev. Lett.}\ }\textbf {\bibinfo {volume} {119}},\ \bibinfo {pages}
  {067201} (\bibinfo {year} {2017})}\BibitemShut {NoStop}%
\bibitem [{\citenamefont {Metavitsiadis}\ \emph {et~al.}(2021)\citenamefont
  {Metavitsiadis}, \citenamefont {Natori}, \citenamefont {Knolle},\ and\
  \citenamefont {Brenig}}]{Metavitsiadis2021}%
  \BibitemOpen
  \bibfield  {author} {\bibinfo {author} {\bibfnamefont {A.}~\bibnamefont
  {Metavitsiadis}}, \bibinfo {author} {\bibfnamefont {W.}~\bibnamefont
  {Natori}}, \bibinfo {author} {\bibfnamefont {J.}~\bibnamefont {Knolle}}, \
  and\ \bibinfo {author} {\bibfnamefont {W.}~\bibnamefont {Brenig}},\ }\href
  {https://arxiv.org/abs/2103.09828} {\  (\bibinfo {year} {2021})}\BibitemShut
  {NoStop}%
\bibitem [{\citenamefont {Fleury}\ and\ \citenamefont {Loudon}(1968)}]{LF1968}%
  \BibitemOpen
  \bibfield  {author} {\bibinfo {author} {\bibfnamefont {P.~A.}\ \bibnamefont
  {Fleury}}\ and\ \bibinfo {author} {\bibfnamefont {R.}~\bibnamefont
  {Loudon}},\ }\href {\doibase 10.1103/PhysRev.166.514} {\bibfield  {journal}
  {\bibinfo  {journal} {Phys. Rev.}\ }\textbf {\bibinfo {volume} {166}},\
  \bibinfo {pages} {514} (\bibinfo {year} {1968})}\BibitemShut {NoStop}%
\bibitem [{\citenamefont {Elliott}\ and\ \citenamefont
  {Loudon}(1963)}]{Elliott1963}%
  \BibitemOpen
  \bibfield  {author} {\bibinfo {author} {\bibfnamefont {R.~J.}\ \bibnamefont
  {Elliott}}\ and\ \bibinfo {author} {\bibfnamefont {R.}~\bibnamefont
  {Loudon}},\ }\href@noop {} {\bibfield  {journal} {\bibinfo  {journal} {Phys.
  Letters}\ }\textbf {\bibinfo {volume} {3}},\ \bibinfo {pages} {189} (\bibinfo
  {year} {1963})}\BibitemShut {NoStop}%
\bibitem [{Yip()}]{Yipingunpublished}%
  \BibitemOpen
  \href@noop {} {}\bibinfo {note} {Private communications with Yiping Wang and
  Kenneth Burch}\BibitemShut {NoStop}%
\bibitem [{\citenamefont {Trebst}(2017)}]{Trebst2017}%
  \BibitemOpen
  \bibfield  {author} {\bibinfo {author} {\bibfnamefont {S.}~\bibnamefont
  {Trebst}},\ }\href {http://arxiv.org/abs/1701.07056} {\bibfield  {journal}
  {\bibinfo  {journal} {arXiv:1701.07056}\ } (\bibinfo {year}
  {2017})}\BibitemShut {NoStop}%
\bibitem [{\citenamefont {Takagi}\ \emph {et~al.}(2019)\citenamefont {Takagi},
  \citenamefont {Takayama}, \citenamefont {Jackeli}, \citenamefont
  {Khaliullin},\ and\ \citenamefont {Nagler}}]{Takagi2019}%
  \BibitemOpen
  \bibfield  {author} {\bibinfo {author} {\bibfnamefont {H.}~\bibnamefont
  {Takagi}}, \bibinfo {author} {\bibfnamefont {T.}~\bibnamefont {Takayama}},
  \bibinfo {author} {\bibfnamefont {G.}~\bibnamefont {Jackeli}}, \bibinfo
  {author} {\bibfnamefont {G.}~\bibnamefont {Khaliullin}}, \ and\ \bibinfo
  {author} {\bibfnamefont {S.~E.}\ \bibnamefont {Nagler}},\ }\href {\doibase
  10.1038/s42254-019-0038-2} {\bibfield  {journal} {\bibinfo  {journal} {Nat.
  Rev. Phys.}\ }\textbf {\bibinfo {volume} {1}},\ \bibinfo {pages} {264}
  (\bibinfo {year} {2019})}\BibitemShut {NoStop}%
\bibitem [{\citenamefont {Motome}\ and\ \citenamefont
  {Nasu}(2020)}]{Motome2020a}%
  \BibitemOpen
  \bibfield  {author} {\bibinfo {author} {\bibfnamefont {Y.}~\bibnamefont
  {Motome}}\ and\ \bibinfo {author} {\bibfnamefont {J.}~\bibnamefont {Nasu}},\
  }\href {\doibase 10.7566/JPSJ.89.012002} {\bibfield  {journal} {\bibinfo
  {journal} {J. Phys. Soc. Jpn.}\ }\textbf {\bibinfo {volume} {89}},\ \bibinfo
  {pages} {012002} (\bibinfo {year} {2020})}\BibitemShut {NoStop}%
\bibitem [{\citenamefont {Singh}\ and\ \citenamefont
  {Gegenwart}(2010)}]{Singh2010}%
  \BibitemOpen
  \bibfield  {author} {\bibinfo {author} {\bibfnamefont {Y.}~\bibnamefont
  {Singh}}\ and\ \bibinfo {author} {\bibfnamefont {P.}~\bibnamefont
  {Gegenwart}},\ }\href {\doibase 10.1103/PhysRevB.82.064412} {\bibfield
  {journal} {\bibinfo  {journal} {Phys. Rev. B}\ }\textbf {\bibinfo {volume}
  {82}},\ \bibinfo {pages} {064412} (\bibinfo {year} {2010})}\BibitemShut
  {NoStop}%
\bibitem [{\citenamefont {Hwan~Chun}\ \emph {et~al.}(2015)\citenamefont
  {Hwan~Chun}, \citenamefont {Kim}, \citenamefont {Kim}, \citenamefont {Zheng},
  \citenamefont {Stoumpos}, \citenamefont {Malliakas}, \citenamefont
  {Mitchell}, \citenamefont {Mehlawat}, \citenamefont {Singh}, \citenamefont
  {Choi}, \citenamefont {Gog}, \citenamefont {Al-Zein}, \citenamefont {Sala},
  \citenamefont {Krisch}, \citenamefont {Chaloupka}, \citenamefont {Jackeli},
  \citenamefont {Khaliullin},\ and\ \citenamefont {Kim}}]{Chun2015}%
  \BibitemOpen
  \bibfield  {author} {\bibinfo {author} {\bibfnamefont {S.}~\bibnamefont
  {Hwan~Chun}}, \bibinfo {author} {\bibfnamefont {J.-W.}\ \bibnamefont {Kim}},
  \bibinfo {author} {\bibfnamefont {J.}~\bibnamefont {Kim}}, \bibinfo {author}
  {\bibfnamefont {H.}~\bibnamefont {Zheng}}, \bibinfo {author} {\bibfnamefont
  {C.~C.}\ \bibnamefont {Stoumpos}}, \bibinfo {author} {\bibfnamefont {C.~D.}\
  \bibnamefont {Malliakas}}, \bibinfo {author} {\bibfnamefont {J.~F.}\
  \bibnamefont {Mitchell}}, \bibinfo {author} {\bibfnamefont {K.}~\bibnamefont
  {Mehlawat}}, \bibinfo {author} {\bibfnamefont {Y.}~\bibnamefont {Singh}},
  \bibinfo {author} {\bibfnamefont {Y.}~\bibnamefont {Choi}}, \bibinfo {author}
  {\bibfnamefont {T.}~\bibnamefont {Gog}}, \bibinfo {author} {\bibfnamefont
  {A.}~\bibnamefont {Al-Zein}}, \bibinfo {author} {\bibfnamefont {M.~M.}\
  \bibnamefont {Sala}}, \bibinfo {author} {\bibfnamefont {M.}~\bibnamefont
  {Krisch}}, \bibinfo {author} {\bibfnamefont {J.}~\bibnamefont {Chaloupka}},
  \bibinfo {author} {\bibfnamefont {G.}~\bibnamefont {Jackeli}}, \bibinfo
  {author} {\bibfnamefont {G.}~\bibnamefont {Khaliullin}}, \ and\ \bibinfo
  {author} {\bibfnamefont {B.~J.}\ \bibnamefont {Kim}},\ }\href
  {http://dx.doi.org/10.1038/nphys3322} {\bibfield  {journal} {\bibinfo
  {journal} {Nat. Phys.}\ }\textbf {\bibinfo {volume} {11}},\ \bibinfo {pages}
  {462 } (\bibinfo {year} {2015})}\BibitemShut {NoStop}%
\bibitem [{\citenamefont {Williams}\ \emph {et~al.}(2016)\citenamefont
  {Williams}, \citenamefont {Johnson}, \citenamefont {Freund}, \citenamefont
  {Choi}, \citenamefont {Jesche}, \citenamefont {Kimchi}, \citenamefont
  {Manni}, \citenamefont {Bombardi}, \citenamefont {Manuel}, \citenamefont
  {Gegenwart},\ and\ \citenamefont {Coldea}}]{Williams2016}%
  \BibitemOpen
  \bibfield  {author} {\bibinfo {author} {\bibfnamefont {S.~C.}\ \bibnamefont
  {Williams}}, \bibinfo {author} {\bibfnamefont {R.~D.}\ \bibnamefont
  {Johnson}}, \bibinfo {author} {\bibfnamefont {F.}~\bibnamefont {Freund}},
  \bibinfo {author} {\bibfnamefont {S.}~\bibnamefont {Choi}}, \bibinfo {author}
  {\bibfnamefont {A.}~\bibnamefont {Jesche}}, \bibinfo {author} {\bibfnamefont
  {I.}~\bibnamefont {Kimchi}}, \bibinfo {author} {\bibfnamefont
  {S.}~\bibnamefont {Manni}}, \bibinfo {author} {\bibfnamefont
  {A.}~\bibnamefont {Bombardi}}, \bibinfo {author} {\bibfnamefont
  {P.}~\bibnamefont {Manuel}}, \bibinfo {author} {\bibfnamefont
  {P.}~\bibnamefont {Gegenwart}}, \ and\ \bibinfo {author} {\bibfnamefont
  {R.}~\bibnamefont {Coldea}},\ }\href {\doibase 10.1103/PhysRevB.93.195158}
  {\bibfield  {journal} {\bibinfo  {journal} {Phys. Rev. B}\ }\textbf {\bibinfo
  {volume} {93}},\ \bibinfo {pages} {195158} (\bibinfo {year}
  {2016})}\BibitemShut {NoStop}%
\bibitem [{\citenamefont {Plumb}\ \emph {et~al.}(2014)\citenamefont {Plumb},
  \citenamefont {Clancy}, \citenamefont {Sandilands}, \citenamefont {Shankar},
  \citenamefont {Hu}, \citenamefont {Burch}, \citenamefont {Kee},\ and\
  \citenamefont {Kim}}]{Plumb2014}%
  \BibitemOpen
  \bibfield  {author} {\bibinfo {author} {\bibfnamefont {K.~W.}\ \bibnamefont
  {Plumb}}, \bibinfo {author} {\bibfnamefont {J.~P.}\ \bibnamefont {Clancy}},
  \bibinfo {author} {\bibfnamefont {L.~J.}\ \bibnamefont {Sandilands}},
  \bibinfo {author} {\bibfnamefont {V.~V.}\ \bibnamefont {Shankar}}, \bibinfo
  {author} {\bibfnamefont {Y.~F.}\ \bibnamefont {Hu}}, \bibinfo {author}
  {\bibfnamefont {K.~S.}\ \bibnamefont {Burch}}, \bibinfo {author}
  {\bibfnamefont {H.-Y.}\ \bibnamefont {Kee}}, \ and\ \bibinfo {author}
  {\bibfnamefont {Y.-J.}\ \bibnamefont {Kim}},\ }\href
  {https://link.aps.org/doi/10.1103/PhysRevB.90.041112} {\bibfield  {journal}
  {\bibinfo  {journal} {Phys. Rev. B}\ }\textbf {\bibinfo {volume} {90}},\
  \bibinfo {pages} {041112} (\bibinfo {year} {2014})}\BibitemShut {NoStop}%
\bibitem [{\citenamefont {Sears}\ \emph {et~al.}(2015)\citenamefont {Sears},
  \citenamefont {Songvilay}, \citenamefont {Plumb}, \citenamefont {Clancy},
  \citenamefont {Qiu}, \citenamefont {Zhao}, \citenamefont {Parshall},\ and\
  \citenamefont {Kim}}]{Sears2015}%
  \BibitemOpen
  \bibfield  {author} {\bibinfo {author} {\bibfnamefont {J.~A.}\ \bibnamefont
  {Sears}}, \bibinfo {author} {\bibfnamefont {M.}~\bibnamefont {Songvilay}},
  \bibinfo {author} {\bibfnamefont {K.~W.}\ \bibnamefont {Plumb}}, \bibinfo
  {author} {\bibfnamefont {J.~P.}\ \bibnamefont {Clancy}}, \bibinfo {author}
  {\bibfnamefont {Y.}~\bibnamefont {Qiu}}, \bibinfo {author} {\bibfnamefont
  {Y.}~\bibnamefont {Zhao}}, \bibinfo {author} {\bibfnamefont {D.}~\bibnamefont
  {Parshall}}, \ and\ \bibinfo {author} {\bibfnamefont {Y.-J.}\ \bibnamefont
  {Kim}},\ }\href {\doibase 10.1103/PhysRevB.91.144420} {\bibfield  {journal}
  {\bibinfo  {journal} {Phys. Rev. B}\ }\textbf {\bibinfo {volume} {91}},\
  \bibinfo {pages} {144420} (\bibinfo {year} {2015})}\BibitemShut {NoStop}%
\bibitem [{\citenamefont {Banerjee}\ \emph {et~al.}(2017)\citenamefont
  {Banerjee}, \citenamefont {Yan}, \citenamefont {Knolle}, \citenamefont
  {Bridges}, \citenamefont {Stone}, \citenamefont {Lumsden}, \citenamefont
  {Mandrus}, \citenamefont {Tennant}, \citenamefont {Moessner},\ and\
  \citenamefont {Nagler}}]{Banerjee2017}%
  \BibitemOpen
  \bibfield  {author} {\bibinfo {author} {\bibfnamefont {A.}~\bibnamefont
  {Banerjee}}, \bibinfo {author} {\bibfnamefont {J.}~\bibnamefont {Yan}},
  \bibinfo {author} {\bibfnamefont {J.}~\bibnamefont {Knolle}}, \bibinfo
  {author} {\bibfnamefont {C.~A.}\ \bibnamefont {Bridges}}, \bibinfo {author}
  {\bibfnamefont {M.~B.}\ \bibnamefont {Stone}}, \bibinfo {author}
  {\bibfnamefont {M.~D.}\ \bibnamefont {Lumsden}}, \bibinfo {author}
  {\bibfnamefont {D.~G.}\ \bibnamefont {Mandrus}}, \bibinfo {author}
  {\bibfnamefont {D.~A.}\ \bibnamefont {Tennant}}, \bibinfo {author}
  {\bibfnamefont {R.}~\bibnamefont {Moessner}}, \ and\ \bibinfo {author}
  {\bibfnamefont {S.~E.}\ \bibnamefont {Nagler}},\ }\href@noop {} {\bibfield
  {journal} {\bibinfo  {journal} {Science}\ }\textbf {\bibinfo {volume}
  {356}},\ \bibinfo {pages} {1055} (\bibinfo {year} {2017})}\BibitemShut
  {NoStop}%
\bibitem [{\citenamefont {Kasahara}\ \emph {et~al.}(2018)\citenamefont
  {Kasahara}, \citenamefont {Ohnishi}, \citenamefont {Mizukami}, \citenamefont
  {Tanaka}, \citenamefont {Ma}, \citenamefont {Sugii}, \citenamefont {Kurita},
  \citenamefont {Tanaka}, \citenamefont {Nasu}, \citenamefont {Motome},
  \citenamefont {Shibauchi},\ and\ \citenamefont {Matsuda}}]{Kasahara2018}%
  \BibitemOpen
  \bibfield  {author} {\bibinfo {author} {\bibfnamefont {Y.}~\bibnamefont
  {Kasahara}}, \bibinfo {author} {\bibfnamefont {T.}~\bibnamefont {Ohnishi}},
  \bibinfo {author} {\bibfnamefont {Y.}~\bibnamefont {Mizukami}}, \bibinfo
  {author} {\bibfnamefont {O.}~\bibnamefont {Tanaka}}, \bibinfo {author}
  {\bibfnamefont {S.}~\bibnamefont {Ma}}, \bibinfo {author} {\bibfnamefont
  {K.}~\bibnamefont {Sugii}}, \bibinfo {author} {\bibfnamefont
  {N.}~\bibnamefont {Kurita}}, \bibinfo {author} {\bibfnamefont
  {H.}~\bibnamefont {Tanaka}}, \bibinfo {author} {\bibfnamefont
  {J.}~\bibnamefont {Nasu}}, \bibinfo {author} {\bibfnamefont {Y.}~\bibnamefont
  {Motome}}, \bibinfo {author} {\bibfnamefont {T.}~\bibnamefont {Shibauchi}}, \
  and\ \bibinfo {author} {\bibfnamefont {Y.}~\bibnamefont {Matsuda}},\ }\href
  {\doibase 10.1038/s41586-018-0274-0} {\bibfield  {journal} {\bibinfo
  {journal} {Nature}\ }\textbf {\bibinfo {volume} {559}},\ \bibinfo {pages}
  {227} (\bibinfo {year} {2018})}\BibitemShut {NoStop}%
\bibitem [{\citenamefont {Biffin}\ \emph
  {et~al.}(2014{\natexlab{a}})\citenamefont {Biffin}, \citenamefont {Johnson},
  \citenamefont {Choi}, \citenamefont {Freund}, \citenamefont {Manni},
  \citenamefont {Bombardi}, \citenamefont {Manuel}, \citenamefont {Gegenwart},\
  and\ \citenamefont {Coldea}}]{Biffin2014a}%
  \BibitemOpen
  \bibfield  {author} {\bibinfo {author} {\bibfnamefont {A.}~\bibnamefont
  {Biffin}}, \bibinfo {author} {\bibfnamefont {R.~D.}\ \bibnamefont {Johnson}},
  \bibinfo {author} {\bibfnamefont {S.}~\bibnamefont {Choi}}, \bibinfo {author}
  {\bibfnamefont {F.}~\bibnamefont {Freund}}, \bibinfo {author} {\bibfnamefont
  {S.}~\bibnamefont {Manni}}, \bibinfo {author} {\bibfnamefont
  {A.}~\bibnamefont {Bombardi}}, \bibinfo {author} {\bibfnamefont
  {P.}~\bibnamefont {Manuel}}, \bibinfo {author} {\bibfnamefont
  {P.}~\bibnamefont {Gegenwart}}, \ and\ \bibinfo {author} {\bibfnamefont
  {R.}~\bibnamefont {Coldea}},\ }\href {\doibase 10.1103/PhysRevB.90.205116}
  {\bibfield  {journal} {\bibinfo  {journal} {Phys. Rev. B}\ }\textbf {\bibinfo
  {volume} {90}},\ \bibinfo {pages} {205116} (\bibinfo {year}
  {2014}{\natexlab{a}})}\BibitemShut {NoStop}%
\bibitem [{\citenamefont {Ruiz}\ \emph {et~al.}(2017)\citenamefont {Ruiz},
  \citenamefont {Frano}, \citenamefont {Breznay}, \citenamefont {Kimchi},
  \citenamefont {Helm}, \citenamefont {Oswald}, \citenamefont {Chan},
  \citenamefont {Birgeneau}, \citenamefont {Islam},\ and\ \citenamefont
  {Analytis}}]{Ruiz2017}%
  \BibitemOpen
  \bibfield  {author} {\bibinfo {author} {\bibfnamefont {A.}~\bibnamefont
  {Ruiz}}, \bibinfo {author} {\bibfnamefont {A.}~\bibnamefont {Frano}},
  \bibinfo {author} {\bibfnamefont {N.~P.}\ \bibnamefont {Breznay}}, \bibinfo
  {author} {\bibfnamefont {I.}~\bibnamefont {Kimchi}}, \bibinfo {author}
  {\bibfnamefont {T.}~\bibnamefont {Helm}}, \bibinfo {author} {\bibfnamefont
  {I.}~\bibnamefont {Oswald}}, \bibinfo {author} {\bibfnamefont {J.~Y.}\
  \bibnamefont {Chan}}, \bibinfo {author} {\bibfnamefont {R.}~\bibnamefont
  {Birgeneau}}, \bibinfo {author} {\bibfnamefont {Z.}~\bibnamefont {Islam}}, \
  and\ \bibinfo {author} {\bibfnamefont {J.~G.}\ \bibnamefont {Analytis}},\
  }\href {\doibase 10.1038/s41467-017-01071-9} {\bibfield  {journal} {\bibinfo
  {journal} {Nat. Commun.}\ }\textbf {\bibinfo {volume} {8}},\ \bibinfo {pages}
  {961} (\bibinfo {year} {2017})}\BibitemShut {NoStop}%
\bibitem [{\citenamefont {Majumder}\ \emph {et~al.}(2019)\citenamefont
  {Majumder}, \citenamefont {Freund}, \citenamefont {Dey}, \citenamefont
  {Prinz-Zwick}, \citenamefont {B\"uttgen}, \citenamefont {Skourski},
  \citenamefont {Jesche}, \citenamefont {Tsirlin},\ and\ \citenamefont
  {Gegenwart}}]{Majumder2019}%
  \BibitemOpen
  \bibfield  {author} {\bibinfo {author} {\bibfnamefont {M.}~\bibnamefont
  {Majumder}}, \bibinfo {author} {\bibfnamefont {F.}~\bibnamefont {Freund}},
  \bibinfo {author} {\bibfnamefont {T.}~\bibnamefont {Dey}}, \bibinfo {author}
  {\bibfnamefont {M.}~\bibnamefont {Prinz-Zwick}}, \bibinfo {author}
  {\bibfnamefont {N.}~\bibnamefont {B\"uttgen}}, \bibinfo {author}
  {\bibfnamefont {Y.}~\bibnamefont {Skourski}}, \bibinfo {author}
  {\bibfnamefont {A.}~\bibnamefont {Jesche}}, \bibinfo {author} {\bibfnamefont
  {A.~A.}\ \bibnamefont {Tsirlin}}, \ and\ \bibinfo {author} {\bibfnamefont
  {P.}~\bibnamefont {Gegenwart}},\ }\href {\doibase
  10.1103/PhysRevMaterials.3.074408} {\bibfield  {journal} {\bibinfo  {journal}
  {Phys. Rev. Materials}\ }\textbf {\bibinfo {volume} {3}},\ \bibinfo {pages}
  {074408} (\bibinfo {year} {2019})}\BibitemShut {NoStop}%
\bibitem [{\citenamefont {Biffin}\ \emph
  {et~al.}(2014{\natexlab{b}})\citenamefont {Biffin}, \citenamefont {Johnson},
  \citenamefont {Kimchi}, \citenamefont {Morris}, \citenamefont {Bombardi},
  \citenamefont {Analytis}, \citenamefont {Vishwanath},\ and\ \citenamefont
  {Coldea}}]{Biffin2014b}%
  \BibitemOpen
  \bibfield  {author} {\bibinfo {author} {\bibfnamefont {A.}~\bibnamefont
  {Biffin}}, \bibinfo {author} {\bibfnamefont {R.~D.}\ \bibnamefont {Johnson}},
  \bibinfo {author} {\bibfnamefont {I.}~\bibnamefont {Kimchi}}, \bibinfo
  {author} {\bibfnamefont {R.}~\bibnamefont {Morris}}, \bibinfo {author}
  {\bibfnamefont {A.}~\bibnamefont {Bombardi}}, \bibinfo {author}
  {\bibfnamefont {J.~G.}\ \bibnamefont {Analytis}}, \bibinfo {author}
  {\bibfnamefont {A.}~\bibnamefont {Vishwanath}}, \ and\ \bibinfo {author}
  {\bibfnamefont {R.}~\bibnamefont {Coldea}},\ }\href
  {https://link.aps.org/doi/10.1103/PhysRevLett.113.197201} {\bibfield
  {journal} {\bibinfo  {journal} {Phys. Rev. Lett.}\ }\textbf {\bibinfo
  {volume} {113}},\ \bibinfo {pages} {197201} (\bibinfo {year}
  {2014}{\natexlab{b}})}\BibitemShut {NoStop}%
\bibitem [{\citenamefont {Modic}\ \emph {et~al.}(2014)\citenamefont {Modic},
  \citenamefont {Smidt}, \citenamefont {Kimchi}, \citenamefont {Breznay},
  \citenamefont {Biffin}, \citenamefont {Choi}, \citenamefont {Johnson},
  \citenamefont {Coldea}, \citenamefont {Watkins-Curry}, \citenamefont
  {McCandless}, \citenamefont {Chan}, \citenamefont {Gandara}, \citenamefont
  {Vishwanath}, \citenamefont {Shekhter}, \citenamefont {Mcdonald},\ and\
  \citenamefont {Analytis}}]{Modic2014}%
  \BibitemOpen
  \bibfield  {author} {\bibinfo {author} {\bibfnamefont {K.}~\bibnamefont
  {Modic}}, \bibinfo {author} {\bibfnamefont {T.}~\bibnamefont {Smidt}},
  \bibinfo {author} {\bibfnamefont {I.}~\bibnamefont {Kimchi}}, \bibinfo
  {author} {\bibfnamefont {N.}~\bibnamefont {Breznay}}, \bibinfo {author}
  {\bibfnamefont {A.}~\bibnamefont {Biffin}}, \bibinfo {author} {\bibfnamefont
  {S.}~\bibnamefont {Choi}}, \bibinfo {author} {\bibfnamefont {R.}~\bibnamefont
  {Johnson}}, \bibinfo {author} {\bibfnamefont {R.}~\bibnamefont {Coldea}},
  \bibinfo {author} {\bibfnamefont {P.}~\bibnamefont {Watkins-Curry}}, \bibinfo
  {author} {\bibfnamefont {G.}~\bibnamefont {McCandless}}, \bibinfo {author}
  {\bibfnamefont {J.}~\bibnamefont {Chan}}, \bibinfo {author} {\bibfnamefont
  {F.}~\bibnamefont {Gandara}}, \bibinfo {author} {\bibfnamefont
  {A.}~\bibnamefont {Vishwanath}}, \bibinfo {author} {\bibfnamefont
  {A.}~\bibnamefont {Shekhter}}, \bibinfo {author} {\bibfnamefont
  {R.}~\bibnamefont {Mcdonald}}, \ and\ \bibinfo {author} {\bibfnamefont
  {J.}~\bibnamefont {Analytis}},\ }\href {\doibase 10.1038/ncomms5203}
  {\bibfield  {journal} {\bibinfo  {journal} {Nat. Commun.}\ }\textbf {\bibinfo
  {volume} {5}},\ \bibinfo {pages} {4203} (\bibinfo {year} {2014})}\BibitemShut
  {NoStop}%
\bibitem [{Note1()}]{Note1}%
  \BibitemOpen
  \bibinfo {note} {In case of Kitaev materials, the super-exchange expansion
  usually does not include processes when two holes meet at the same oxygen
  site in the intermediate state since these intermediate states are higher
  energy states and their inclusion only slightly modifies the effective
  couplings but does not change the picture qualitatively. We note, however,
  that in systems with small spin-orbit coupling, such as the cuprates, such
  processes can also lead to small anisotropic interactions \cite
  {Yushankhai1999}.}\BibitemShut {Stop}%
\bibitem [{\citenamefont {Katukuri}\ \emph {et~al.}(2014)\citenamefont
  {Katukuri}, \citenamefont {Nishimoto}, \citenamefont {Yushankhai},
  \citenamefont {Stoyanova}, \citenamefont {Kandpal}, \citenamefont {Choi},
  \citenamefont {Coldea}, \citenamefont {Rousochatzakis}, \citenamefont
  {Hozoi},\ and\ \citenamefont {van~den Brink}}]{Katukuri2014}%
  \BibitemOpen
  \bibfield  {author} {\bibinfo {author} {\bibfnamefont {V.~M.}\ \bibnamefont
  {Katukuri}}, \bibinfo {author} {\bibfnamefont {S.}~\bibnamefont {Nishimoto}},
  \bibinfo {author} {\bibfnamefont {V.}~\bibnamefont {Yushankhai}}, \bibinfo
  {author} {\bibfnamefont {A.}~\bibnamefont {Stoyanova}}, \bibinfo {author}
  {\bibfnamefont {H.}~\bibnamefont {Kandpal}}, \bibinfo {author} {\bibfnamefont
  {S.}~\bibnamefont {Choi}}, \bibinfo {author} {\bibfnamefont {R.}~\bibnamefont
  {Coldea}}, \bibinfo {author} {\bibfnamefont {I.}~\bibnamefont
  {Rousochatzakis}}, \bibinfo {author} {\bibfnamefont {L.}~\bibnamefont
  {Hozoi}}, \ and\ \bibinfo {author} {\bibfnamefont {J.}~\bibnamefont {van~den
  Brink}},\ }\href {http://stacks.iop.org/1367-2630/16/i=1/a=013056} {\bibfield
   {journal} {\bibinfo  {journal} {New J. Phys.}\ }\textbf {\bibinfo {volume}
  {16}},\ \bibinfo {pages} {013056} (\bibinfo {year} {2014})}\BibitemShut
  {NoStop}%
\bibitem [{\citenamefont {Rau}\ \emph {et~al.}(2014)\citenamefont {Rau},
  \citenamefont {Lee},\ and\ \citenamefont {Kee}}]{Rau2014}%
  \BibitemOpen
  \bibfield  {author} {\bibinfo {author} {\bibfnamefont {J.~G.}\ \bibnamefont
  {Rau}}, \bibinfo {author} {\bibfnamefont {E.~K.-H.}\ \bibnamefont {Lee}}, \
  and\ \bibinfo {author} {\bibfnamefont {H.-Y.}\ \bibnamefont {Kee}},\ }\href
  {\doibase 10.1103/PhysRevLett.112.077204} {\bibfield  {journal} {\bibinfo
  {journal} {Phys. Rev. Lett.}\ }\textbf {\bibinfo {volume} {112}},\ \bibinfo
  {pages} {077204} (\bibinfo {year} {2014})}\BibitemShut {NoStop}%
\bibitem [{\citenamefont {Lee}\ and\ \citenamefont {Kim}(2015)}]{Lee2015}%
  \BibitemOpen
  \bibfield  {author} {\bibinfo {author} {\bibfnamefont {E.~K.-H.}\
  \bibnamefont {Lee}}\ and\ \bibinfo {author} {\bibfnamefont {Y.~B.}\
  \bibnamefont {Kim}},\ }\href {\doibase 10.1103/PhysRevB.91.064407} {\bibfield
   {journal} {\bibinfo  {journal} {Phys. Rev. B}\ }\textbf {\bibinfo {volume}
  {91}},\ \bibinfo {pages} {064407} (\bibinfo {year} {2015})}\BibitemShut
  {NoStop}%
\bibitem [{\citenamefont {Lee}\ \emph {et~al.}(2016)\citenamefont {Lee},
  \citenamefont {Rau},\ and\ \citenamefont {Kim}}]{Lee2016}%
  \BibitemOpen
  \bibfield  {author} {\bibinfo {author} {\bibfnamefont {E.~K.-H.}\
  \bibnamefont {Lee}}, \bibinfo {author} {\bibfnamefont {J.~G.}\ \bibnamefont
  {Rau}}, \ and\ \bibinfo {author} {\bibfnamefont {Y.~B.}\ \bibnamefont
  {Kim}},\ }\href {\doibase 10.1103/PhysRevB.93.184420} {\bibfield  {journal}
  {\bibinfo  {journal} {Phys. Rev. B}\ }\textbf {\bibinfo {volume} {93}},\
  \bibinfo {pages} {184420} (\bibinfo {year} {2016})}\BibitemShut {NoStop}%
\bibitem [{\citenamefont {Rousochatzakis}\ and\ \citenamefont
  {Perkins}(2017)}]{IoannisGamma}%
  \BibitemOpen
  \bibfield  {author} {\bibinfo {author} {\bibfnamefont {I.}~\bibnamefont
  {Rousochatzakis}}\ and\ \bibinfo {author} {\bibfnamefont {N.~B.}\
  \bibnamefont {Perkins}},\ }\href {\doibase 10.1103/PhysRevLett.118.147204}
  {\bibfield  {journal} {\bibinfo  {journal} {Phys. Rev. Lett.}\ }\textbf
  {\bibinfo {volume} {118}},\ \bibinfo {pages} {147204} (\bibinfo {year}
  {2017})}\BibitemShut {NoStop}%
\bibitem [{\citenamefont {Winter}\ \emph {et~al.}(2017)\citenamefont {Winter},
  \citenamefont {Tsirlin}, \citenamefont {Daghofer}, \citenamefont {van~den
  Brink}, \citenamefont {Singh}, \citenamefont {Gegenwart},\ and\ \citenamefont
  {Valenti­}}]{Winter2017}%
  \BibitemOpen
  \bibfield  {author} {\bibinfo {author} {\bibfnamefont {S.~M.}\ \bibnamefont
  {Winter}}, \bibinfo {author} {\bibfnamefont {A.~A.}\ \bibnamefont {Tsirlin}},
  \bibinfo {author} {\bibfnamefont {M.}~\bibnamefont {Daghofer}}, \bibinfo
  {author} {\bibfnamefont {J.}~\bibnamefont {van~den Brink}}, \bibinfo {author}
  {\bibfnamefont {Y.}~\bibnamefont {Singh}}, \bibinfo {author} {\bibfnamefont
  {P.}~\bibnamefont {Gegenwart}}, \ and\ \bibinfo {author} {\bibfnamefont
  {R.}~\bibnamefont {Valenti­}},\ }\href
  {http://stacks.iop.org/0953-8984/29/i=49/a=493002} {\bibfield  {journal}
  {\bibinfo  {journal} {J. Phys.: Condens. Matter}\ }\textbf {\bibinfo {volume}
  {29}},\ \bibinfo {pages} {493002} (\bibinfo {year} {2017})}\BibitemShut
  {NoStop}%
\bibitem [{\citenamefont {Jackeli}\ and\ \citenamefont
  {Khaliullin}(2009)}]{Jackeli2009}%
  \BibitemOpen
  \bibfield  {author} {\bibinfo {author} {\bibfnamefont {G.}~\bibnamefont
  {Jackeli}}\ and\ \bibinfo {author} {\bibfnamefont {G.}~\bibnamefont
  {Khaliullin}},\ }\href {\doibase 10.1103/PhysRevLett.102.017205} {\bibfield
  {journal} {\bibinfo  {journal} {Phys. Rev. Lett.}\ }\textbf {\bibinfo
  {volume} {102}},\ \bibinfo {pages} {017205} (\bibinfo {year}
  {2009})}\BibitemShut {NoStop}%
\bibitem [{\citenamefont {Chaloupka}\ \emph {et~al.}(2010)\citenamefont
  {Chaloupka}, \citenamefont {Jackeli},\ and\ \citenamefont
  {Khaliullin}}]{Jackeli2010}%
  \BibitemOpen
  \bibfield  {author} {\bibinfo {author} {\bibfnamefont {J.}~\bibnamefont
  {Chaloupka}}, \bibinfo {author} {\bibfnamefont {G.}~\bibnamefont {Jackeli}},
  \ and\ \bibinfo {author} {\bibfnamefont {G.}~\bibnamefont {Khaliullin}},\
  }\href {\doibase 10.1103/PhysRevLett.105.027204} {\bibfield  {journal}
  {\bibinfo  {journal} {Phys. Rev. Lett.}\ }\textbf {\bibinfo {volume} {105}},\
  \bibinfo {pages} {027204} (\bibinfo {year} {2010})}\BibitemShut {NoStop}%
\bibitem [{\citenamefont {Perkins}\ \emph {et~al.}(2014)\citenamefont
  {Perkins}, \citenamefont {Sizyuk},\ and\ \citenamefont
  {W\"olfle}}]{Perkins2014}%
  \BibitemOpen
  \bibfield  {author} {\bibinfo {author} {\bibfnamefont {N.~B.}\ \bibnamefont
  {Perkins}}, \bibinfo {author} {\bibfnamefont {Y.}~\bibnamefont {Sizyuk}}, \
  and\ \bibinfo {author} {\bibfnamefont {P.}~\bibnamefont {W\"olfle}},\ }\href
  {\doibase 10.1103/PhysRevB.89.035143} {\bibfield  {journal} {\bibinfo
  {journal} {Phys. Rev. B}\ }\textbf {\bibinfo {volume} {89}},\ \bibinfo
  {pages} {035143} (\bibinfo {year} {2014})}\BibitemShut {NoStop}%
\bibitem [{\citenamefont {Sizyuk}\ \emph {et~al.}(2014)\citenamefont {Sizyuk},
  \citenamefont {Price}, \citenamefont {W\"olfle},\ and\ \citenamefont
  {Perkins}}]{Sizyuk2014}%
  \BibitemOpen
  \bibfield  {author} {\bibinfo {author} {\bibfnamefont {Y.}~\bibnamefont
  {Sizyuk}}, \bibinfo {author} {\bibfnamefont {C.}~\bibnamefont {Price}},
  \bibinfo {author} {\bibfnamefont {P.}~\bibnamefont {W\"olfle}}, \ and\
  \bibinfo {author} {\bibfnamefont {N.~B.}\ \bibnamefont {Perkins}},\ }\href
  {\doibase 10.1103/PhysRevB.90.155126} {\bibfield  {journal} {\bibinfo
  {journal} {Phys. Rev. B}\ }\textbf {\bibinfo {volume} {90}},\ \bibinfo
  {pages} {155126} (\bibinfo {year} {2014})}\BibitemShut {NoStop}%
\bibitem [{\citenamefont {Klein}(1974)}]{Klein1974}%
  \BibitemOpen
  \bibfield  {author} {\bibinfo {author} {\bibfnamefont {D.~J.}\ \bibnamefont
  {Klein}},\ }\href {\doibase 10.1063/1.1682018} {\bibfield  {journal}
  {\bibinfo  {journal} {The Journal of Chemical Physics}\ }\textbf {\bibinfo
  {volume} {61}},\ \bibinfo {pages} {786} (\bibinfo {year} {1974})},\ \Eprint
  {http://arxiv.org/abs/https://doi.org/10.1063/1.1682018}
  {https://doi.org/10.1063/1.1682018} \BibitemShut {NoStop}%
\bibitem [{\citenamefont {Ducatman}\ \emph {et~al.}(2018)\citenamefont
  {Ducatman}, \citenamefont {Rousochatzakis},\ and\ \citenamefont
  {Perkins}}]{Ducatman2018}%
  \BibitemOpen
  \bibfield  {author} {\bibinfo {author} {\bibfnamefont {S.}~\bibnamefont
  {Ducatman}}, \bibinfo {author} {\bibfnamefont {I.}~\bibnamefont
  {Rousochatzakis}}, \ and\ \bibinfo {author} {\bibfnamefont {N.~B.}\
  \bibnamefont {Perkins}},\ }\href
  {https://link.aps.org/doi/10.1103/PhysRevB.97.125125} {\bibfield  {journal}
  {\bibinfo  {journal} {Phys. Rev. B}\ }\textbf {\bibinfo {volume} {97}},\
  \bibinfo {pages} {125125} (\bibinfo {year} {2018})}\BibitemShut {NoStop}%
\bibitem [{\citenamefont {Rousochatzakis}\ and\ \citenamefont
  {Perkins}(2018)}]{Rousochatzakis2018}%
  \BibitemOpen
  \bibfield  {author} {\bibinfo {author} {\bibfnamefont {I.}~\bibnamefont
  {Rousochatzakis}}\ and\ \bibinfo {author} {\bibfnamefont {N.~B.}\
  \bibnamefont {Perkins}},\ }\href
  {https://link.aps.org/doi/10.1103/PhysRevB.97.174423} {\bibfield  {journal}
  {\bibinfo  {journal} {Phys. Rev. B}\ }\textbf {\bibinfo {volume} {97}},\
  \bibinfo {pages} {174423} (\bibinfo {year} {2018})}\BibitemShut {NoStop}%
\bibitem [{\citenamefont {Li}\ \emph {et~al.}(2020{\natexlab{a}})\citenamefont
  {Li}, \citenamefont {Rousochatzakis},\ and\ \citenamefont
  {Perkins}}]{Li2019}%
  \BibitemOpen
  \bibfield  {author} {\bibinfo {author} {\bibfnamefont {M.}~\bibnamefont
  {Li}}, \bibinfo {author} {\bibfnamefont {I.}~\bibnamefont {Rousochatzakis}},
  \ and\ \bibinfo {author} {\bibfnamefont {N.~B.}\ \bibnamefont {Perkins}},\
  }\href {\doibase 10.1103/PhysRevResearch.2.013065} {\bibfield  {journal}
  {\bibinfo  {journal} {Phys. Rev. Research}\ }\textbf {\bibinfo {volume}
  {2}},\ \bibinfo {pages} {013065} (\bibinfo {year}
  {2020}{\natexlab{a}})}\BibitemShut {NoStop}%
\bibitem [{\citenamefont {Li}\ \emph {et~al.}(2020{\natexlab{b}})\citenamefont
  {Li}, \citenamefont {Rousochatzakis},\ and\ \citenamefont
  {Perkins}}]{Li2020}%
  \BibitemOpen
  \bibfield  {author} {\bibinfo {author} {\bibfnamefont {M.}~\bibnamefont
  {Li}}, \bibinfo {author} {\bibfnamefont {I.}~\bibnamefont {Rousochatzakis}},
  \ and\ \bibinfo {author} {\bibfnamefont {N.~B.}\ \bibnamefont {Perkins}},\
  }\href {\doibase 10.1103/PhysRevResearch.2.033328} {\bibfield  {journal}
  {\bibinfo  {journal} {Phys. Rev. Research}\ }\textbf {\bibinfo {volume}
  {2}},\ \bibinfo {pages} {033328} (\bibinfo {year}
  {2020}{\natexlab{b}})}\BibitemShut {NoStop}%
\bibitem [{\citenamefont {Ruiz}\ \emph {et~al.}(2020)\citenamefont {Ruiz},
  \citenamefont {Nagarajan}, \citenamefont {Vranas}, \citenamefont {Lopez},
  \citenamefont {McCandless}, \citenamefont {Kimchi}, \citenamefont {Chan},
  \citenamefont {Breznay}, \citenamefont {Fra\~n\'o}, \citenamefont
  {Frandsen},\ and\ \citenamefont {Analytis}}]{Ruiz2020}%
  \BibitemOpen
  \bibfield  {author} {\bibinfo {author} {\bibfnamefont {A.}~\bibnamefont
  {Ruiz}}, \bibinfo {author} {\bibfnamefont {V.}~\bibnamefont {Nagarajan}},
  \bibinfo {author} {\bibfnamefont {M.}~\bibnamefont {Vranas}}, \bibinfo
  {author} {\bibfnamefont {G.}~\bibnamefont {Lopez}}, \bibinfo {author}
  {\bibfnamefont {G.~T.}\ \bibnamefont {McCandless}}, \bibinfo {author}
  {\bibfnamefont {I.}~\bibnamefont {Kimchi}}, \bibinfo {author} {\bibfnamefont
  {J.~Y.}\ \bibnamefont {Chan}}, \bibinfo {author} {\bibfnamefont {N.~P.}\
  \bibnamefont {Breznay}}, \bibinfo {author} {\bibfnamefont {A.}~\bibnamefont
  {Fra\~n\'o}}, \bibinfo {author} {\bibfnamefont {B.~A.}\ \bibnamefont
  {Frandsen}}, \ and\ \bibinfo {author} {\bibfnamefont {J.~G.}\ \bibnamefont
  {Analytis}},\ }\href {\doibase 10.1103/PhysRevB.101.075112} {\bibfield
  {journal} {\bibinfo  {journal} {Phys. Rev. B}\ }\textbf {\bibinfo {volume}
  {101}},\ \bibinfo {pages} {075112} (\bibinfo {year} {2020})}\BibitemShut
  {NoStop}%
\bibitem [{\citenamefont {Kim}\ \emph {et~al.}(2015)\citenamefont {Kim},
  \citenamefont {Lee},\ and\ \citenamefont {Kim}}]{Kim2015}%
  \BibitemOpen
  \bibfield  {author} {\bibinfo {author} {\bibfnamefont {H.-S.}\ \bibnamefont
  {Kim}}, \bibinfo {author} {\bibfnamefont {E.~K.-H.}\ \bibnamefont {Lee}}, \
  and\ \bibinfo {author} {\bibfnamefont {Y.~B.}\ \bibnamefont {Kim}},\ }\href
  {\doibase 10.1209/0295-5075/112/67004} {\bibfield  {journal} {\bibinfo
  {journal} {{EPL} (Europhysics Letters)}\ }\textbf {\bibinfo {volume} {112}},\
  \bibinfo {pages} {67004} (\bibinfo {year} {2015})}\BibitemShut {NoStop}%
\bibitem [{Note2()}]{Note2}%
  \BibitemOpen
  \bibinfo {note} {The zero-field ground state of $\beta $-$\protect \text
  {Li}_2\protect \text {IrO}_3$~\cite {Biffin2014a,Ruiz2017,Majumder2019}
  breaks some of the symmetries of the lattice~\cite {Ducatman2018,Li2019}, but
  the point group of this state is isomorphic to $D_{2h}$ so we can still use
  this group for the analysis of the Raman scattering channels.}\BibitemShut
  {Stop}%
\bibitem [{\citenamefont {Perreault}(2016)}]{Brentthesis}%
  \BibitemOpen
  \bibinfo {editor} {\bibfnamefont {B.}~\bibnamefont {Perreault}},\ ed.,\
  \href@noop {} {\emph {\bibinfo {title} {{Thesis: Identifying a Kitaev Spin
  Liquid}}}}\ (\bibinfo  {publisher} {UMN},\ \bibinfo {year}
  {2016})\BibitemShut {NoStop}%
\bibitem [{Note3()}]{Note3}%
  \BibitemOpen
  \bibinfo {note} {This peak has, in fact, been observed experimentally \cite
  {Yipingunpublished} and will be discussed elsewhere.}\BibitemShut {Stop}%
\bibitem [{\citenamefont {Natori}\ \emph {et~al.}(2019)\citenamefont {Natori},
  \citenamefont {Moessner},\ and\ \citenamefont {Knolle}}]{Natori2019}%
  \BibitemOpen
  \bibfield  {author} {\bibinfo {author} {\bibfnamefont {W.~M.~H.}\
  \bibnamefont {Natori}}, \bibinfo {author} {\bibfnamefont {R.}~\bibnamefont
  {Moessner}}, \ and\ \bibinfo {author} {\bibfnamefont {J.}~\bibnamefont
  {Knolle}},\ }\href {\doibase 10.1103/PhysRevB.100.144403} {\bibfield
  {journal} {\bibinfo  {journal} {Phys. Rev. B}\ }\textbf {\bibinfo {volume}
  {100}},\ \bibinfo {pages} {144403} (\bibinfo {year} {2019})}\BibitemShut
  {NoStop}%
\bibitem [{\citenamefont {Arakawa}\ and\ \citenamefont
  {Yonemitsu}(2021)}]{Arakawa2021}%
  \BibitemOpen
  \bibfield  {author} {\bibinfo {author} {\bibfnamefont {N.}~\bibnamefont
  {Arakawa}}\ and\ \bibinfo {author} {\bibfnamefont {K.}~\bibnamefont
  {Yonemitsu}},\ }\href {\doibase 10.1103/PhysRevB.103.L100408} {\bibfield
  {journal} {\bibinfo  {journal} {Phys. Rev. B}\ }\textbf {\bibinfo {volume}
  {103}},\ \bibinfo {pages} {L100408} (\bibinfo {year} {2021})}\BibitemShut
  {NoStop}%
\bibitem [{\citenamefont {Winter}\ \emph {et~al.}(2016)\citenamefont {Winter},
  \citenamefont {Li}, \citenamefont {Jeschke},\ and\ \citenamefont
  {Valent\'{\i}}}]{Winter2016}%
  \BibitemOpen
  \bibfield  {author} {\bibinfo {author} {\bibfnamefont {S.~M.}\ \bibnamefont
  {Winter}}, \bibinfo {author} {\bibfnamefont {Y.}~\bibnamefont {Li}}, \bibinfo
  {author} {\bibfnamefont {H.~O.}\ \bibnamefont {Jeschke}}, \ and\ \bibinfo
  {author} {\bibfnamefont {R.}~\bibnamefont {Valent\'{\i}}},\ }\href {\doibase
  10.1103/PhysRevB.93.214431} {\bibfield  {journal} {\bibinfo  {journal} {Phys.
  Rev. B}\ }\textbf {\bibinfo {volume} {93}},\ \bibinfo {pages} {214431}
  (\bibinfo {year} {2016})}\BibitemShut {NoStop}%
\bibitem [{\citenamefont {Yushankhai}\ and\ \citenamefont
  {Hayn}(1999)}]{Yushankhai1999}%
  \BibitemOpen
  \bibfield  {author} {\bibinfo {author} {\bibfnamefont {V.~Y.}\ \bibnamefont
  {Yushankhai}}\ and\ \bibinfo {author} {\bibfnamefont {R.}~\bibnamefont
  {Hayn}},\ }\href {https://doi.org/10.1209/epl/i1999-00360-9} {\bibfield
  {journal} {\bibinfo  {journal} {Europhys. Lett.}\ }\textbf {\bibinfo {volume}
  {47}},\ \bibinfo {pages} {116} (\bibinfo {year} {1999})}\BibitemShut
  {NoStop}%
\end{thebibliography}%

\end{document}